\documentclass[iop,apj,numberedappendix,appendixfloats,twocolappendix]{emulateapj}
\usepackage{mathtools}
\usepackage{graphicx}
\usepackage{physics}
\usepackage{xcolor}
\usepackage{hyperref}

\newcommand{\fid}{H1}
\newcommand{\ampup}{H2}
\newcommand{\ampdown}{H3}
\newcommand{\normup}{H4}
\newcommand{\comp}{H5}
\newcommand{\hm}{H7}
\newcommand{\uvb}{H6}

\newcommand{\HI}{H~\textsc{i} }

\newcommand{\HeII}{He~\textsc{ii} }
\newcommand{\HeIII}{He~\textsc{iii} }

\begin{document}

\shorttitle{Helium Reionization Simulations II}
\shortauthors{La Plante et al.}

\title{Helium Reionization Simulations. II. Signatures of Quasar Activity on the IGM}
\author{Paul La Plante\altaffilmark{1,2}\altaffilmark{*},
  Hy Trac\altaffilmark{1},
  Rupert Croft\altaffilmark{1},
  and Renyue Cen\altaffilmark{3}}
\altaffiltext{1}{McWilliams Center for Cosmology,
  Department of Physics,
  Carnegie Mellon University,
  Pittsburgh, PA 15213, USA}
\altaffiltext{2}{Center for Particle Cosmology,
  Department of Physics and Astronomy,
  University of Pennsylvania,
  Philadelphia, PA 19104, USA}
\altaffiltext{3}{Department of Astrophysical Science,
  Princeton University,
  Princeton NJ 08544, USA}
\altaffiltext{*}{plaplant@andrew.cmu.edu}

\begin{abstract}
  We have run a new suite of simulations that solve hydrodynamics and radiative
  transfer simultaneously to study helium~\textsc{ii} reionization. Our suite of
  simulations employs various models for populating quasars inside of dark
  matter halos, which affect the \HeII\ reionization history. In particular, we
  are able to explore the impact that differences in the timing and duration of
  reionization have on observables. We examine the thermal signature that
  reionization leaves on the IGM, and measure the temperature-density
  relation. As previous studies have shown, we confirm that the photoheating
  feedback from helium~\textsc{ii} reionization raises the temperature of the
  intergalactic medium (IGM) by several thousand kelvin. To compare against
  observations, we generate synthetic Ly$\alpha$ forest sightlines
  on-the-fly and match the observed effective optical depth
  $\tau_{\mathrm{eff}}(z)$ of hydrogen to recent observations. We show that when
  the simulations have been normalized to have the same values of
  $\tau_\mathrm{eff}$, the effect that helium~\textsc{ii} reionization has on
  observations of the hydrogen Ly$\alpha$ forest is minimal. Specifically,
  the flux PDF and the one-dimensional power spectrum are sensitive to the
  thermal state of the IGM, but do not show direct evidence for the ionization
  state of helium. We show that the peak temperature of the IGM typically
  corresponds to the time of 90-95\% helium ionization by volume, and is a
  relatively robust indicator of the timing of reionization. Future observations
  of helium reionization from the hydrogen Ly$\alpha$ forest should thus
  focus on measuring the temperature of the IGM, especially at mean
  density. Detecting the peak in the IGM temperature would provide valuable
  information about the timing of the end of helium~\textsc{ii} reionization.
\end{abstract}


\section{Introduction}
\addtocounter{footnote}{-1} 
\addtocounter{Hfootnote}{-1}
\label{sec:intro}

Helium~\textsc{ii} reionization is a fascinating portion of the Universe's
history and is the last major phase change of the intergalactic medium
(IGM). After hydrogen reionization at high redshift ($z \gtrsim 6$) from the
first stars and galaxies, helium was singly ionized. However, the second
ionization of helium requires significantly more energy (54.4 eV vs. 24.6 eV for
the first ionization). The stars providing photons for hydrogen reionization did
not emit a significant number of these high-energy photons. Thus, helium was not
doubly ionized until later in the Universe's evolution, when quasars produced
enough high-energy photons to significantly change the ionization level of
helium. Following the formation of quasars at redshifts $6 \geq z \geq 2$, the
helium of the IGM became totally ionized, leaving an imprint on the IGM.

The process of helium~\textsc{ii} reionization leaves important observational
signatures on the Ly$\alpha$ forest, which is a measure of the relative
amount of photon absorption due to gas in the IGM. The Ly$\alpha$ forest can
be observed most readily for neutral hydrogen and has been observed at medium
resolution (\textit{e.g.}, the Baryon Oscillation Spectroscopic Survey, BOSS,
\citealt{mcdonald_etal2006,lee_etal2015}) and high resolution (\textit{e.g.},
Keck-HIRES and Magellan-MIKE, \citealt{lu_etal1996,viel_etal2013}). To date,
there have been more than 150,000 Ly$\alpha$ forest spectra measured from
BOSS alone \citep{dawson_etal2013}, and the number of systems is expected to
increase by almost an order of magnitude after the deployment of the next
generation of telescopes \citep{myers_etal2015}. This rich observational data
set contains much information about the IGM, most notably the abundance of
neutral hydrogen and its temperature.

A related measurement to the hydrogen Ly$\alpha$ forest is the analogous
feature for \HeII. However, to date, there have been only about 50 systems for
which the \HeII\ measurement has been made
\citep{syphers_etal2009a,syphers_etal2009b,syphers_etal2012}. The reason for the
comparative lack of \HeII\ measurements is due to the presence of Lyman-limit
systems (LLS), which are optically thick and lead to large absorption
features. This absorption contaminates the signal, and makes detection of \HeII\
signatures difficult \citep{moller_jakobsen1990,zheng_etal2005}. Nevertheless,
the detection of the helium analog of the Gunn-Peterson trough
\citep{gunn_peterson1965} offers an indication of when helium~\textsc{ii}
reionization ended. Recent observations have shown a Gunn-Peterson trough for
helium at redshifts $z > 3$
\citep{jakobsen_etal1994,zheng_etal2008,syphers_shull2014}, which shows the
\HeII\ volume fraction must have been greater than
$f_\mathrm{HeII} \gtrsim 10^{-3}$ along these sightlines. Helium absorption then
becomes patchy, with extended regions of absorption and transmission in the
\HeII\ Ly$\alpha$ forest \citep{reimers_etal1997}, and seems to be completed
by $z\sim2.7$ \citep{dixon_furlanetto2009,worseck_etal2011}. However, the
comparatively low number of sightlines that show the Ly$\alpha$ forest
signature for \HeII\ leaves much statistical uncertainly about the exact timing
and nature of the reionization process.

In order to better explore some of the signatures that helium~\textsc{ii}
reionization leaves on the IGM, we have run a new suite of simulations that
include simultaneously solved hydrodynamics and radiative transfer. These
simulations represent the first efforts to incorporate all of the relevant
physics together using a spatially varying radiation field sourced by quasars,
in order to better predict the impact on observations. Previous studies
typically incorporated different degrees of coupling different
schemes. Typically, radiative transfer is solved in post-processing of $N$-body
or hydrodynamic simulations (\textit{e.g.},
\citealt{mcquinn_etal2009,mcquinn_etal2011,compostella_etal2013,compostella_etal2014}),
which does not incorporate the effect of photoheating on the IGM that
accompanies reionization. Alternatively, previous studies have included
radiative transfer by using a uniform ionization background (\textit{e.g.},
\citealt{theuns_etal1998,jena_etal2005,viel_etal2013,puchwein_etal2015,bolton_etal2016}),
an approach that does not capture the large-scale inhomogeneities of the
radiation field. Notably, the study of \citet{meiksin_tittley2012} does feature
hydrodynamics and radiative transfer coupled together, though for a smaller box
size (25 Mpc $h^{-1}$) than the one discussed here. We note, however, that the
radiative transfer in these simulations was only computed on a relatively narrow
slice (about 100 kpc $h^{-1}$). Still other previous studies use semi-analytic
models to understand the contribution of quasars to the ionizing background of
the IGM at these redshifts \citep{daloisio_etal2016}, although they do not feature
all of the physics incorporated here. Thus, the simulations presented here
represent a step forward in accurately modeling the reionization process, and
capture the effects of heating from sources and the inhomogeneous and
anisotropic aspects of sources.

This work represents the second paper in a series on helium~\textsc{ii}
reionization simulations. \citealt{laplante_trac2015} (hereafter
Paper~I) outlined a method whereby dark matter halos from $N$-body simulations
are populated with quasars such that the quasar luminosity function (QLF) from
the SDSS and COSMOS surveys
\citep{masters_etal2012,mcgreer_etal2013,ross_etal2013} and the two-point
autocorrelation function from BOSS \citep{white_etal2012} are reproduced. This
ensures that our radiation sources match the latest observational constraints in
terms of their number density and topology.

We organize the rest of this paper as follows. In Section~\ref{sec:radhydro} we
discuss our simulation technique and describe the method by which we include
sources of ionization. In Section~\ref{sec:heiii} we discuss in more detail the
individual models explored here, and the differences apparent in the helium
ionization fraction. In Section~\ref{sec:igmtemp} we explore impacts of
reionization on the thermal history of the IGM. In Section~\ref{sec:lya} we
discuss generating synthetic Ly$\alpha$ sightlines from the simulations, and
compare them with recent observations. In Section~\ref{sec:conclusion} we
summarize and explore avenues for future research. Throughout this work, we
assume a $\Lambda$CDM cosmology with $\Omega_m = 0.27$, $\Omega_\Lambda = 0.73$,
$\Omega_b = 0.045$, $h = 0.7$, $\sigma_8 = 0.8$, and $Y_\mathrm{He} =
0.24$. These values are consistent with the \textit{WMAP}-9 year results
\citep{hinshaw_etal2013}.

\section{Radiation-hydrodynamic Simulations}
\label{sec:radhydro}
To faithfully capture helium~\textsc{ii} reionization, the ideal simulations
should include dark matter, baryonic matter, and radiation coupled together. The
dark matter is necessary for establishing the large-scale structure of the
Universe, and the baryonic matter captures the distribution of neutral and
ionized gas in the IGM. By coupling radiation to this gas as the simulation is
proceeding, a more accurate state of the IGM is calculated. As mentioned above,
owing to the large degree of photoheating of the IGM induced by the energetic
photons from quasars, the thermal state of the mean-density IGM is dominated by
quasars and the reionization of helium. Furthermore, the clustered nature of
quasars argues for simulations in which the radiation sources are tracked
explicitly, instead of incorporating them as a uniform background. Thus, these
simulations are able to capture many of the features important to
helium~\textsc{ii} reionization, and generate predictions that can be readily
compared with observations.

\subsection{Populating Simulations with Quasars}
\label{sec:quasar_sources}
The simulations presented here have been run using the RadHydro code, which
includes $N$-body, hydrodynamics, and radiative transfer calculations. The code
employs a particle mesh (PM) solver for gravity calculations, a fixed-grid
Eulerian code for solving hydrodynamics, and a ray-tracing scheme for computing
radiative transfer. The radiative transfer calculations use a non-equilibrium
solver for the photoionization balance equations, and use many time steps per
hydro step to ensure accurate calculation of the thermal state. The code has
been used to study hydrogen reionization
\citep{trac_cen2007,trac_etal2008,battaglia_etal2013a}, and has been modified
extensively for the current application to helium~\textsc{ii} reionization.

Our simulation strategy is as follows. As a result of the requirement of a large
box size to capture relatively rare objects, the simulation does not resolve the
galaxy-scale physics (and by extension, quasar-scale physics). It is therefore
necessary to populate the volume with sources using an alternative method. To
this end, we perform the simulation in two steps: a first pass to generate a
catalog of quasar sources, and a second pass that uses the sources to perform
full reionization simulations. We first run a P$^3$M $N$-body simulation
including only dark matter \citep{trac_etal2015}. Initial conditions for these
simulations are generated at $z = 150$ using transfer functions generated by
CAMB \citep{lewis_etal2000}. These $N$-body simulations are run at high
resolution, where for our fiducial simulations we use a simulation volume of
size $L=200$ $h^{-1}$Mpc with 2048$^3$ particles. This yields a particle mass of
$m_p = 6.98 \times 10^7$ $h^{-1}M_\odot$. Halo-finding is done on-the-fly using
a friend-of-friends (FoF) algorithm with mean inter-particle spacing of $b=0.08$
to find halo members. This value avoids the overbridging problem in standard
FoF with $b=0.2$. The halo finder is used to locate all halos with 50 or more
members. Once the FoF halos are found, a spherical overdensity algorithm is
used to create a corresponding halo catalog. These halo catalogs are produced
every 20 Myr in cosmological time while the simulation is running. The halos
from the catalogs are then treated as candidate hosts for the quasars to be used
in our simulations.

With the halo catalogs from the high-resolution simulation in hand, the halos
can be populated with quasars, the sources of helium-ionizing
radiation. Following the procedure outlined in Paper~I, we populate these halos
with quasars that reproduce the observed QLF and clustering
measurements. Briefly, the model uses the technique of abundance matching in
order to populate potential quasar hosts (\textit{i.e.}, dark matter halos) with
quasars in order to reproduce a specified QLF. We should mention that abundance
matching is not the only method by which dark matter halos can be populated with
quasars, and alternative methods exist.  See
\citet{cen_safarzadeh2015a,cen_safarzadeh2015b,cen_safarzadeh2016} for
alternative methods of populating halos with quasars, and discussion of
observables related to the clustering, quasar lifetimes, and the tSZ effect. The
method allows the user to specify the QLF to use, and either a lightbulb or
exponential model for the quasar light curve. By construction, the method will
reproduce the desired QLF at all redshifts (starting at $z \sim 6$, the earliest
redshift at which we include quasar sources), provided the quasar lifetime (and
time between halo catalog snapshots) is small compared to the Hubble time. The
fiducial QLF used in the work presented here combines the results of several
different luminosity functions at different redshifts: at high redshift
($z \gtrsim 5$), the QLF reproduces the observations of
\citet{mcgreer_etal2013}. At intermediate redshift ($z \sim 4$), the QLF
reproduces the observations of \citet{masters_etal2012}. At lower redshift
($z \lesssim 3.5$), the QLF parameters used are those from
\citet{ross_etal2013}. Combining the measurements of the QLF at multiple epochs
ensures that the number density sources of helium-ionizing radiation found in
the simulations are observationally accurate. Since the timing of reionization
is determined by a large part by the abundance of sources, having an
observationally accurate quasar number density is of the utmost importance. The
simulations run here use two slightly different methods for combining the
different measurements, which we call Q1 and Q2. See Appendix~\ref{appendix:qlf}
for further discussion on the details of the QLF used in these simulations.

In addition to matching the number density of quasar sources, the method of
Paper~I also matches the observed clustering of quasars. Using the abundance
matching technique leaves the lifetime of quasars unconstrained, which affects
the bias of quasars. Reproducing the bias of quasars ensures that simulations
reproduce the topology of reionization: although the number of sources is fixed
by the QLF, the clustering of quasars will affect the size and shape of ionized
regions. In general, since quasars are known to be highly biased
\citep{white_etal2012}, they are found to be strongly clustered, which leads to
early overlap of doubly ionized regions \citep{mcquinn_etal2009}. In Paper~I, we
use a suite of $N$-body simulations to study how the lifetime of quasars affects
their clustering. We identify a set of parameters that reproduce the clustering
as measured in \citet{white_etal2012} at redshift $z \sim 2.4$. The model
developed in Paper~I allows for the lifetime of quasars $t_q$ to change as a
function of luminosity following a power-law relation, parameterized as
$t_q(L) = t_0(L/L_0)^\gamma$, where $L$ is the peak luminosity of the quasar,
and $t_0$ and $\gamma$ are two parameters allowed to vary. Unless otherwise
noted, the models discussed in these simulations used an exponential light
curve, with $\gamma=-0.1$. As discussed below, in instances where the QLF is
modified to explore a different reionization history, the quasar lifetime $t_0$
is modified to match the clustering measurements.

\subsection{Quasar Properties}
\label{sec:quasar_properties}
For individual quasar objects, there are two components of the spectral energy
distribution (SED) that must be specified: the normalization, and the spectral
index. The QLF is typically reported in terms of magnitude, rather than
luminosity. Specifically, the convention used when reporting the QLF in
\citet{ross_etal2013} is to use the absolute $i$-band magnitude at $z=2$. In
order to determine the energy output of a quasar, we convert from magnitude into
luminosity using Equation~(4) of \citet{richards_etal2006}:
\begin{multline}
\log_{10}\qty(\frac{L_{2500\ \mathrm{\AA}}}{4\pi d^2}) = \\
 -0.4\qty[M_i(z=2) + 48.60 + 2.5\log_{10}(1+2)], \label{eqn:lofm}
\end{multline}
where $d=10$ pc $= 3.08 \times 10^{19}$ cm. This formula converts the magnitude
of the QLF into a specific luminosity at 2500 \AA. Once this specific luminosity
has been found, the specific luminosity in the extreme ultraviolet (EUV) region
must be calculated to determine the output of radiation relevant to helium
reionization. For the purposes of this calculation, we use the quasar SED
template of \citet{lusso_etal2015}. This template assumes a power-law form for
the SED with a spectral index of $\alpha = 0.61$ ($f_\nu \propto \nu^{-\alpha}$)
for $\lambda \geq 912$ \AA\ and $\alpha = 1.7$ for shorter wavelengths. The
number of photons is then computed in seven different frequency bins for the
radiative transfer calculation, spanning photon energies from $h \nu = 13.6$ eV
to 1 keV (see Appendix~\ref{sec:nfreq} for further discussion). At energies
higher than this, the mean free path of photons interacting with singly ionized
helium becomes comparable to the Hubble scale, and as a practical matter, much
larger than the box size of the simulation.

As discussed in Paper~I, there is a moderate degree of uncertainty in the
systematic effects of the quasar population. For instance, reddening of quasars
due to dust, obscured quasars, contamination of non-quasar objects in
photometric surveys, and poor knowledge of the intrinsic colors of quasars could
all systematically shift the normalization of the QLF. In order to marginalize
over some of this uncertainty, we have conducted several simulations with the
same underlying gas distribution and large-scale structure, but with different
quasar populations. Specifically, we modify the normalization of the QLF and the
normalization of the SED. These different simulations allow us to explore some
of the effect that these systematic uncertainties generate, and how they might
impact different observations of the IGM. We further discuss all of the models
explored below in Section~\ref{sec:models}.

\subsection{Simulation Features}
\label{sec:features}
Although the main focus of this study is to understand the impact of helium
reionization, an accurate treatment of hydrogen reionization is nevertheless
important. In some sense, the initial conditions of helium~\textsc{ii}
reionization (especially with respect to the temperature of the IGM) are set by
the timing of hydrogen reionization and the inside-out nature of denser regions
undergoing reionization earlier than less dense ones.

In order to capture the inhomogeneous effects that hydrogen reionization has on
the IGM, the method of ``patchy reionization'' developed in
\citet{battaglia_etal2013a} is applied to the simulation volume, which predicts
a redshift of reionization based on the density field from a dark-matter-only
simulation. A mean redshift of reionization $z_\mathrm{re} = 8$ was used for
these simulations, with the fiducial values for the other parameters in the
model that control the duration of reionization. The application of this method
better captures the thermal state of the IGM following hydrogen reionization
than using a uniform radiation background.

The radiative transfer is calculated using explicit ray tracing of photons from
quasars, using the scheme described in \citet{trac_etal2008}. However, tracking
rays from galaxies in addition to those from quasars would be prohibitively
expensive. The stellar content of galaxies does not produce an appreciable
number of photons with $h\nu > 54.4$ eV, and they are thus largely unimportant
for helium~\textsc{ii} reionization \citep{furlanetto_oh2008a}. However,
galaxies do produce photons that contribute to hydrogen ionization. The
ionization balance equation for hydrogen can be written as
\begin{equation}
\dv{n_\mathrm{HI}}{t} = - \Gamma_\mathrm{tot} n_\mathrm{HI} + \alpha_\mathrm{HII} n_\mathrm{HII} n_e,
\label{eqn:gamma_gal}
\end{equation}
where $\Gamma_\mathrm{tot}$ is the total photoionization rate per atom in
s$^{-1}$, $\alpha_\mathrm{HII}$ is the recombination coefficient, and $n_i$ is
the comoving number density of species $i$. For the case of hydrogen, there are
contributions from both quasars and galaxies, which can be expressed as
$\Gamma_\mathrm{tot} = \Gamma_\mathrm{qso} + \Gamma_\mathrm{gal}$. The
computation of $\Gamma_\mathrm{qso}$ is computed explicitly via ray tracing, but
the value of $\Gamma_\mathrm{gal}$ must be specified. The photoheating rates are
computed for each frequency bin based on the photoionization rates. Cooling
rates are included for recombination, collisional ionization and excitation,
free-free interactions, and inverse Compton processes. Additional heating from
supernova feedback is added for the highest density cells ($\Delta \geq
200$). The feedback is added purely as thermal energy rather than as thermal and
kinetic, and so this feedback may be underestimated
\citep{kimm_cen2014}. However, since this affects only the high-density cells
and not the bulk of the volume relevant for the observables discussed later,
this difference is not significant for the results. For the purposes of running
the simulation, the value of $\Gamma_\mathrm{gal}$ is assumed to be a uniform
value. For late times ($z \lesssim 6$), the hydrogen in the IGM is highly
ionized and hence optically thin, and so treating the UV background as uniform
is a valid approximation. One approach is to use a value based on a
semi-analytic model (\textit{e.g.}, \citealt{haardt_madau2012}, hereafter
HM12). However, this approach relies on the specifics of the model chosen and
does not account for other details in the simulation (such as the quasar
contribution to hydrogen ionization, patchy hydrogen reionization, etc.).

In order to circumvent some of these issues, we choose to set the value of
$\Gamma_\mathrm{gal}$ to match the observed effective optical depth
$\tau_\mathrm{eff}$ measured by \citet{lee_etal2015}. This evolution of
$\tau_\mathrm{eff}$ is based primarily on measurements from SDSS DR7, presented
by \citet{becker_etal2013}. We generate Ly$\alpha$ sightlines on-the-fly
while the simulation is running, and modify the value of $\Gamma_\mathrm{gal}$
in order to match $\tau_\mathrm{eff}(z)$.  Instead of generating the full number
of sightlines available to us ($N_\mathrm{grid}^2$), we reduce the number of
sightlines drawn by a factor of four in each dimension for a total of
$N_\mathrm{grid}^2/16$. In comparisons performed between using the full sample
and this reduced subset, we did not find significant differences in the
calculated value of $\tau_\mathrm{eff}$, and therefore inferred the same target
value of $\Gamma_\mathrm{gal}$.
By matching the value of $\tau_\mathrm{eff}$ by construction, we are better able
to compare between simulations and against observation. This also avoids
renormalizing the Ly$\alpha$ forest in post-processing, which is the usual
approach taken in simulations comparing against the Ly$\alpha$ forest
(\textit{e.g.}, \citealt{bolton_etal2009b}). In other words,
$\Gamma_\mathrm{gal}$ becomes a free parameter that we adjust at every time step
in the simulation in order to match the value of $\tau_\mathrm{eff}$ specified
by \citet{lee_etal2015}, such that $\Gamma_\mathrm{gal} + \Gamma_\mathrm{qso}$
reproduces the proper optical depth.

Below in Section~\ref{sec:models}, we discuss the simulations performed in our
simulation suite. Some of the models have an increased number of photons
produced by quasars, above the fiducial values assumed by the quasar properties
as discussed in Section~\ref{sec:quasar_properties}. For these models with an
increased number of photons, the contribution of $\Gamma_\mathrm{qso}$ is large
enough that even if $\Gamma_\mathrm{gal} = 0$, the IGM becomes too highly
ionized, and the value of $\tau_\mathrm{eff}$ is lower than that of
\citet{lee_etal2015}. Accordingly, it becomes impossible to match the value of
$\tau_\mathrm{eff}$ because of the increased radiation output of quasars.

Given the fact that $\tau_\mathrm{eff}$ from simulations is lower than that of
\citet{lee_etal2015}, the value of $\Gamma_\mathrm{tot}$ must be decreased in
order to match the target value. As stated above, the radiation from quasars is
more than sufficient to match the value of $\tau_\mathrm{eff}$, so the value of
$\Gamma_\mathrm{qso}$ must be decreased. Therefore, it becomes necessary to
choose a minimum value of $\Gamma_\mathrm{gal}$, below which the radiation
output of quasars must be decreased to agree with observations. We choose to
have a finite value of $\Gamma_\mathrm{gal}$ for these simulations, since the
stellar output of galaxies still provide a contribution to the hydrogen
ionization level at these redshifts. Most models
\citep{haardt_madau1996,haardt_madau2012} or measurements that infer this value
\citep{becker_etal2007,bolton_haehnelt2007,faucher-giguere_etal2008a,becker_bolton2013}
of the UV background at these redshifts have a contribution from galaxies of
$10^{-13}$ s$^{-1} \lesssim \Gamma_\mathrm{gal} \lesssim 10^{-12}$
s$^{-1}$.

\begin{deluxetable*}{ccccccccccc}
  \tablecaption{List of the parameters of the simulations presented in this
    work. \label{table:sim-summary}} \tablewidth{0pt}
  \tablehead{\colhead{Simulation} & \colhead{Box Size\tablenotemark{a}} &
    \colhead{$N_\mathrm{grid}$} & \colhead{$z_\mathrm{50}$\tablenotemark{b}} &
    \colhead{$z_\mathrm{99}$} & \colhead{$\Delta z_{50}$\tablenotemark{c}} &
    \colhead{$\Delta z_{90}$} & \colhead{Quasar Model\tablenotemark{d}} &
    \colhead{$t_0$\tablenotemark{e}} & \colhead{QLF Amplitude} & \colhead{SED
      Amplitude}} \startdata
  \fid          & 200 & 2048$^3$ & 3.34 & 2.69 & 0.80 & 2.31 & Q1  & 30.9 & 1    & 1   \\
  \ampup        & 200 & 2048$^3$ & 3.96 & 2.73 & 0.90 & 2.73 & Q1  & 40   & 2    & 1   \\
  \ampdown      & 200 & 2048$^3$ & 2.96 & 2.23 & 0.79 & 2.71 & Q1  & 20   & 0.5  & 1   \\
  \normup       & 200 & 2048$^3$ & 4.22 & 2.71 & 1.83 & 2.92 & Q1  & 30.9 & 1    & 2   \\
  \comp         & 200 & 2048$^3$ & 3.65 & 2.84 & 1.06 & 2.25 & Q2  & 30   & 1.67 & 1.5 \\
  \uvb          & 200 & 2048$^3$ & 4.14 & 3.16 & 0.58 & 1.51 & UVB & $\cdot\cdot\cdot$  & $\cdot\cdot\cdot$ & $\cdot\cdot\cdot$ \\
  \enddata
  \tablenotetext{a}{In comoving $h^{-1}$Mpc}
  \tablenotetext{b}{Redshift when
    $x_\mathrm{HeIII} = 0.50$ (defined in Equation~(\ref{eqn:xfrac})) or
    $x_\mathrm{HeIII} = 0.99$ by volume}
  \tablenotetext{c}{Duration in redshift of
    the central 50\% change in ionization fraction (defined in
    Equation~(\ref{eqn:z50}))}
  \tablenotetext{d}{See Appendix~\ref{appendix:qlf} for
    the differences between quasar models Q1 and Q2} \tablenotetext{e}{$t_0$ as
    defined in Paper~I, measured in Myr}
\end{deluxetable*}

Following the models and measurements, we require for our simulations that
$\Gamma_\mathrm{gal} \geq 10^{-13}$ s$^{-1}$. If $\tau_\mathrm{eff}$ is still
too low given this minimum value of $\Gamma_\mathrm{gal}$, the value of
$\Gamma_\mathrm{qso}$ must be decreased. Because this value is only set
indirectly by the number of photons produced by quasars in the ray-tracing
scheme, the total output of radiation from quasars is decreased to match
$\tau_\mathrm{eff}$. This approach ensures that all of the simulations match the
measured value of \citet{lee_etal2015}. As the simulation progresses, if the
ionization level needs to be increased to match the desired value, then the
photon production of quasars in increased back to its default value before
increasing $\Gamma_\mathrm{gal}$. Further details of the renormalization process
can be found in Appendix~\ref{appendix:gamma_gal}.

This approach of modifying the value of $\Gamma_\mathrm{gal}$ on-the-fly to
match the values of $\tau_\mathrm{eff}$ is, to our knowledge, unique to the
simulations presented here. In addition to facilitating the comparison between
the simulations and observations, this approach has several other benefits. For
instance, by ensuring that we have the proper thermal state of the IGM, the
pressure smoothing of the gas is more accurate. This property has implications
for measurements related to the Ly$\alpha$ forest, discussed more fully in
Section~\ref{sec:lya}. Additionally, observations that depend on the value of
$\tau_\mathrm{eff}$ are true apples-to-apples comparisons, and isolate the
effect of the differences in the timing of helium~\textsc{ii}
reionization. Thus, this suite of reionization simulations allows for a
straightforward determination of effects directly attributable to quasar
activity as it pertains to helium reionization.

Table~\ref{table:sim-summary} summarizes the properties of the simulations
examined in this paper. All simulations are conducted with a box size of $L=200$
$h^{-1}$Mpc, which is large enough to include a number of high-luminosity
quasars that are important for helium~\textsc{ii} reionization. Our default
resolution for the gas grid uses $N_\mathrm{g}=2048^3$ resolution elements. For
dark matter, we use $N_\mathrm{dm} = 2048^3$ particles as well. The grid on
which the equations of radiative transfer are solved is coarser by a factor of
4, \textit{i.e.}, $N_\mathrm{rt} = N_\mathrm{g}/64$. For all of the simulations
in the suite, the same initial conditions for the dark matter particles and the
gas cells are used, so that the only difference is the helium~\textsc{ii}
reionization history sourced by quasars. This allows us to isolate the impact
that varying helium~\textsc{ii} reionization has on measurements from our
simulations, since the gas and matter distributions are largely the
same. Indeed, the power spectra for dark matter in the simulations is
effectively identical in all of the simulations, and the gas power spectra only
show differences on small scales ($k~\gtrsim~10$~Mpc$^{-1}$~$h$).

\subsection{Details of the Simulation Suite}
\label{sec:models}
We now discuss in detail some of the differences between the various simulations
run. All of the simulations use the same set of initial conditions for dark
matter and baryons, and the halo catalogs from the corresponding $N$-body
simulation are therefore the same. (See Section~\ref{sec:quasar_sources} for
more information.) Furthermore, all of the simulations use the patchy hydrogen
reionization discussed in Section~\ref{sec:features} at high redshift before
helium~\textsc{ii} reionization. The one exception to this is the simulation
that uses a uniform UV background, Simulation \uvb, which uses the
photoionization and photoheating rates from HM12. Additionally, also as
discussed in Section~\ref{sec:features}, the simulations feature a dynamic
renormalization of $\Gamma_\mathrm{gal}$ to match the reported value of
$\tau_\mathrm{HI}$ as provided by \citet{lee_etal2015}. This renormalization
applies to almost all of the simulations, including \uvb, where all of the
photoheating and photoionization rates are scaled to match
$\tau_\mathrm{eff}$. As a point of comparison, we have run an additional
simulation that purposely does not match the functional form of
$\tau_\mathrm{eff}$ in order to test for features that may appear as the result
of helium~\textsc{ii} reionization. We will discuss this simulation further in
Appendix~\ref{appendix:hm}.

\begin{enumerate}
\item The simulation \fid\ is one which uses a QLF that is generated in the
  manner discussed in Section~\ref{sec:quasar_sources}. In general, the
  amplitude of the QLF is low at early times, but has a relatively steep
  low-luminosity slope. This leads to a quasar population that features a large
  number of low-luminosity objects. Since the effective lifetime of quasars is
  generally proportional to their luminosity, these sources are also relatively
  short-lived. As the Universe evolves, the amplitude of the QLF becomes
  greater, and the faint-end slope becomes shallower. This leads to a similar
  number of objects overall, but with larger, more luminous sources being the
  primary drivers of reionization. As we show in Section~\ref{sec:heiii}, large
  objects also tend to have larger regions of doubly ionized helium, since the
  longer lifetimes lead to larger reionization regions. This evolution becomes
  clear when visualizing the reionization process (see Figure~\ref{fig:panel2}).
\item As mentioned in Section~\ref{sec:intro}, there is some uncertainty in the
  overall amplitude of the QLF. In order to explore this uncertainty, we have
  run simulations \ampup\ and \ampdown, which use the same input QLF as \fid,
  but with a change to the QLF amplitude. In \ampup\ the amplitude of the QLF is
  increased by a factor of 2 at all redshifts, and in \ampdown, the amplitude is
  decreased by a factor of 2. In both cases, the lifetime of quasars is modified
  in order to reproduce the quasar clustering measurements of
  \citet{white_etal2012}, as discussed in
  Section~\ref{sec:quasar_sources}. Although the statistical uncertainty of the
  QLF is lower than this amount at low redshift (\textit{i.e.}, the data from
  \citet{ross_etal2013} have errors that are better than 10\%), there are
  considerable uncertainties at high redshift. Furthermore, there are potential
  sources of systematic uncertainty (\textit{e.g.}, reddening of objects due to
  dust, obscured sources, or mischaracterization of potential sources as
  stars). By exploring changes in the amplitude of the QLF, we are better able
  to characterize the impact that different redshifts of helium~\textsc{ii}
  reionization can have on observables.
\item A separate source of uncertainty related to the quasar sources is the
  normalization of individual quasar objects given a specific luminosity. As
  explained in Section~\ref{sec:quasar_sources}, we use
  Equation~(\ref{eqn:lofm}) to convert from the observed magnitude into the
  specific luminosity at 2500 \AA\ $L_{2500}$, and the SED template of
  \citet{lusso_etal2015} to determine the EUV radiation. The statistical
  uncertainties of \citet{lusso_etal2015} are very small for the UV portion of
  the SED (wavelengths where $\lambda > 912$ \AA), although differences arise
  when comparing the spectral indices between different SEDs (\textit{e.g.},
  \citealt{richards_etal2006,hopkins_etal2007,shang_etal2011}). To explore some
  of the uncertainty associated with the SED, we have run Simulation \normup\
  with a quasar model that has the same QLF amplitude as \fid, but in which the
  photon number count has been increased by a factor of 2. This results in a
  comparable number of photons being produced as in \ampup, but with the same
  number of objects and topology as in \fid. As a result, we expect the regions
  of doubly ionized helium to be larger than those found in \fid, which would
  lead to patchier reionization. We would also expect the timing of reionization
  to be similar to \ampup.
\item As mentioned above, an additional uncertainty related to the observed QLF
  involves the method by which observations from different redshift ranges are
  incorporated into one single QLF that evolves with redshift. We present two
  alternative methods of performing this combination in
  Appendix~\ref{appendix:qlf}. We call the two models Q1 and Q2. Simulation
  \comp\ uses a method slightly different from the fiducial one of Simulation
  \fid. As with the uncertainties explored in Simulations \ampup\ and \ampdown,
  this comparison underlines the importance of accurately determining the QLF at
  all redshifts to better understand helium~\textsc{ii} reionization.  When
  creating this QLF, several of the parameters of the QLF were modified in an
  effort to better reproduce the timing of the reionization found in Simulation
  \fid.

\begin{figure}[t]
  \centering
  \includegraphics[width=0.5\textwidth]{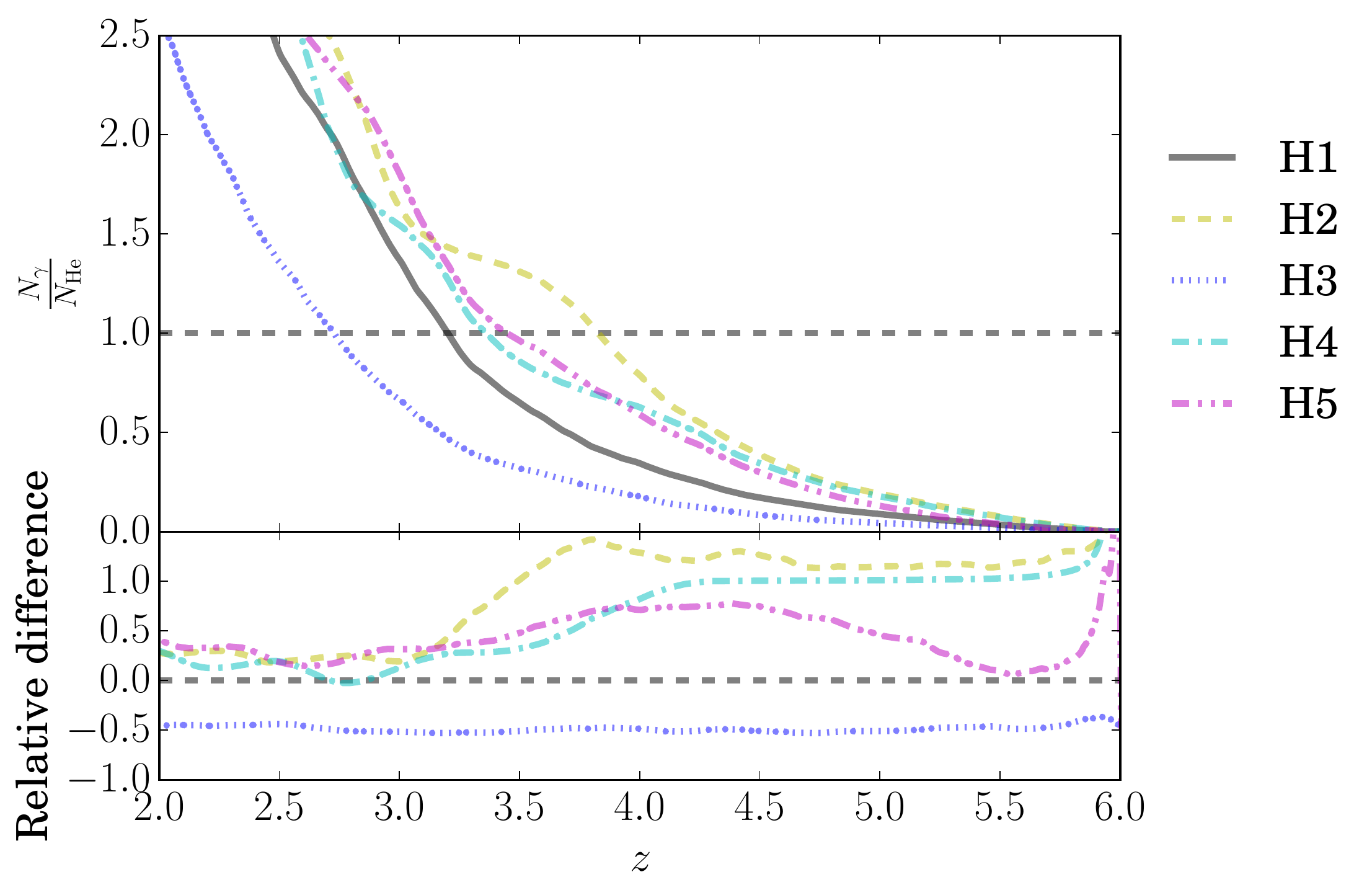}
  \caption{Comparison of the number of helium-ionizing photons ($h\nu \geq 54.4$
    eV) produced by quasars in each of the simulation models as a function of
    redshift. Top: the cumulative number of helium-ionizing photons in the
    simulation volume relative to the number of helium atoms. If all photons
    produced ionized helium with no recombinations, then helium reionization
    would be completed by the intersection with this line. Bottom: the number of
    photons produced relative to Simulation \fid. The simulations are described
    in detail in Section~\ref{sec:models}. Note that Simulation \ampup\ and
    Simulation \normup\ in principle produce a comparable number of photons as a
    function of redshift. Nevertheless, the two simulations have different
    reionization histories, as well as different reionization topologies.}
  \label{fig:photons}
\end{figure}

\item Finally, as a point of comparison, we have run a simulation that does not
  include explicit quasar sources and instead features a uniform UV
  background. The photoionization and photoheating rates are given by those in
  HM12. This allows for a comparison with other studies that employ a uniform UV
  background \citep{becker_etal2011a,puchwein_etal2015}. However, for a fair
  comparison with the other simulations presented here, we have renormalized
  these rates to match $\tau_{\mathrm{eff}}$ as outlined in
  Section~\ref{sec:features}. Although only the value of $\Gamma_\mathrm{HI}$
  affects the observed $\tau_\mathrm{eff}$, we apply the same renormalization to
  all of the photoionization and photoheating rates. Simulation \uvb\ uses this
  uniform background, and can be thought of as the limiting case of having many
  low-luminosity ($\order{10^9-10^{10}\ L_\odot}$) objects drive helium
  reionization, rather than comparatively few high-luminosity
  ($\order{10^{12}-10^{13}\ L_\odot}$) ones.
\end{enumerate}

Figure~\ref{fig:photons} shows the cumulative number of photons capable of
ionizing helium ($h\nu \geq 54.4$ eV) as a function of redshift for each of the
simulations presented here. The top panel shows as a point of comparison the
total number of helium atoms in the volume. At early times, there are noticeable
differences between Simulations~\ampup\ and \normup, which in principle should
both have twice as many photons as Simulation \fid. These variations are likely
due to shot-noise introduced by the relatively rare quasars, which becomes less
extreme at later times. For redshifts $z \lesssim 4$, Simulations~\ampup\ and
\normup\ no longer have produced twice as many photons as Simulation~\fid. This
is due to the renormalization process of changing the output of quasars
on-the-fly to match $\tau_\mathrm{HI,eff}$, as described in
Section~\ref{sec:features}. For further details, see
Appendix~\ref{appendix:gamma_gal}. If all of the photons produced by quasars
were absorbed by helium atoms and there were no recombinations, then
helium~\textsc{ii} reionization would be completed when equality is
reached. Nevertheless, not all photons are absorbed (especially for the
highest-energy frequency bin, because of the very low cross-section of helium at
these frequencies), and recombination is prevalent, especially in dense
regions. Thus, the actual timing of reionization can be significantly different
from when photon-helium atom equality is reached.

\section{Helium III Ionization Fraction}
\label{sec:heiii}

\begin{figure}[t]
  \centering
  \includegraphics[width=0.5\textwidth]{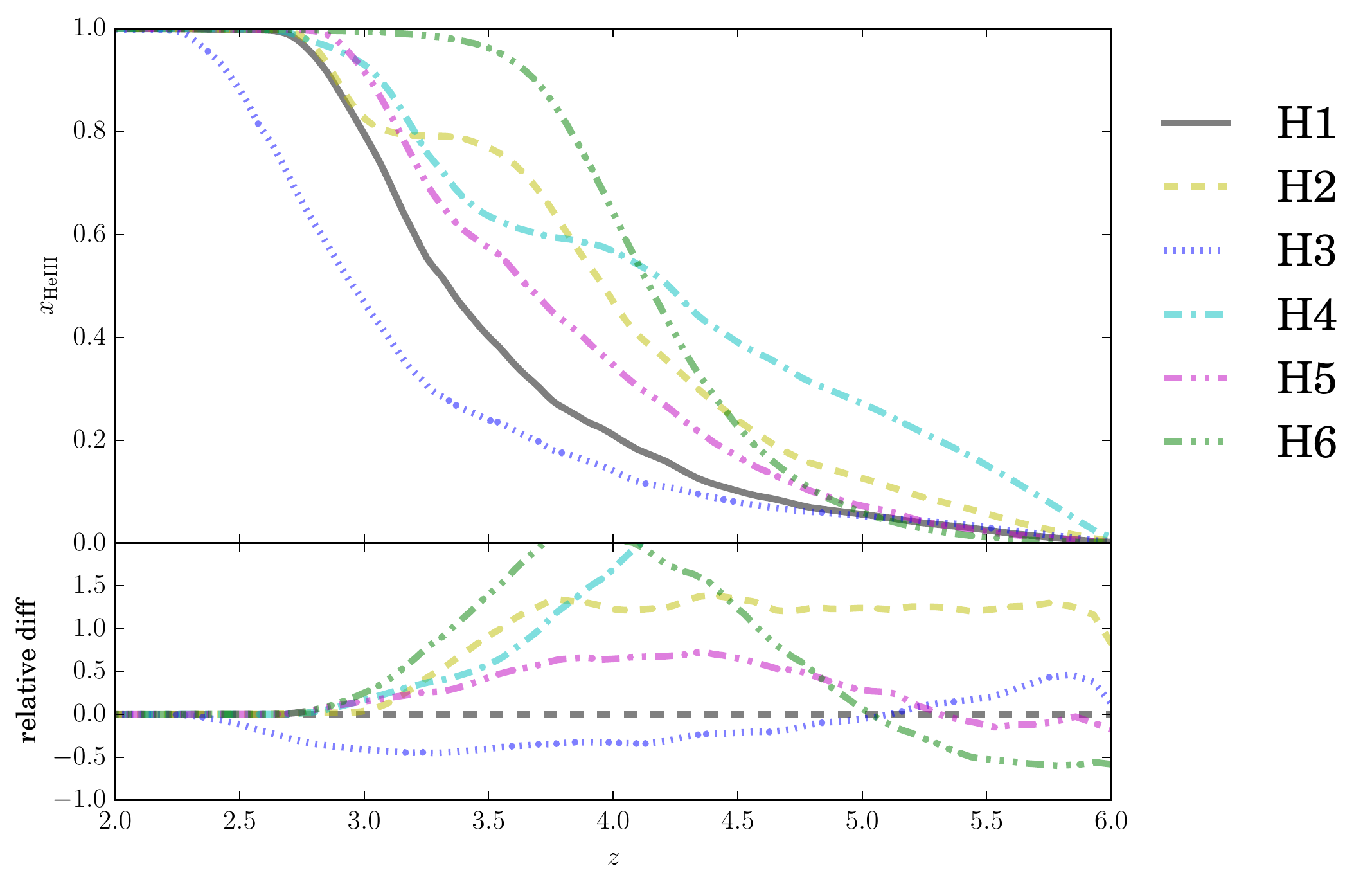}
  \caption{Helium ionization fraction $x_\mathrm{HeIII}$ as a function of
    redshift for the different quasar models explored in this work. The top
    panel shows the ionization fraction, and the bottom panel shows the relative
    difference compared to the fiducial simulation (\fid). The different models
    presented here are described in Section~\ref{sec:features} and summarized in
    Table~\ref{table:sim-summary}. Among the models shown for comparison is a
    simulation with a uniform UV background from \citet{haardt_madau2012}
    (Simulation \uvb). Note that most of the simulations reach a 99\% ionization
    fraction in the range of $2.7 \lesssim z \lesssim 3$, which is consistent
    with observational findings
    \citep{dixon_furlanetto2009,worseck_etal2011}. In all simulations, the
    duration of helium~\textsc{ii} reionization is typically
    $0.8 \lesssim \Delta_{50} \lesssim 1$, with the notable exceptions of
    Simulations \normup\ and \uvb. The durations for these simulations are
    significantly longer and shorter, respectively, than the other
    simulations. The former has a long duration due to the early onset of
    relatively massive quasars, while the latter assumes a quasar emissivity
    that rises sharply, starting at redshift $z \sim 5$. See the text in
    Section~\ref{sec:heiii} for additional discussion.}
  \label{fig:fHe}
\end{figure}

One of the most basic results from the simulations is the calculation of the
\HeIII\ ionization fraction as a function of redshift. We define the ionization
fraction $x_\mathrm{HeIII}$ as the (volume-weighted) amount of doubly ionized
helium relative to the total amount for all cells $i$ in the volume:
\begin{equation}
x_\mathrm{HeIII} \equiv \sum_i \frac{n_{\mathrm{HeIII},i}}{n_{\mathrm{He},i}}.
\label{eqn:xfrac}
\end{equation}
Given a particular model for the quasar sources, the ionization fraction
reflects the impact of these sources on the IGM. For instance, the duration of
reionization gives some information about the important sources: a relatively
long reionization argues for more sources that are fainter, and a shorter
reionization is driven by a few large sources. When comparing features in
observables produced from simulations, it is usually more important to compare
results at the same ionization fraction than at the same redshift. We refer to
different redshifts related to an ionization fraction with a subscript, such
that $z_{n} \Rightarrow x_\mathrm{HeIII} = n\%$. For instance,
$z_{50} \Rightarrow x_\mathrm{HeIII} = 50\% = 0.5$. In addition to finding the
redshift corresponding to different ionization fractions, we are also interested
in quantifying the duration of reionization. To this end, we define
\begin{equation}
\Delta z_{50} \equiv z_{25} - z_{75},
\label{eqn:z50}
\end{equation}
which corresponds to the duration in redshift of the central 50\% change in
ionization fraction. We also define a similar quantity $\Delta z_{90}$, which
represents the difference between $z_{5} - z_{95}$. We report the redshifts
associated with certain ionization fractions, as well as $\Delta z_{50}$ and
$\Delta z_{90}$, in Table~\ref{table:sim-summary}, which summarizes the main
results of the simulations. As a reference for converting $\Delta z$ into time
units, the shortest reionization scenario, Simulation \uvb, has a central
duration of $\Delta z_{50} = 0.58 = 252$ Myr, whereas the longest reionization
scenario, Simulation \normup, has a duration of $\Delta z_{50} = 1.83 = 834$
Myr. These reionization scenarios take place over a relatively extended portion
of the Universe's history, and leave a lasting impression on the IGM.

\begin{figure*}[t]
  \centering
  \includegraphics[width=0.9\textwidth]{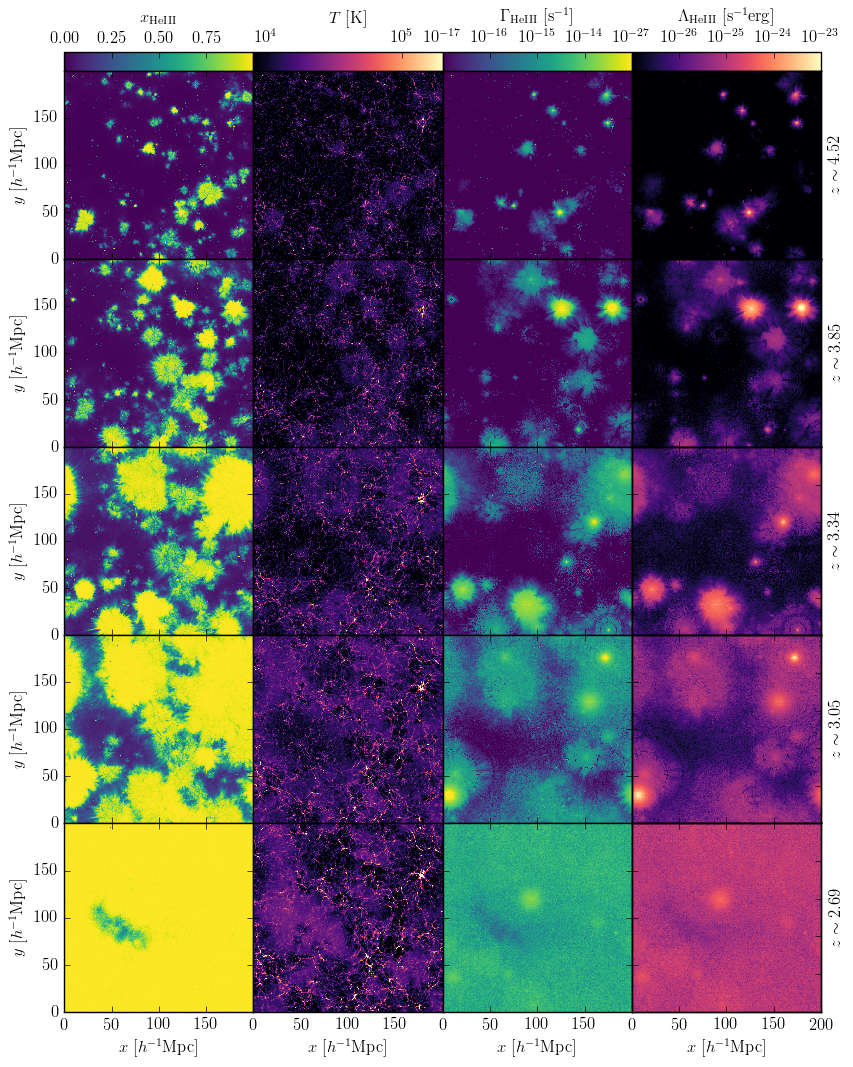}
  \caption{Comparison of different properties in Simulation \fid. Each panel
    shows a 2-dimensional slice through the simulation volume with the thickness
    of a single RT cell ($\sim$400 $h^{-1}$kpc). The columns, from left to
    right, show the \HeIII\ ionization fraction, the gas temperature, the
    \HeIII\ photoionization rate $\Gamma_{\mathrm{HeIII}}$, and the \HeIII\
    photoheating rate $\Lambda_{\mathrm{HeIII}}$. The third and fourth columns
    only include the contribution to the photoionization and photoheating from
    the quasar sources, and do not include other sources of ionization and
    heating (\textit{e.g.}, collisional ionization or heating). The different
    rows show redshift snapshots corresponding to volume-average ionization
    fractions of $x_{\mathrm{HeIII}} = 0.1, 0.25, 0.5, 0.75,$ and 0.99, from top
    to bottom. Note that early on in the reionization process, the average
    \HeIII\ bubble size is small ($\sim$5 $h^{-1}$Mpc), but later on in
    reionization, the size of ionized regions becomes much larger ($\sim$50
    $h^{-1}$Mpc in some cases). This change in bubble size is due to relatively
    long lifetimes of luminous quasars. The grainy appearance in Columns 3 and 4
    is primarily caused by subtle details of the RT implementation and
    visualization process and is not representative of the accuracy of the
    calculation.}
  \label{fig:panel1}
\end{figure*}

\begin{figure*}[t!]
  \centering
  \includegraphics[width=0.9\textwidth]{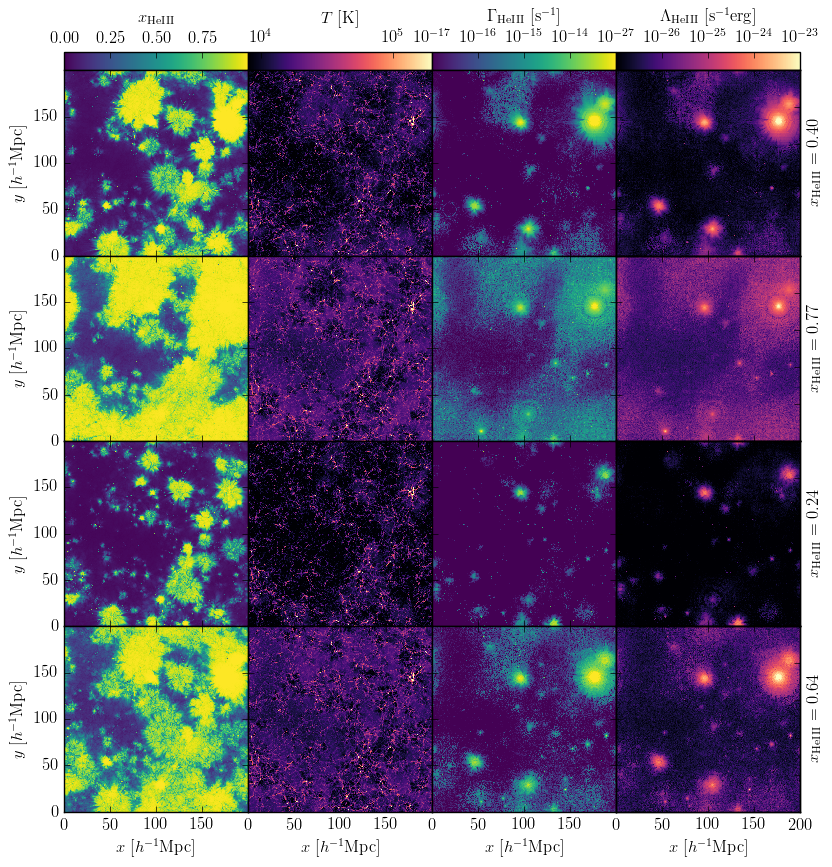}
  \caption{Similar plot to Figure~\ref{fig:panel1}, but comparing different
    simulations at redshift $z \sim 3.5$. The rows show from top to bottom
    Simulations \fid, \ampup, \ampdown, and \normup. The value of
    $x_\mathrm{HeIII}$ is shown to the right of each row. In addition to the
    obvious differences in helium ionization level morphology, the temperature
    of the IGM (second column) is also very different for the different
    simulations. There is also an apparent difference between the quasar models
    used in Simulation \ampup\ (second row) and Simulation \normup\ (fourth
    row), which in principle have similar photon counts, but are at different
    ionization levels. The reasons for these differences are discussed in
    Section~\ref{sec:ion_frac}.}
  \label{fig:panel2}
\end{figure*}

\subsection{Ionization Fraction Evolution}
\label{sec:ion_frac}
Figure~\ref{fig:fHe} shows the volume-averaged ionization fraction of the
different simulations as a function of redshift. We define the quantities
$\Delta z_{50}$ and $\Delta z_{90}$ as the duration, in redshift, for the volume
to transition from 25-75\% ionized (by volume) and 5-95\% ionized,
respectively. In general, helium~\textsc{ii} reionization is a very extended
process, with $\Delta z_{90} \gtrsim 2$ for almost all of the reionization
scenarios, with Simulation \normup\ having very extended reionization times of
$\Delta z_{90} \sim 2.9$. However, there is a large variation in the timing of
reionization. The earliest simulation to reach 50\% ionization is \normup, which
occurs at $z_{50} \sim 4.22$. The latest simulation is \ampdown, which occurs at
$z_{50} \sim 2.96$. The fiducial reionization scenario, \fid, is 50\% ionized at
$z_{50} \sim 3.34$. As pointed out below in Sections~\ref{sec:igmtemp} and
\ref{sec:lya}, in general observations are more sensitive to the end of helium
reionization, when the volume becomes 90-95\% doubly ionized. The main exception
to this result is Simulation \ampup, which reaches a maximum temperature at
$z \sim 3.41$, which corresponds to an ionized fraction of 80\%. The reason for
the difference is related to the method by which the quasar emission is modified
to match $\tau_\mathrm{eff}$, as outlined in
Section~\ref{sec:features}. Nevertheless, knowing the full reionization history
has important implications on the thermal history of the IGM.

Figure~\ref{fig:panel1} shows visualizations of Simulation \fid. The four
columns, from left to right, show the \HeIII\ ionization fraction
$x_\mathrm{HeIII}$, the gas temperature, the \HeII\ photoionization rate
$\Gamma_\mathrm{HeII}$, and the \HeII\ photoheating rate
$\Lambda_\mathrm{HeII}$. The rows show the same slice of the simulation at
increasing values of ionization fraction, which from top to bottom are
$x_\mathrm{HeIII} = 0.1$, 0.25, 0.5, 0.75, and 0.99. The corresponding redshift
is shown on the right side of the panels. These slices show a segment of the
$yz$-plane of the simulation, with a thickness of one radiative transfer cell in
the $x$-direction. This width corresponds to a comoving distance of $\sim$ 400
$h^{-1}$kpc. In a loose sense, the first and second columns are integrated
quantities corresponding to the third and fourth columns, respectively. In both
cases, the figure shows only photoionization and photoheating rates, which in
particular does not include collisional ionization and heating prevalent in
regions of high density. Nevertheless, the photoionization and photoheating
rates are dominated by the contribution of photons from quasars in the
volume. Furthermore, for the temperature of the IGM (Column 2), the hottest
regions are found along filaments and other dense regions of cosmic
structure. Although these regions are the hottest, photons from quasars
dramatically heat the low-density IGM by several thousand kelvin. See
Section~\ref{sec:igmtemp} for further discussion of the IGM temperature.

As discussed in Paper~I, in our model the clustering of quasars indirectly
affects their lifetimes. Because the lifetimes of quasars affect the size of
reionized regions (visible in Figures~\ref{fig:panel1} and \ref{fig:panel2}),
the proper clustering affects the coherent scale of reionization. The size of
reionized regions also affects the heating of the IGM, as larger reionization
regions encompass moderate- to low-density regions earlier than smaller
regions. The reason for this is that the relatively fast timing of recombination
means that moderate- to high-density gas quickly recombines and requires
additional radiation in order to re-reionize. For comparatively large regions,
more of the gas that is ionized is low-density, so there is less
recombination. Also worth noting is that the relatively high clustering leads to
an early overlap of reionized regions, which again reflects the timing of
ionization reaching regions of low density.

Figure~\ref{fig:panel2} shows visualizations of Simulations \fid, \ampup,
\ampdown, and \normup, all at redshift $z \sim 3.5$. Although the underlying gas
and large-scale structure is largely similar (as can be seen by comparing Column
2 of the different rows), the ionization and temperature distributions are very
different for the different simulations. The differences are driven by the
different quasar models used in the simulations. Of particular interest is the
difference between Simulations \ampup\ and \normup\ (Rows 2 and 4). When
performing simple photon-counting calculations, as seen in
Figure~\ref{fig:photons}, both of these simulations should produce a similar
number: Simulation \ampup\ increases by a factor of 2 in the total number of
quasars at a given epoch, whereas Simulation \normup\ increases by a factor of 2
in the number of photons produced per quasar.

Despite this similarity, there are significant differences between the
simulations, most notably the ionization fraction (Column 1). Additionally,
Columns 3 and 4 show that Simulation \ampup\ has a greater quasar activity at a
given redshift. Part of the differences between the simulations can be
attributed to the method by which quasars are populated in the volume: as
explained in Section~\ref{sec:quasar_sources} (and more in depth in Paper~I),
quasars are placed in halos using abundance matching. Thus, when the amplitude
of the luminosity function is increased, sources of the same luminosity are
placed in lower-mass halos. In addition to making rare objects more common,
there are more sources in general. This feature leads to a greater number of
photons intersecting gas cells that have not previously been exposed to quasar
radiation.

\begin{figure*}[t]
  \centering
  \resizebox{0.6\textwidth}{!}{\includegraphics{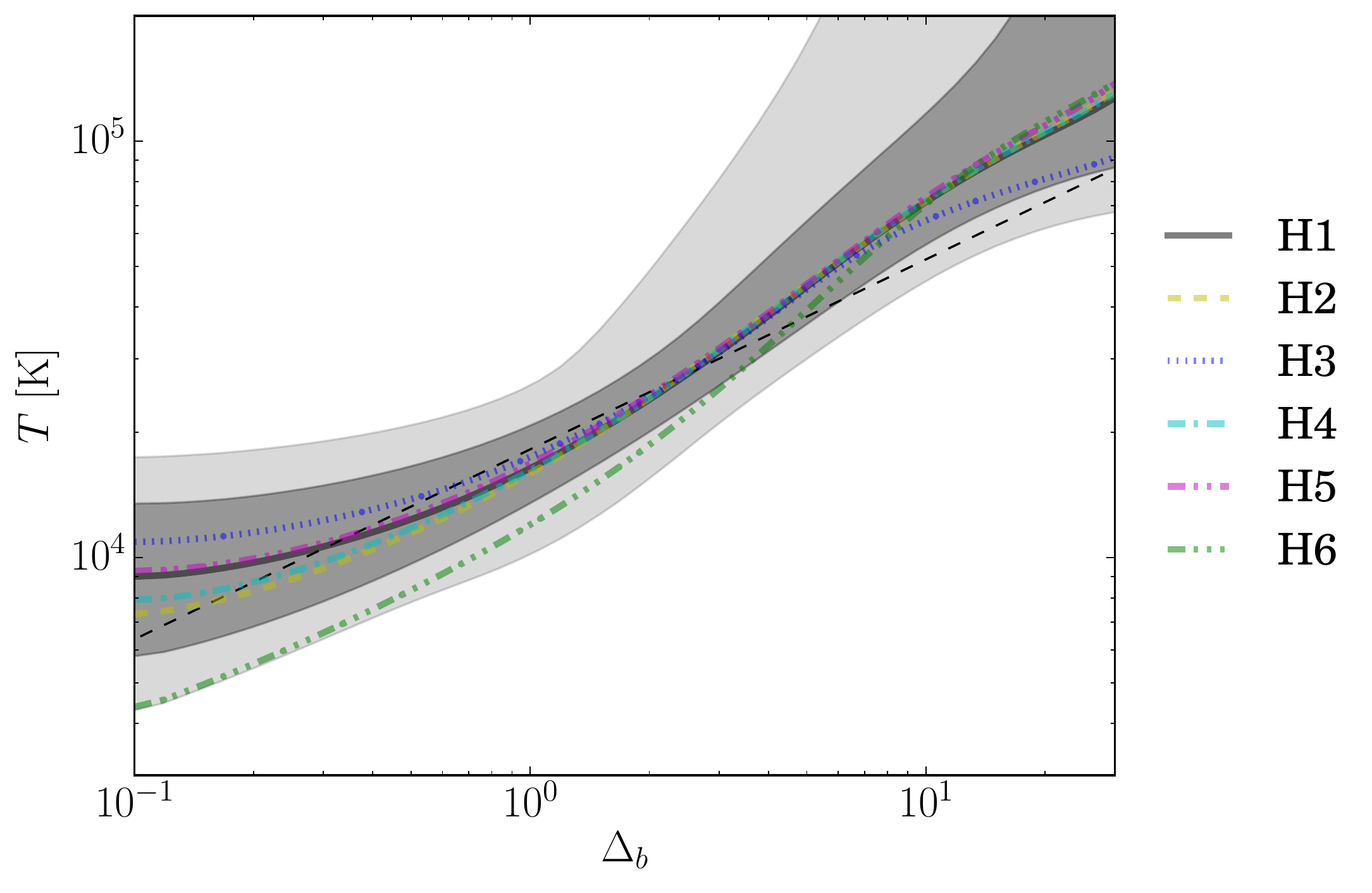}}\\
  \resizebox{0.48\textwidth}{!}{\includegraphics{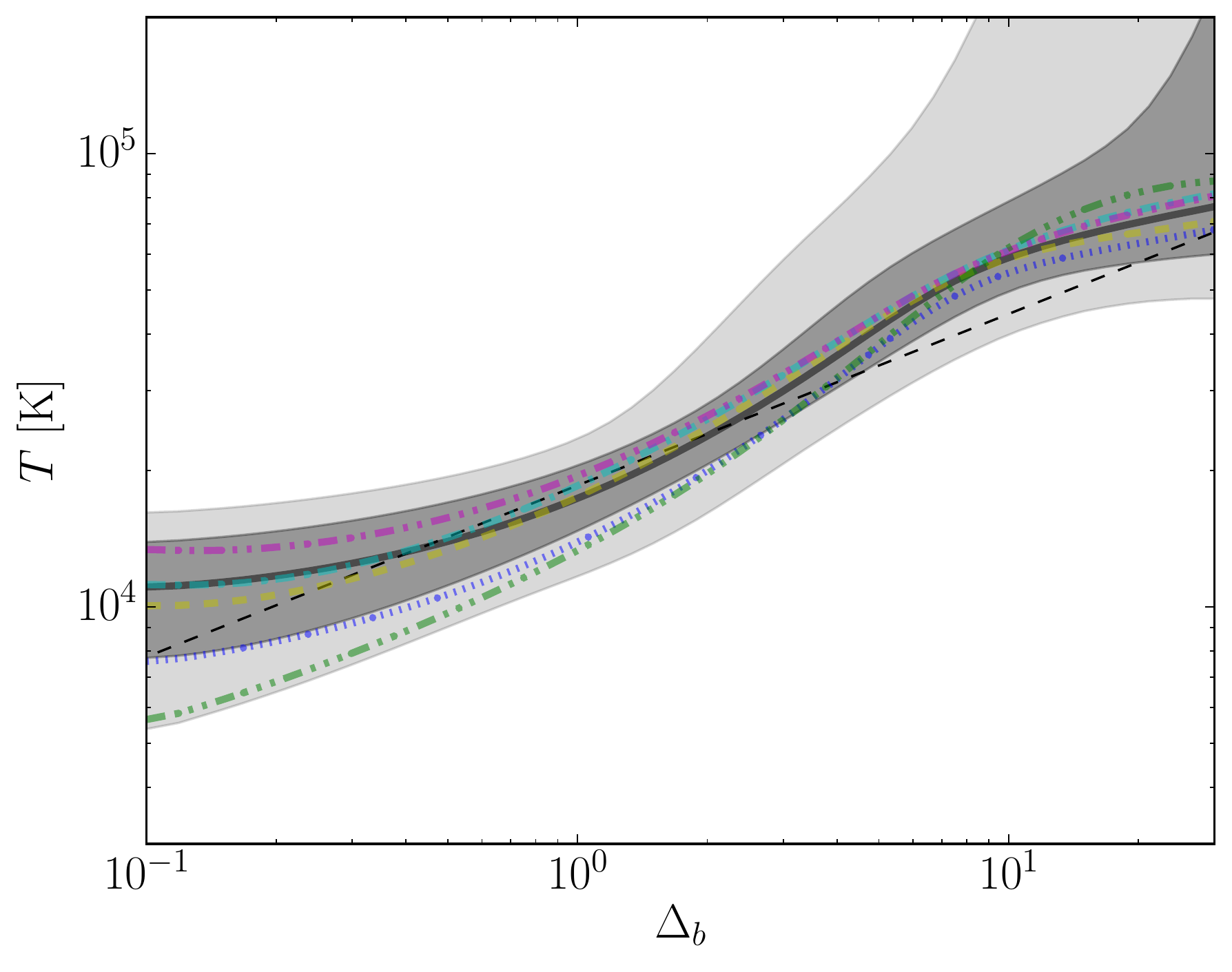}}%
  \resizebox{0.48\textwidth}{!}{\includegraphics{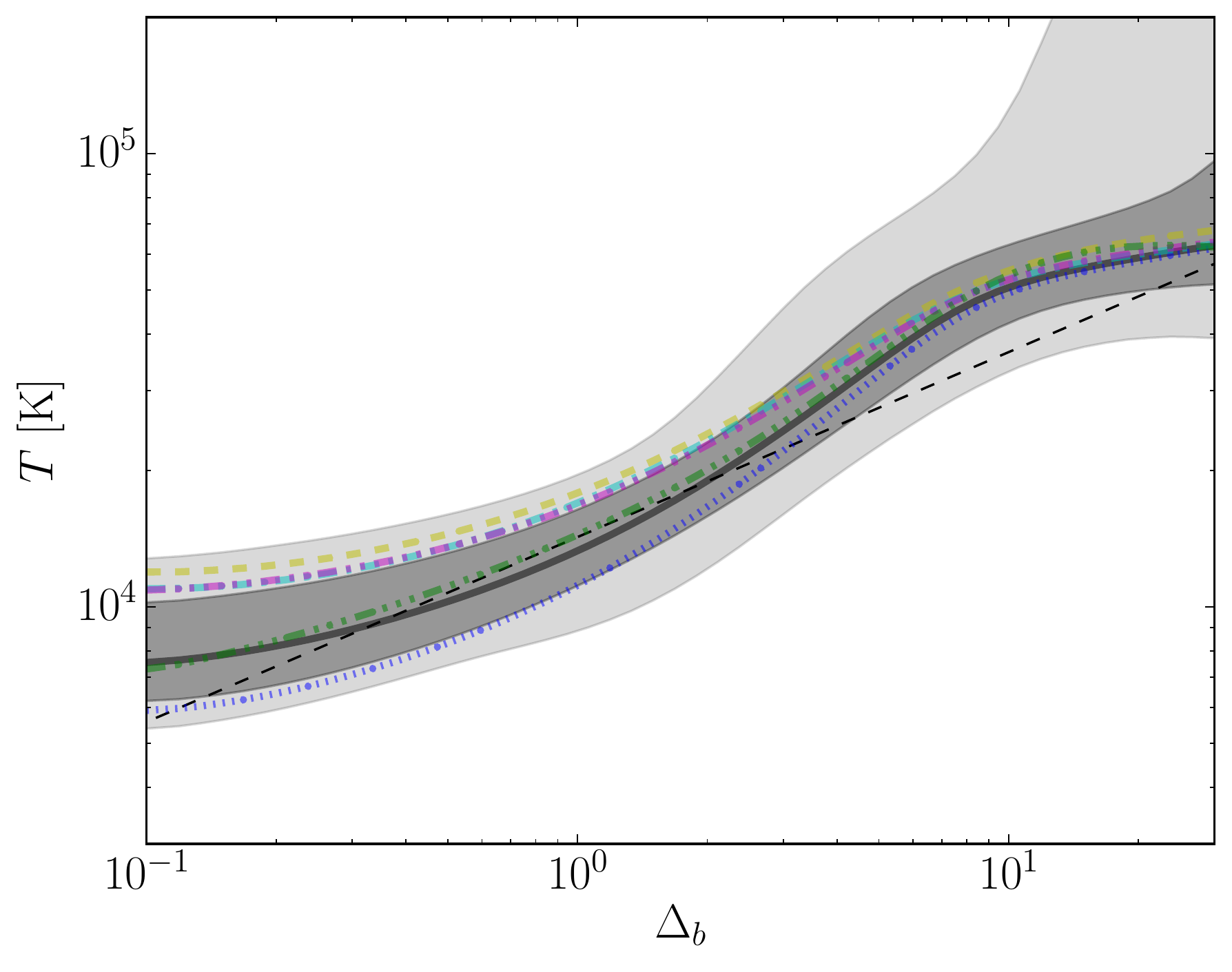}}\\
  \caption{Comparison of the gas temperature $T$ as a function of baryon density
    $\Delta$. The panels are the temperature-density relation as measured from
    the simulations at $z \sim 2.5$ (top), $z \sim 3$ (bottom left), and
    $z \sim 3.5$ (bottom right). The gray shaded regions correspond to the 68th
    and 95th percentiles of Simulation \fid. The overall amplitude of the
    relation rises as redshift decreases, showing that the overall temperature
    of the IGM increases as helium~\textsc{ii} reionization heats up the
    volume. In addition, the slope of the relation becomes steeper as the
    simulations evolve. The temperature of relatively dense regions
    ($\Delta \gtrsim 10$) continues to rise even after helium~\textsc{ii}
    reionization is largely finished. This is due to recombination of the gas,
    followed by additional reionization, adding more heat to the
    gas. Conversely, once ionization is completed, the low-density regions
    ($\Delta < 1$) cool adiabatically, with little heat input. The dashed black
    line in each figure is the best-fit power-law relation given by
    Equation~(\ref{eqn:tdelta}) for Simulation \fid. This should be compared
    with the gray solid line, which follows the relationship for each density
    value $\Delta$. Note that in general, the simple power law does not
    accurately capture the relationship between density and temperature. See the
    text for additional discussion.}
  \label{fig:rhotemp}
\end{figure*}

Conversely, in Simulation \normup, the number of photons produced per source is
increased, but the total number of sources is the same as in Simulation
\fid. (Indeed, the same quasar catalog is used in the two simulations, and only
the normalization of quasar radiation is changed between the two. The general
morphology of ionized regions in Column 1 and the instantaneous quasar activity
in Columns 3 and 4 are very similar in Rows 1 and 4.)  Although twice as many
photons are produced per source, the long mean free path of helium-reionizing
photons means that not all photons are absorbed. Furthermore, as a result of
spectral filtering of the radiation from quasars, the photons with energy
$h\nu \sim 54.4$ eV will be readily absorbed before more energetic photons,
changing the effective SED of the quasar sources \citep{meiksin_etal2010}. The
higher energy photons typically are not absorbed, leading to the large
discrepancy in neutral fraction observed between these simulations. Thus,
although a simple semi-analytic calculation would yield the same reionization
time for these two simulations, we can see that a full treatment leads to
important differences between the two cases.

The ionization fraction observed in our simulations is worth comparing with the
results of \citet{mcquinn_etal2009} and \citet{compostella_etal2013}, hereafter
M09 and C13. The duration of reionization in our simulations is comparable to
the models explored in M09 (as seen in their Figure~3). However, the
reionization histories in C13 are much briefer than those seen here. This is
largely because the quasar population in their fiducial reionization model does
not include sources for $z > 4$. The authors include an additional ``extended''
model that includes sources beginning at $z = 5$, which shows a duration of
reionization more comparable to those in M09 and this work. Observations from
\citet{mcgreer_etal2013} show a non-negligible population of high-redshift
quasars, which in Simulation \fid\ causes the ionization fraction of helium to
have a value of a few percent at $z \sim 5$, with the volume being nearly a
quarter ionized by $z \sim 4$. Thus, future studies should include high-redshift
quasars as an important part of helium~\textsc{ii} reionization.

\section{The Temperature History of the IGM}
\label{sec:igmtemp}

One important impact of helium~\textsc{ii} reionization on the IGM is the
temperature feedback. Since quasars emit a hard spectrum with many energetic
photons and the IGM is in a highly ionized state, the excess energy remaining
after photoionization is converted into heat in the gas. Although secondary
ionizations are possible (\textit{e.g.},
\citealt{shull1979,furlanetto_stoever2010}), their impact is negligible for
helium reionization because of the ionization level of the IGM
\citep{mcquinn_etal2009}. Photoheating from radiation from quasars increases the
average temperature of the IGM by $\sim$10,000 K, and as we show in
Sections~\ref{sec:rhot} and \ref{sec:medt}, contains important information about
the history of helium~\textsc{ii} reionization.

\subsection{Temperature-Density relation}
\label{sec:rhot}
The relationship between the temperature of the IGM $T$ and the baryon
overdensity $\Delta \equiv \Delta_b$ is an important measure of the state of the
IGM, and it is intimately related to the reionization process. One can write the
relationship between temperature and density as a power law and fit for the two
parameters that define it \citep{hui_gnedin1997}:
\begin{equation}
T(\Delta) = T_0 \Delta^{\gamma-1},
\label{eqn:tdelta}
\end{equation}
where $T$ is the gas temperature, and $T_0$ and $\gamma$ define the power-law
relation between the gas density and temperature. This is the so-called
temperature-density relation, also sometimes called the equation of state of the
IGM (although we note that it is not a true equation of
state). \citet{hui_gnedin1997} showed that at late times following hydrogen
reionization, the slope of the relation approaches $\gamma = 1.62$. In general,
this relationship should hold for the low-density gas in the IGM where adiabatic
cooling or heating and a uniform radiation field following reionization are the
dominant sources of temperature change. However, the addition of heat from
helium~\textsc{ii} reionization changes the slope of this relation, as well as
the overall amplitude.

Figure~\ref{fig:rhotemp} shows the temperature-density relation for the gas in
the different simulations. The relationship is shown at several different
redshifts, in order to demonstrate several different effects that reionization
has on the IGM temperature. In particular, the general trend is indicative of an
``inside-out'' reionization scenario. In such a scenario, the radiation from
sources (quasars, in this case) propagate outward, and are absorbed in
high-density regions near sources before low-density ones, depositing heat as
the radiation is absorbed. Because the gas is reionized at different times and
is dominated by adiabatic cooling following reionization, the relative
temperature between different gas densities reflects the reionization
history. In particular, the temperature of underdense regions can in fact be
higher than mean-density regions because the radiation from quasars tends to
reach these regions at a later redshift. In the meantime, the gas from
high-density regions has additional time to cool adiabatically. Because the
amount of heat deposited in the gas from photoionization does not depend on the
density, the gas from higher density regions may be at a lower temperature than
the low-density gas when the low-density gas is reionized. Thus, the
temperature-density relation can be relatively flat for medium- to low-density
gas, and even turn over such that low-density regions have a \textit{higher}
temperature than mean-density ones (\textit{e.g.}, as in \citealt{trac_etal2008}
for hydrogen reionization). The simulations presented here do not exhibit this
inversion because of both the longer mean free path of helium-ionizing photons
and the relatively smaller amount of adiabatic cooling experienced by gas at
this redshift.\footnote{The adiabatic cooling of gas causes the temperature to
  decrease as $T \propto (1+z)^2$; thus, a duration of reionization in redshift
  space of $\Delta z \sim 1$ at the higher redshift of hydrogen reionization
  leads to a larger relative change in temperature than the lower redshift of
  helium~\textsc{ii} reionization.} Nevertheless, several of our simulations,
and Simulation \comp\ at $z \sim 3$ in particular, show a relatively flat
relation for underdense regions.

Another feature in Figure~\ref{fig:rhotemp} is the evolution of regions of high
density ($\Delta \gtrsim 10$). In these regions, the density of gas is high
enough that an appreciable fraction of the doubly ionized helium can recombine
with electrons to form singly ionized helium. Once the gas has recombined, it
can undergo an additional reionization event, which will deposit additional heat
into the gas. As can be seen in the Figure, the higher density regions show
higher temperatures as redshift decreases, even after helium~\textsc{ii}
reionization is nominally completed. Thus, the temperature of these different
regions at the same redshift can somewhat break the degeneracy between the
different reionization scenarios. Since these differences are visible in higher
density gas, it may be possible to observe these differences in the
Ly$\beta$ forest, since these observations saturate at higher densities than
Ly$\alpha$ \citep{dijkstra_etal2004,irsic_viel2014}.

\begin{figure}[t]
  \centering
  \includegraphics[width=0.45\textwidth]{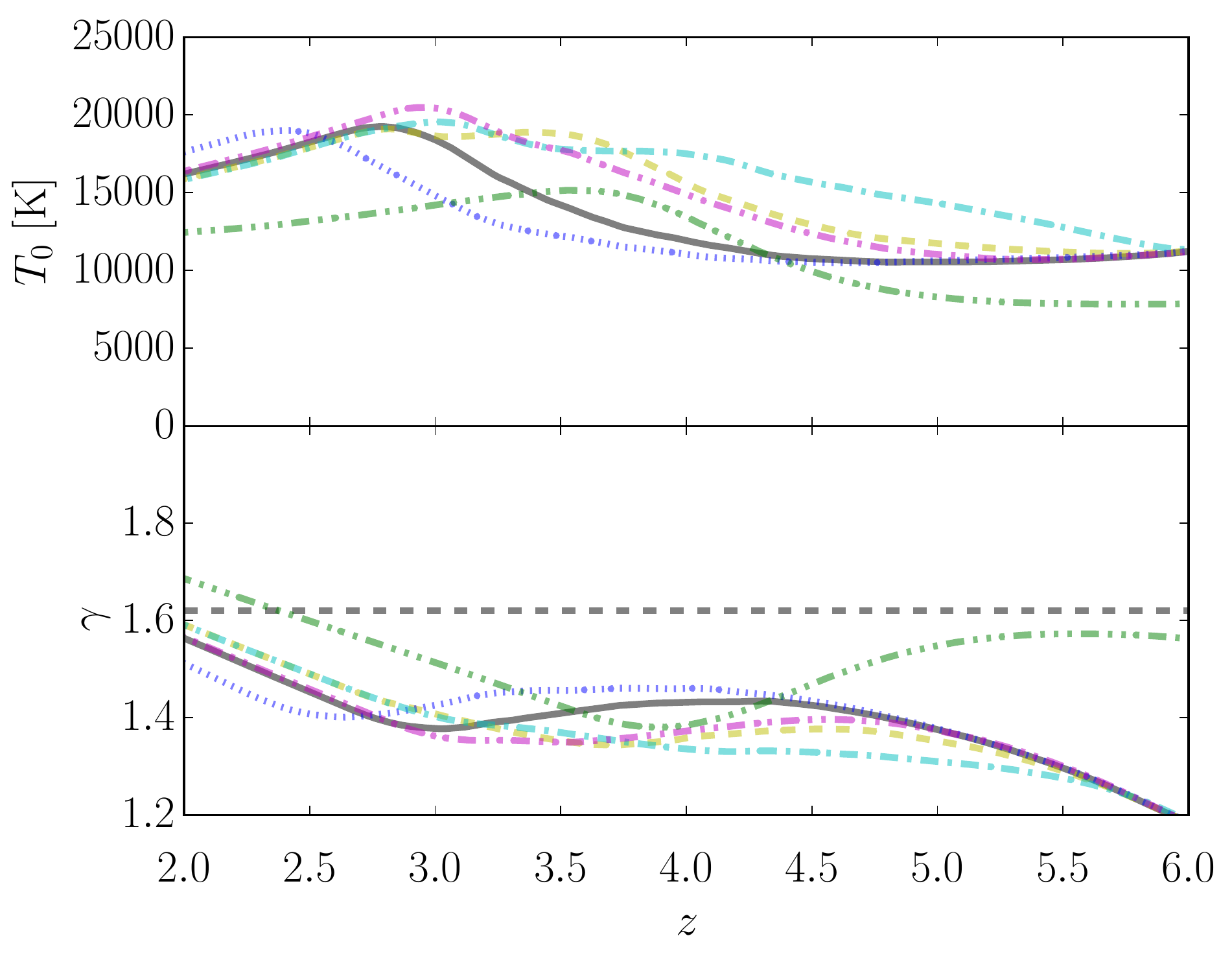}
  \caption{Parameters of the IGM power law temperature-density relation in
    Equation~(\ref{eqn:tdelta}) as a function of redshift for the different
    simulations. The top panel shows the temperature parameter $T_0$, and the
    bottom panel shows $\gamma$. In the panel for $\gamma$, we have shown the
    line of $\gamma = 1.62$, the predicted slope from \citet{hui_gnedin1997} for
    the relation following hydrogen reionization. At early times, Simulation
    \uvb\ approaches this value, but then deviates from it following helium
    reionization. The other simulations do not approach this value, probably
    because helium~\textsc{ii} reionization begins before a steady state can be
    established. As discussed in the text, the initial flattening of the slope
    ($\gamma < 1.6$) is due to the inside-out nature of reionization, and the
    later steepening ($\gamma \sim 1.62$) is due to establishing equilibrium
    with the radiation field. For the $T_0$ parameter, the rise and fall of the
    value is consistent with the rise and fall seen in
    Figure~\ref{fig:rhotemp}. See the text for further discussion.}
  \label{fig:t0gamma}
\end{figure}

Figure~\ref{fig:t0gamma} shows the evolution of the parameters of the
temperature-density relation given in Equation~(\ref{eqn:tdelta}) as a function
of redshift for the different simulations. We find a linear fit for the
parameters after applying a log-transform to the temperature and density for
each gas cell, and volume-weight the results. \footnote{The fit was performed
  using a simple linear regression of the log-transformed temperature-density
  relation. All cells in the volume were used to generate the fit. Restricting
  the fit to cells where $\Delta \leq 3$, as discussed in other works, can
  change the value of $T_0$ by up to 10\%, although the value of $\gamma$ does
  not change significantly. Because of the ambiguities associated with these
  choices, we believe that the median temperature at mean density (discussed in
  Section~\ref{sec:medt}) is a more robust measure of the ``average
  temperature'' of the IGM.} As can be seen by the general structure of
Figure~\ref{fig:rhotemp} and as was noted in C13, fitting the entire
temperature-density relation to a single power law may not be the optimal
parameterization because of the wide dispersion of temperatures at a given
density value. We should note that part of the difficulty in fitting the result
to a power law comes from the approximate nature of the relation: for high
values of $\Delta$, the approximation breaks down. Furthermore, the resolution
of the simulations does not capture all of the structure of the IGM, which leads
to smoothing at certain scales. Nevertheless, we present these results for the
sake of comparison.

In general, we see a similar trend to Figure~\ref{fig:rhotemp}, where the
temperature value at mean density $T_0$ increases as reionization proceeds,
reaches a peak value, and then decreases again. This is a general trend seen in
the thermal evolution of the IGM and is explored more below in
Section~\ref{sec:medt}. Another general trend is the evolution of the power-law
index $\gamma$ which is roughly consistent between simulations. We reproduce the
observation of M09 that $\gamma \sim 1.3$ during the bulk of helium reionization
for our different scenarios. In the lower panel of Figure~\ref{fig:t0gamma} we
show the value of $\gamma = 1.62$, which is the asymptotic value of the IGM from
\citet{hui_gnedin1997} following hydrogen reionization without additional
sources of photoheating.

As can be seen from Figure~\ref{fig:t0gamma}, simulations that include a patchy
hydrogen reionization are not consistent with this value, although Simulation
\uvb, which features a significantly earlier hydrogen reionization epoch,
approaches this value. However, once helium~\textsc{ii} reionization begins,
there is a notable flattening of the temperature-density relation (where
$\gamma = 1$ represents the limit of an isothermal gas). As a larger portion of
the volume becomes ionized, denser regions will recombine and undergo additional
reionization events, leading to additional heat being deposited at these
densities. Conversely, low-density regions are dominated by adiabatic
cooling. This leads to an overall steepening of the slope $\gamma$, a trend seen
at low redshifts following the completion of helium~\textsc{ii} reionization.
These trends are also visible in Figure~\ref{fig:rhotemp}. In particular at
$z \sim 3$, most of the simulations have a comparable value of $\gamma$. Indeed,
the shape of these temperature-density relations in the central panel of
Figure~\ref{fig:rhotemp} is similar, albeit with different vertical offsets.

The results of M09 and C13 are largely consistent with the findings presented
here. Before helium~\textsc{ii} reionization begins, the temperature-density
relation tightly follows a power-law expression. Once helium~\textsc{ii}
reionization begins, the distribution of temperature as a function of density
becomes highly variable, with a large dispersion forming for a given density
value. This dispersion signifies the inhomogeneous reionization process and is
a general feature of helium~\textsc{ii} reionization. Additionally, as in C13,
we find that the overall relation between temperature and density is ill-fit by
a single power law. C13 finds that the temperature-density relation flattens out
and begins to turn over at $\Delta \sim 10$ (\textit{cf}. their
Figure~8). Although we do not see a turn-over in our measurements, it is still
clear that using a single power law to characterize the relationship between
temperature and density is insufficient for the IGM following reionization.

\subsection{Temperature at mean density}
\label{sec:medt}

\begin{figure}[t]
  \centering
  \includegraphics[width=0.45\textwidth]{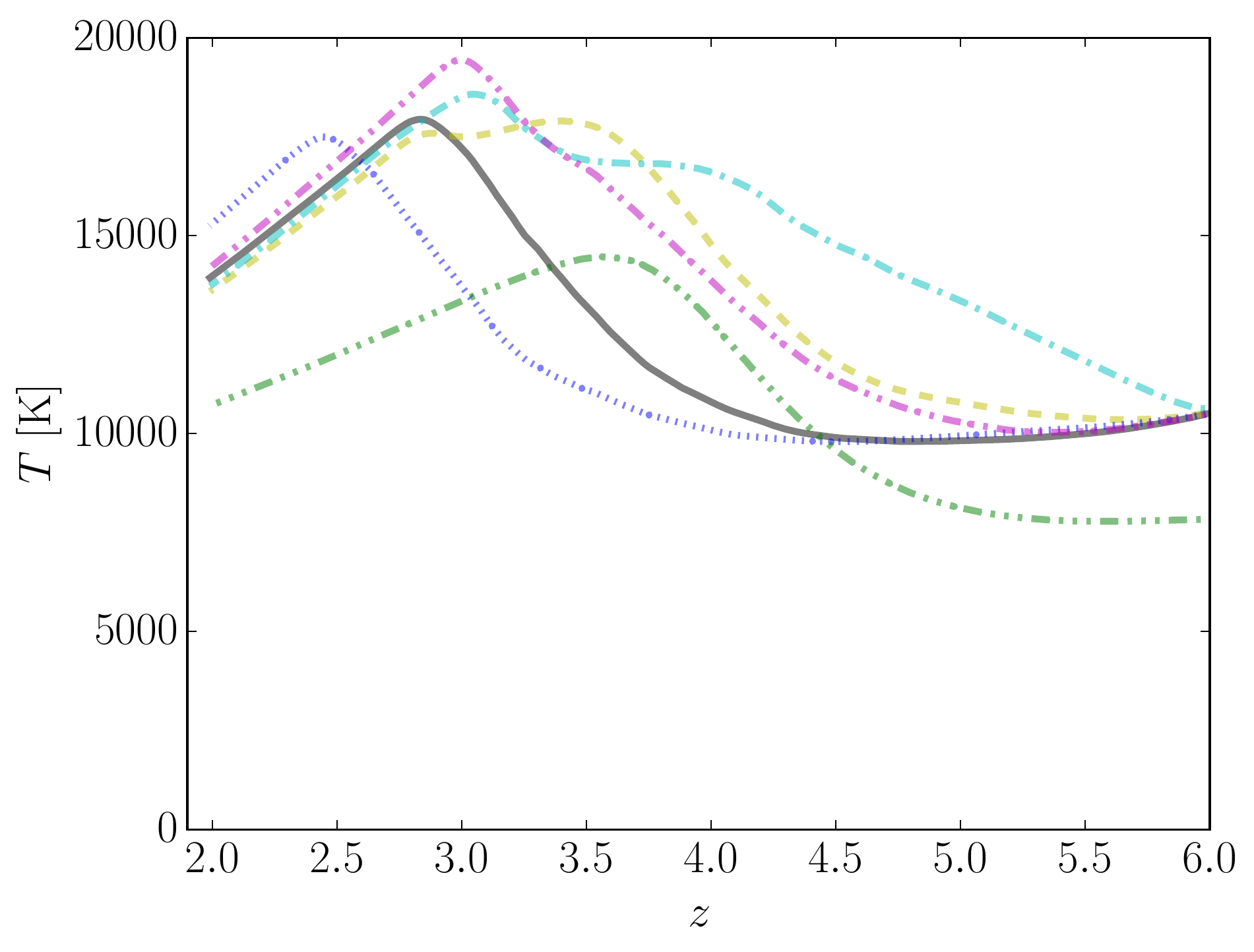}
  \caption{Median temperature at mean density ($0.95 \leq \Delta \leq 1.05$) of
    the IGM as a function of redshift. The temperature at mean density is
    significantly higher for the patchy hydrogen reionization scenarios than for
    the one with a uniform UVB (Simulation \uvb), because the hydrogen
    reionization occurs significantly earlier. Once quasar activity begins, the
    temperature of the IGM rises as a result of photoheating. The peak
    corresponds to an ionization fraction of $x_\mathrm{HeIII} \sim 0.90$-0.95,
    marking the tail-end of reionization. Following helium~\textsc{ii}
    reionization, the mean-density gas begins to adiabatically cool again,
    leading to the peak structure seen in the Figure. See the text for
    additional discussion.}
  \label{fig:temp}
\end{figure}

An important marker of the progress of helium~\textsc{ii} reionization is the
temperature at mean density of the simulation ($\Delta \sim 1$), since the
temperature in these regions is dominated by adiabatic cooling of the Universe
and heating from radiative transfer \citep{hui_gnedin1997}. The interplay of
these two factors determines the temperature of these regions of average
density. The average temperature of these regions show two characteristic
bumps as a function of redshift: one initial increase from $T \sim 200$ K to
$T \sim 10^4$ K as a result of hydrogen reionization at
$8 \lesssim z \lesssim 10$, and a subsequent increase in temperature from
$T \sim 10^4$ K to $T \sim 2 \times 10^4$ K at $2 \lesssim z \lesssim 3.5$ as a
result of helium reionization
\citep{furlanetto_oh2008a,puchwein_etal2015,upton-sanderbeck_etal2016}. In
between the two epochs of reionization, and following helium~\textsc{ii}
reionization, adiabatic cooling dominates, and so the average temperature
decreases. The locations and widths of these features can provide valuable
insight into the timing and duration of reionization.

Previous studies of the mean temperature of the IGM, both semi-analytic
\citep{furlanetto_oh2008a} and using simulations with a uniform UVB
\citep{puchwein_etal2015,bolton_etal2016} have shown that the general picture of
the IGM temperature should hold, and it can therefore be used to extract
information about reionization. For our purposes here, we concern ourselves
primarily with this second epoch of heating in the IGM, corresponding to
helium~\textsc{ii} reionization.

\begin{figure*}[t]
  \centering
  \includegraphics[width=0.8\textwidth]{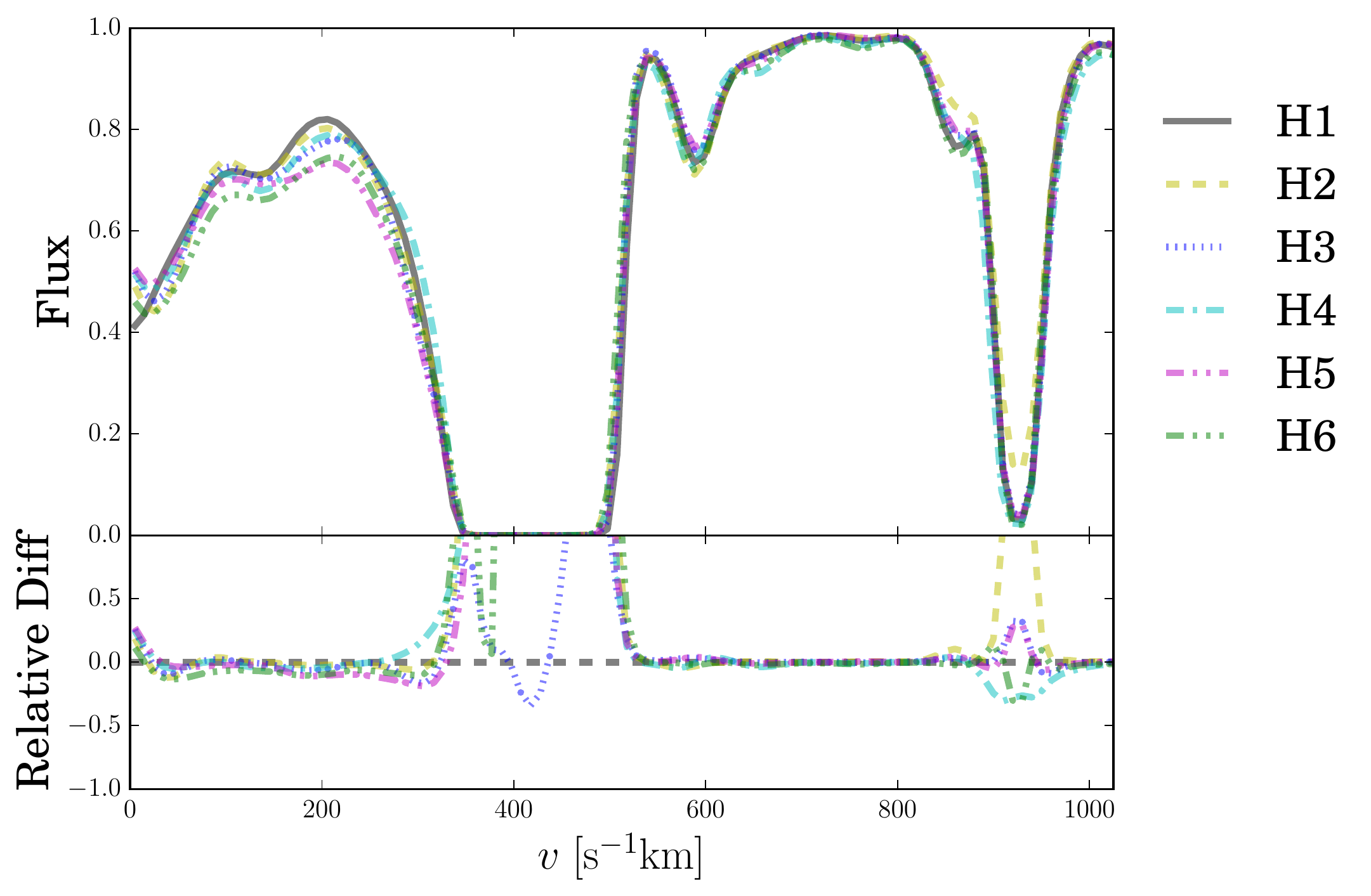}\\
  \includegraphics[width=0.8\textwidth]{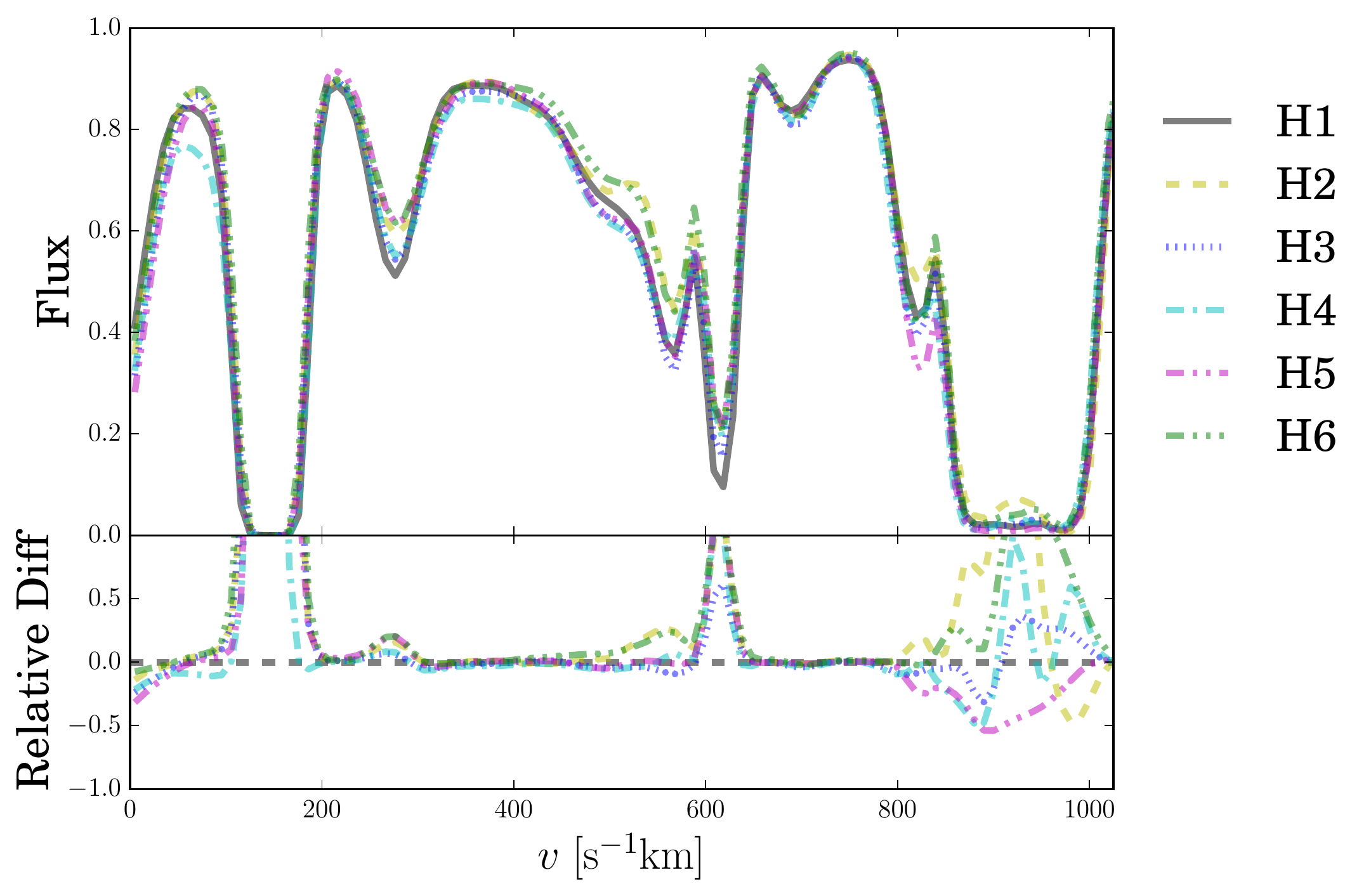}\\
  \caption{Visualization of two typical sightlines through the simulation volume
    at $z \sim 2.7$. The sightlines shown in each figure are taken from the same
    physical location in the volume from all simulations and therefore encounter
    similar large-scale structure. The segments shown here are about 10 comoving
    $h^{-1}$ Mpc in size. The differences in flux between simulations can be
    significant, although not always in the same direction. This behavior argues
    for calculating flux explicitly using Equation~\ref{eqn:lya} rather than
    using an approximate expression for $\tau$. Relatively large differences in
    regions of low flux ($F \lesssim 0.1$) are due more to the low values of
    flux than to significant differences between the simulations. See the text
    in Section~\ref{sec:lya} for further discussion.}
  \label{fig:sightlines}
\end{figure*}

Figure~\ref{fig:temp} shows the median temperature at mean density of the
different simulations. In order to compute the temperature at mean density, at
each time step in the simulation we find the median temperature (as well as the
$\pm$68th and 95th percentiles) of all gas cells that have
$0.95 \leq \Delta \leq 1.05$. At high redshift ($z \gtrsim 6$), the simulations
have largely the same temperature because the IGM temperature is dominated by
hydrogen reionization. As explained in Section~\ref{sec:features}, all of the
simulations with explicit quasar sources use a semi-analytic method for
calculating patchy hydrogen reionization. The exception to this is Simulation
\uvb, which uses the uniform UV background of HM12 for both hydrogen and helium
reionization. Notably, the timing of hydrogen reionization is significantly
earlier than for the patch hydrogen method used ($z_\mathrm{re} \sim 13$ for
HM12 compared to $z_\mathrm{re} \sim 8$ for the patchy hydrogen), so the IGM has
had additional time to adiabatically cool. This leads to the lower initial
temperature at $z \sim 6$ seen in Figure~\ref{fig:temp}.

We also note that in Figure~\ref{fig:temp} the temperature of the IGM peaks at a
redshift that corresponds to 90-95\% of the helium \textsc{iii} ionization
level. This is consistent with the idea that the gas at mean density composes a
large fraction of the volume of the simulation volume and so will
preferentially reionize later than regions of high density. Following this peak
in the IGM temperature, the adiabatic cooling of the Universe becomes the
dominant mechanism because this comparatively low-density gas generally does
not recombine (because recombination is $\propto \rho_g^2$, as shown in
Equation~(\ref{eqn:gamma_gal})).

\section{Measurements of the Ly$\alpha$ Forest}
\label{sec:lya}
An important observational tool used to understand helium~\textsc{ii}
reionization is the Ly$\alpha$ forest. Observationally, there have been many
rich data sets using the Ly$\alpha$ forest, especially for cosmological
measurements. The BOSS sample \citep{lee_etal2013} has been used to observe the
baryon acoustic oscillation (BAO) feature
\citep{busca_etal2013,slosar_etal2013}, as well as generate one-dimensional
power spectra \citep{palanque-delabrouille_etal2013}, which have been used to
constrain neutrino masses and other cosmological parameters
\citep{palanque-delabrouille_etal2015}. High-resolution measurements from
Keck-HIRES and Magellan-MIKE
\citep{lu_etal1996,becker_etal2007,becker_etal2011b,calverley_etal2011} have
given us information about the temperature history of the IGM.

Synthetic Ly$\alpha$ spectra can be created for the \HI\ and \HeII\
densities. (See Paper~III of this series for further discussion of the \HeII\
Ly$\alpha$ forest). In the following analysis, we have drawn the spectra along
the $x$-axis of the simulation, although we find nearly identical results when
projecting along different axes. Once these spectra have been calculated, they
can be used to measure the effective optical depth $\tau_{\mathrm{eff}}$ of the
volume, compute the flux PDF, and calculate one-dimensional power spectra. To
generate a synthetic sightline, we define a set of pixels along a line of sight
in the simulation volume, such that the number of pixels is equal to the number
of grid cells. For the resolution level discussed in these simulations, this
means $N_\mathrm{pix} = 2048$. The resulting resolution of the Ly$\alpha$ forest
is about 98~kpc~$h^{-1}$ comoving. \citet{lukic_etal2015} showed that at a
comparatively high-redshift ($z \gtrsim 3.5$) resolution of $\sim 20$
kpc~$h^{-1}$ comoving was required to resolve all features of the Ly$\alpha$
forest to sub-percent level accuracy, especially at small scales. Thus, some
inaccuracies may be introduced at small scales and high redshift from the
resolution of the simulations. At the same time, at mean density at $z \sim 3$
the Jeans scale is typically 500 kpc~$h^{-1}$ comoving, so the simulation has
sufficient resolution to capture features introduced by Jeans smoothing.

For each pixel $i$, the optical depth of the pixel $\tau_i$ is calculated
through the contributions of every other pixel according to the formula
\citep{bolton_etal2009b}
\begin{equation}
\tau_i = \frac{c \sigma_\alpha \dd{R}}{\pi^{1/2}} \sum_{j=1}^{N_\mathrm{pix}} %
\frac{n_\mathrm{HI}(j)}{b_\mathrm{HI}(j)} H(a,x),
\label{eqn:lya}
\end{equation}
where $\sigma_\alpha = 4.479 \times 10^{-18}$ cm$^{-2}$ is the cross-section of
the Ly$\alpha$ transition, $b_\mathrm{HI} = \sqrt{2k_B T/m_\mathrm{H}}$ is
the Doppler parameter, $\dd{R}$ is the (physical) width of the pixel, and
$H(a,x)$ is the Voigt-Hjerting function \citep{hjerting1938}:
\begin{equation}
H(a,x) = \frac{a}{\pi} \int_{-\infty}^\infty \frac{e^{-y^2}}{a^2 + (x-y)^2} \dd{y},
\label{eqn:voigt-hjerting}
\end{equation}
where $x = [v_\mathrm{H}(i) - u(j)]/b_\mathrm{HI}(j)$ is the difference in
redshift space between pixels $i$ and $j$ relative to the Doppler broadening,
$u(j) = v_\mathrm{H}(j) + v_\mathrm{pec}(j)$ is the total velocity difference of
Hubble flow plus peculiar velocity,
$a = \Lambda_\alpha \lambda_\alpha/4\pi b_\mathrm{HI}(j)$ represents the gas
damping, where $\Lambda_\alpha = 6.265 \times 10^8$ s$^{-1}$ is the damping
constant and $\lambda_\alpha = 1215.67$ \AA\ is the wavelength corresponding to
the Ly$\alpha$ transition. In order to efficiently compute the
Voigt-Hjerting function, we use the analytic approximation provided by
\citet{tepper-garcia2006}.

As can be seen from Equations~(\ref{eqn:lya}-\ref{eqn:voigt-hjerting}), the
thermal properties of the gas enter in the form of the Doppler parameter
$b$. This term increases as the temperature of the gas increases and serves to
broaden the apparent width in velocity space of a particular gas parcel. The
tendency of absorption features to widen in velocity space as the temperature
increases can be used to learn about the thermal state of the IGM. More
approximately, the local optical depth of the IGM will depend on the average
temperature of the volume. We will further discuss some of the implications of
this process below in Section~\ref{sec:fluxpdf}. The temperature of the gas can
also affect the hydrogen ionization level, since the recombination coefficient
$\alpha_\mathrm{HII}$ in Equation~(\ref{eqn:gamma_gal}) decreases with increased
temperature.

Figure~\ref{fig:sightlines} shows two typical sightline sections generated from
the gas properties in the simulations at $z \sim 2.7$. These synthetic spectra
have a comoving size of about 10 $h^{-1}$ Mpc. The sightlines in different
simulations are drawn from the same location in the simulation volume, which
means that the sightlines encounter similar large-scale structure of the
underlying gas. Accordingly, the differences in flux observed can be traced to
local differences in the radiation field.

As described in Section~\ref{sec:features}, all of the simulations have been
renormalized such that the overall effective optical depth $\tau_\mathrm{eff}$
is consistent across simulations. This allows for more straightforward
comparison between simulations. It also allows for us to determine which
statistical differences observed in the simulations can be attributed to the
timing of helium~\textsc{ii} reionization.

We note that the general large-scale absorption is similar across
simulations. Also worth mentioning is the fact that differences between
simulations are not always in the same direction. For instance, in the bottom
panel of Figure~\ref{fig:sightlines}, Simulation~\comp\ shows higher flux than
Simulation~\fid\ in an absorption feature at $v \sim 250$ km~s$^{-1}$, and more
absorption at $v \sim 800$ km~s$^{-1}$. These differences are due to relatively
small-scale effects of being nearby quasars that are active at different times
in some simulations, or perhaps not at all in others. There are also relatively
large differences in regions of low flux. These differences are driven primarily
by there being little overall flux, rather than truly having large deviations
between simulations.

\subsection{Effective Optical Depth}
\label{sec:taueff}

Once the optical depth for each pixel has been calculated, the corresponding
flux is given simply by $F_i = \exp(-\tau_i)$. We can then define the effective
optical depth of the volume by averaging over all values of the flux:
\begin{equation}
\ev{F} = \exp(-\tau_\mathrm{eff}).
\label{eqn:tau_eff}
\end{equation}
In general $\tau_\mathrm{eff} \neq \ev{\tau}$. The effective optical depth as a
function of redshift has been measured to high precision as a volume-averaged
quantity for the \HI\ forest \citep{lee_etal2015} and for individual objects of
the \HeII\ forest \citep{worseck_etal2014}.  \citet{lee_etal2015} reported that
the BOSS survey measures more than 50,000 quasar spectra at intermediate-to-high
redshift and has a formula for the evolution of the effective optical depth as a
function of redshift $\tau_\mathrm{eff}(z)$.

In general, cosmological simulations of the Ly$\alpha$ forest must renormalize
the flux level measured in order to match the observed optical depth
measurements (see, \textit{e.g.}, \citealt{bolton_etal2009b}). The reason is
that the resolution of these simulations is typically not high enough to capture
the small-scale high-absorption LLSs and damped Ly$\alpha$ systems that can lead
to cosmological simulations predicting too high of a value of
$\tau_\mathrm{eff}$ (although see \citealt{mcquinn_etal2009} for attempts to
account for these systems in simulations). Typically, this renormalization of
Ly$\alpha$ spectra is done in post-processing when the sightlines are generated.

\begin{figure}[t]
  \centering
  \includegraphics[width=0.45\textwidth]{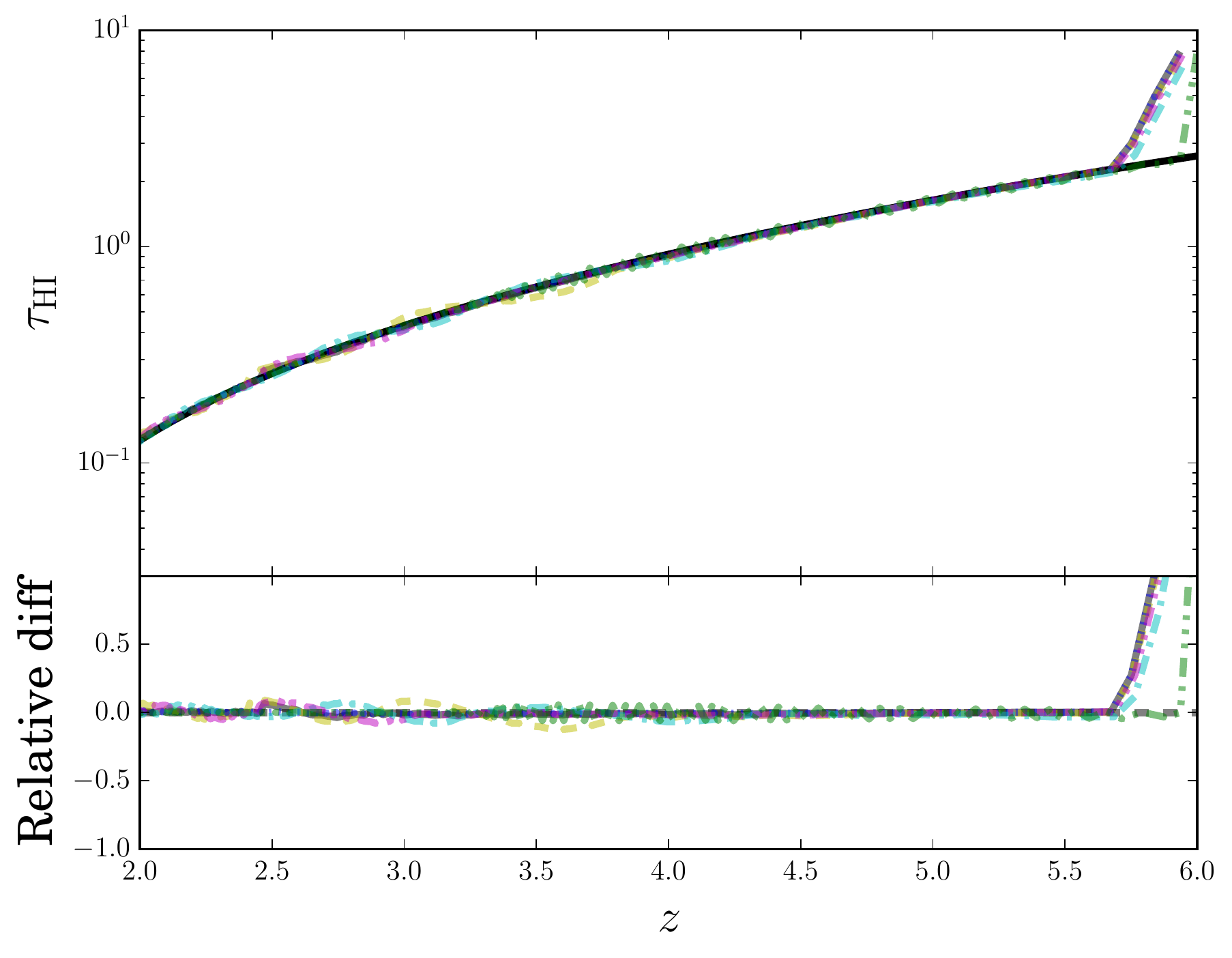}
  \caption{Effective optical depth of hydrogen $\tau_\mathrm{eff,HI}$ as a
    function of redshift for the different simulations. The solid black line
    shows the observational data from \citet{lee_etal2015}. The line styles for
    the simulations are the same as in Figure~\ref{fig:fHe}. As discussed in
    Section~\ref{sec:taueff}, the amount of hydrogen-ionizing radiation from
    galaxies $\Gamma_\mathrm{gal}$ is modified while the simulation is running,
    so that this quantity is matched by construction. This avoids the
    requirement of renormalizing the simulations in post-processing.}
  \label{fig:teffHI}
\end{figure}

\begin{figure*}[t]
  \centering
  \resizebox{0.6\textwidth}{!}{\includegraphics{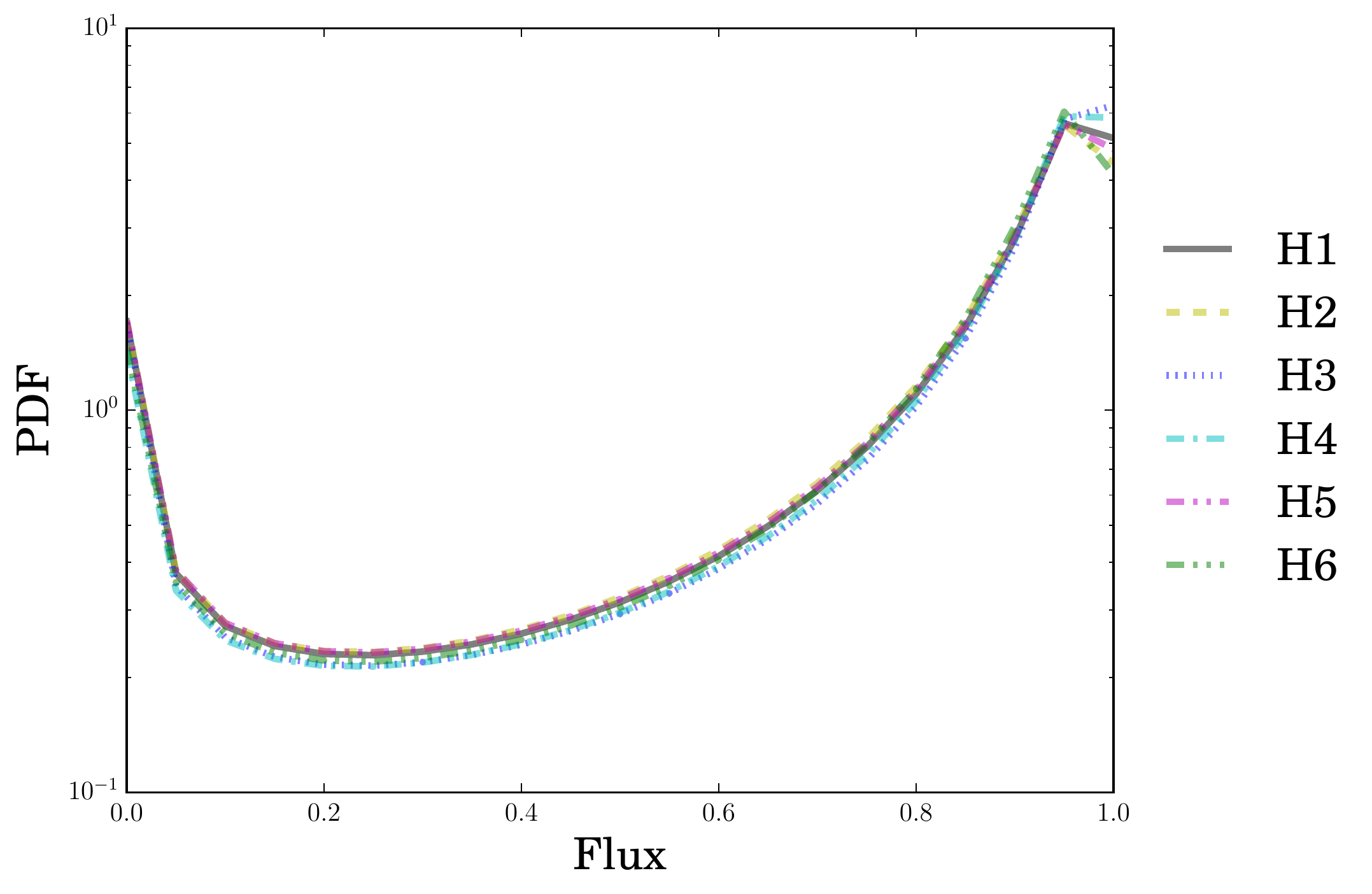}} \\
  \resizebox{0.48\textwidth}{!}{\includegraphics{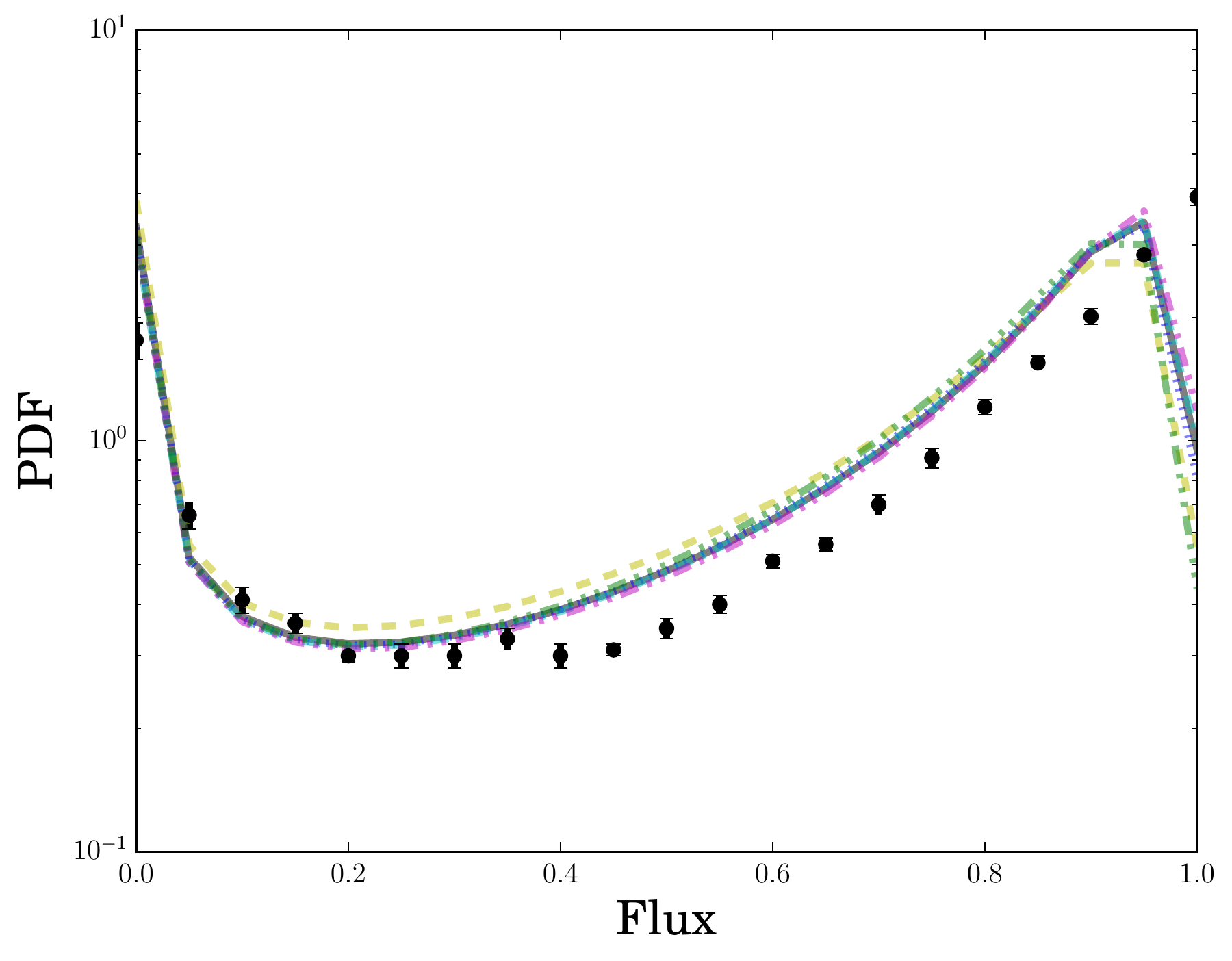}}%
  \resizebox{0.48\textwidth}{!}{\includegraphics{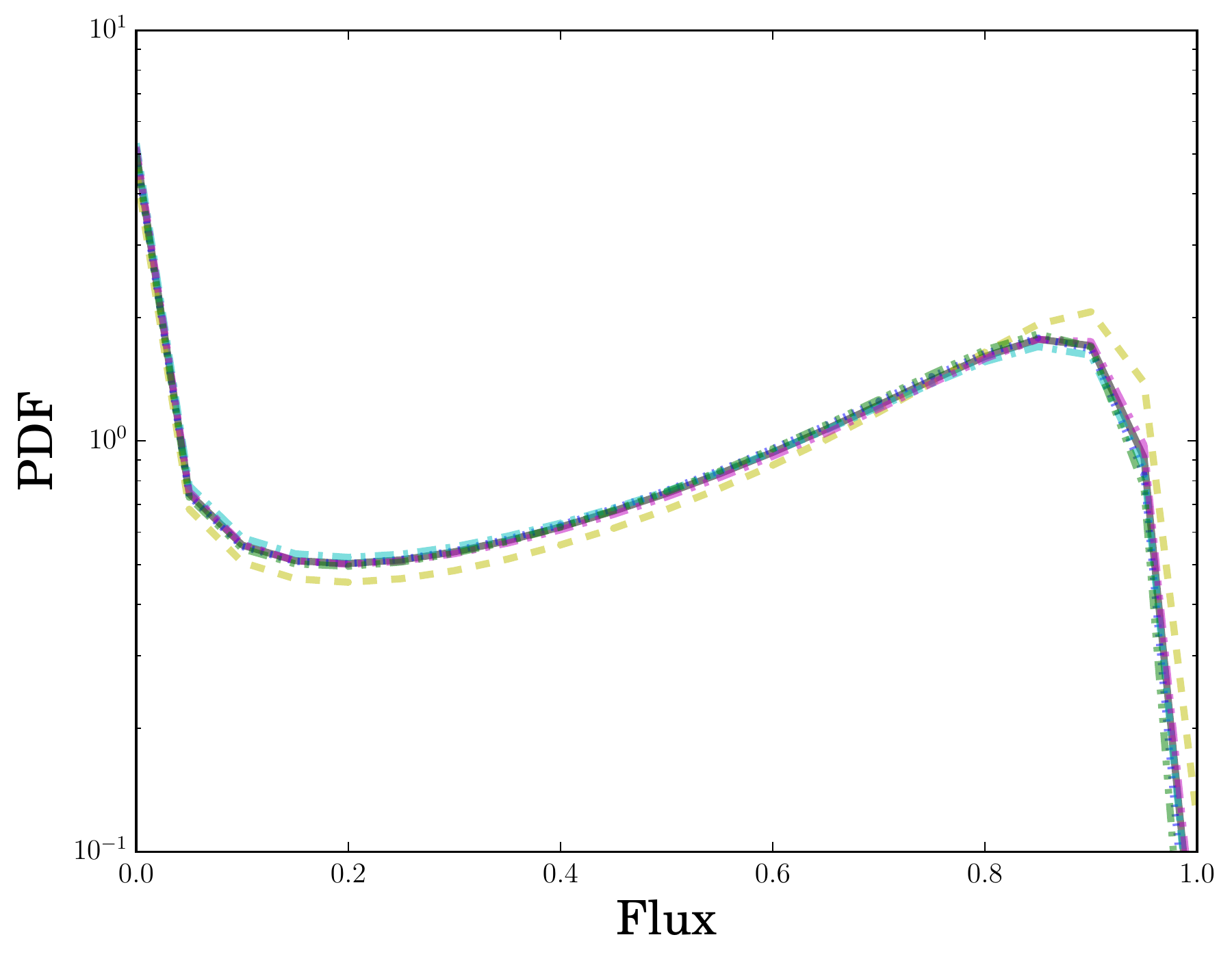}} \\
  \caption{Comparison of the flux PDF of the \HI\ Ly$\alpha$ forest at
    $z \sim 2.5$ (top),$z \sim 3$ (bottom left), and $z \sim 3.5$ (bottom
    right). All of the simulations have the same color scheme as in
    Figure~\ref{fig:fHe}. In the bottom left panel, the data points are taken
    from the results of \citet{calura_etal2012}, at $z \sim 2.9$. All of the
    simulations show a generally similar distribution of fluxes. This result
    implies that the flux PDF is only weakly sensitive to the temperature
    information of the IGM, since the only main difference between the
    simulations (except for the helium ionization fraction) is the
    temperature. The flux PDF is instead more sensitive to $\tau_\mathrm{eff}$
    and observationally, to the continuum-level uncertainty of the Ly$\alpha$
    forest. See the text in Section~\ref{sec:fluxpdf} and
    Appendix~\ref{appendix:fluxpdf} for additional details.}
  \label{fig:fluxpdf_HI}
\end{figure*}

Figure~\ref{fig:teffHI} shows $\tau_\mathrm{eff}$ for all of the simulations
presented in this work. As noted in Section~\ref{sec:features}, this quantity is
matched by construction for all of the simulations. In general the agreement is
excellent. For redshifts $z \lesssim 6$ (the nominal end of hydrogen
reionization, after which $\tau \lesssim 1$), all of the simulations match the
observed value from \citet{lee_etal2015} to within a few percent. This matching
allows for a more straightforward comparison between the simulations and
observations.

As explained in Sections~\ref{sec:features} and \ref{sec:models}, our
simulations change the value of $\Gamma_\mathrm{gal}$ on-the-fly in order to
match the value of $\tau_\mathrm{eff}$ as specified by \citet{lee_etal2015}. By
ensuring that all of our simulations match the same value of
$\tau_\mathrm{eff}$, we are better able to compare them with each other and with
the observations. Previous studies of the Ly$\alpha$ forest
\citep{theuns_etal2002,ciardi_etal2003,dallaglio_etal2008,faucher-giguere_etal2008a}
have reported a dip in $\tau_\mathrm{eff}$ at $z \sim 3.2$. In some of these
works, the authors cited this dip as evidence of helium~\textsc{ii}
reionization because an increased IGM temperature decreases the optical
depth. By matching the $\tau_\mathrm{eff}$ of \citet{lee_etal2015}, which does
not contain this dip, it is possible that we would miss this feature. We explore
this possibility in more detail in Appendix~\ref{appendix:hm}.

When we compare our results with the simulations of M09 and C13, in all cases,
$\tau_\mathrm{eff}$ is comparable to the most recent determinations of the \HI\
Ly$\alpha$ forest for the then state-of-the-art measurements. Our simulations
are the only ones that renormalize $\Gamma_\mathrm{HI}$ in real time, so we are
able to match the value of $\tau_\mathrm{eff}$ by construction. Nevertheless,
our values of $\Gamma_\mathrm{HI}$ are comparable to those in M09 and C12, as
well as HM12. We again note that the relative uncertainty on
$\Gamma_\mathrm{HI}$ is much larger than that of $\tau_\mathrm{HI}$, and so to
generate more realistic comparisons with measurements of the \HI\ forest, we
advocate matching the value of $\tau_\mathrm{eff}$ by construction, as we have
done here.

\subsection{Flux PDF}
\label{sec:fluxpdf}

Another statistic related to the Ly$\alpha$ forest is the flux PDF. This
measurement is carried out by taking the flux value of each of the pixels in the
sightlines of the Ly$\alpha$ forest and creating a normalized PDF of their
values. The result gives additional information about the distribution of gas in
the IGM. The flux PDF is also dependent on the resolution of the
measurement. For instance, compare the results from a relatively high-resolution
measurement \citep{calura_etal2012} with that of a relatively low-resolution
measurement \citep{lee_etal2015}. In the lower resolution case, the pixels of
extreme absorption or emission become averaged, and the flux PDF tends toward
the mean. Thus, the measured PDF is resolution dependent.

\begin{figure*}[t]
  \centering
  \resizebox{0.6\textwidth}{!}{\includegraphics{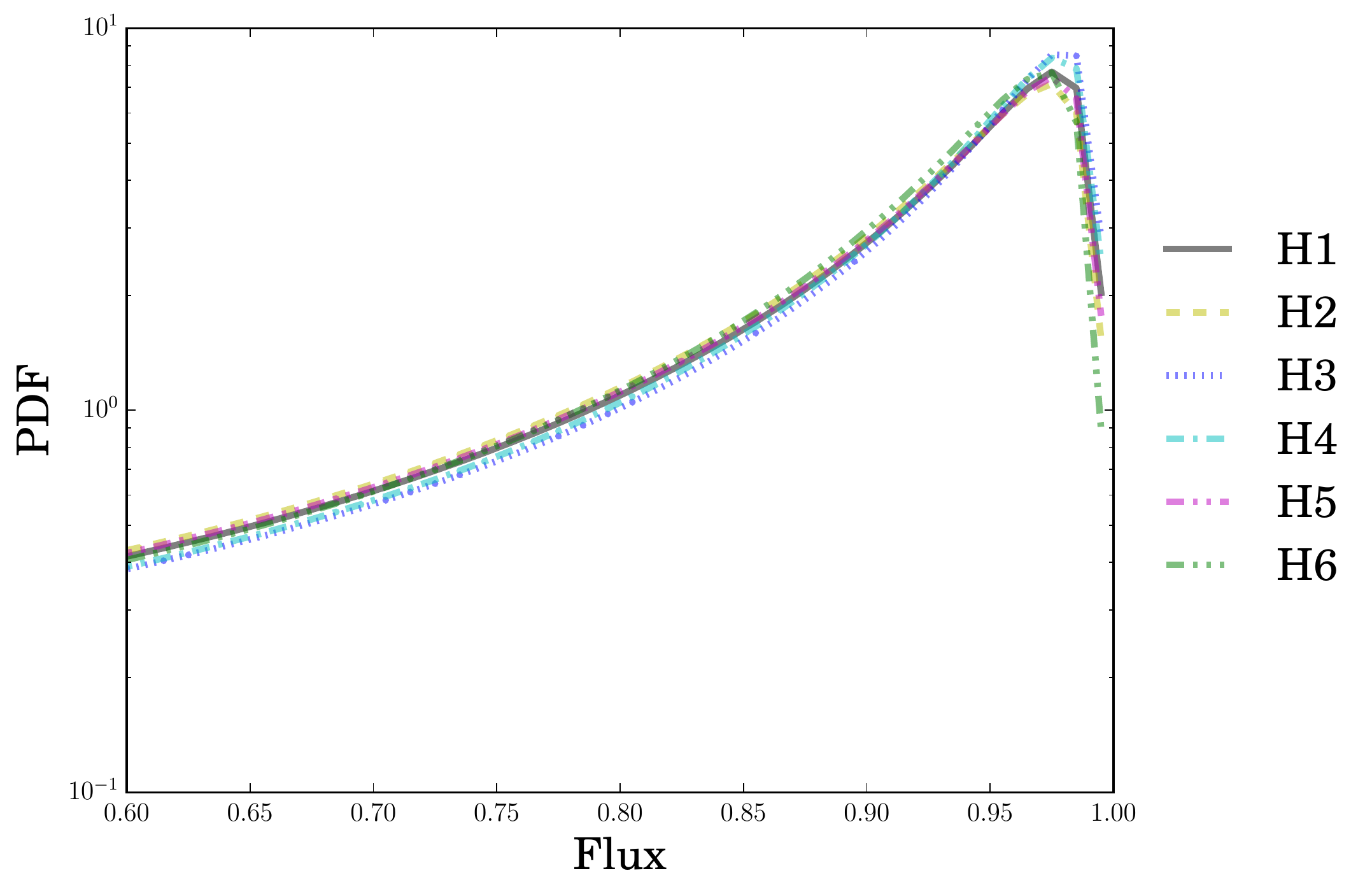}} \\
  \resizebox{0.48\textwidth}{!}{\includegraphics{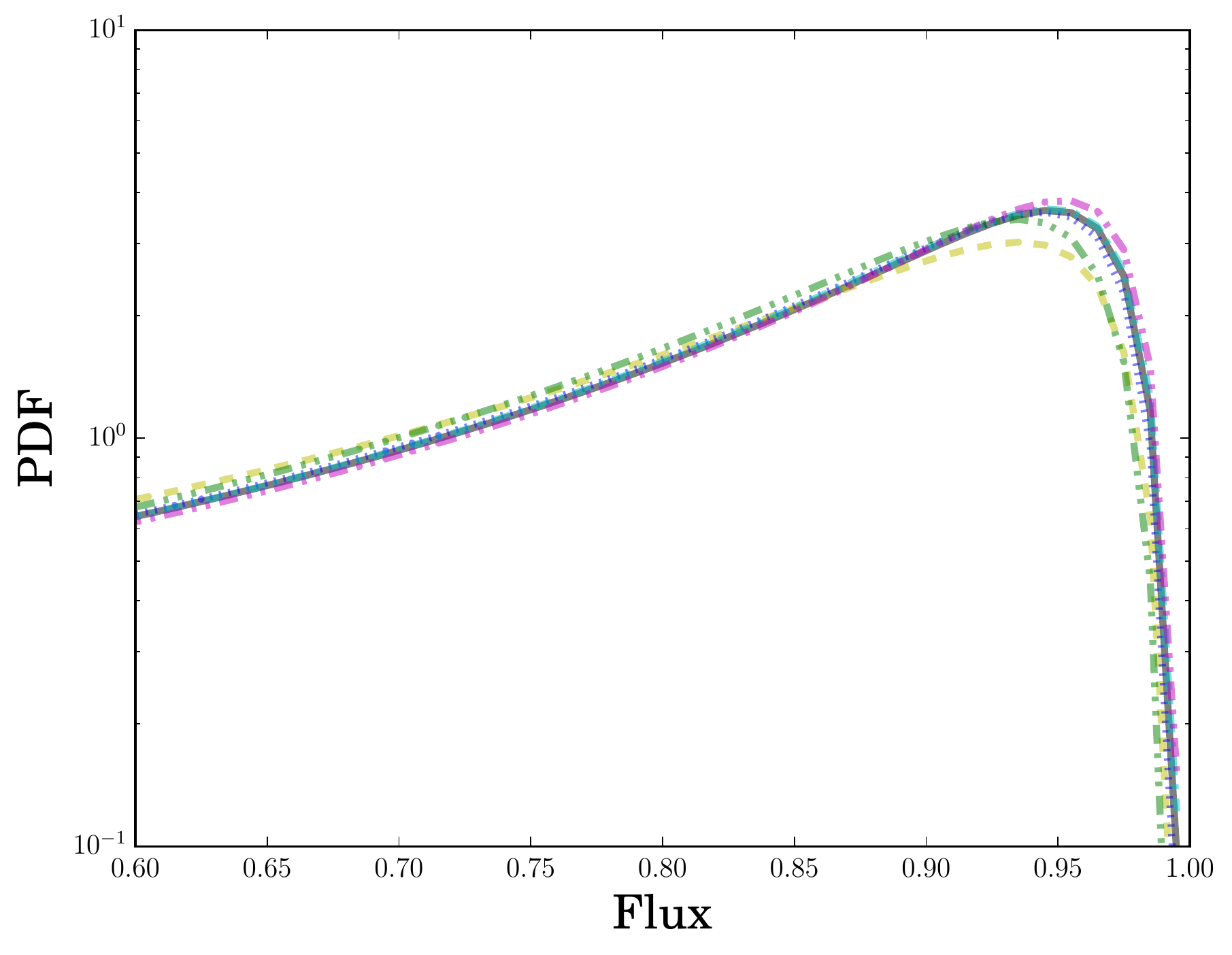}}%
  \resizebox{0.48\textwidth}{!}{\includegraphics{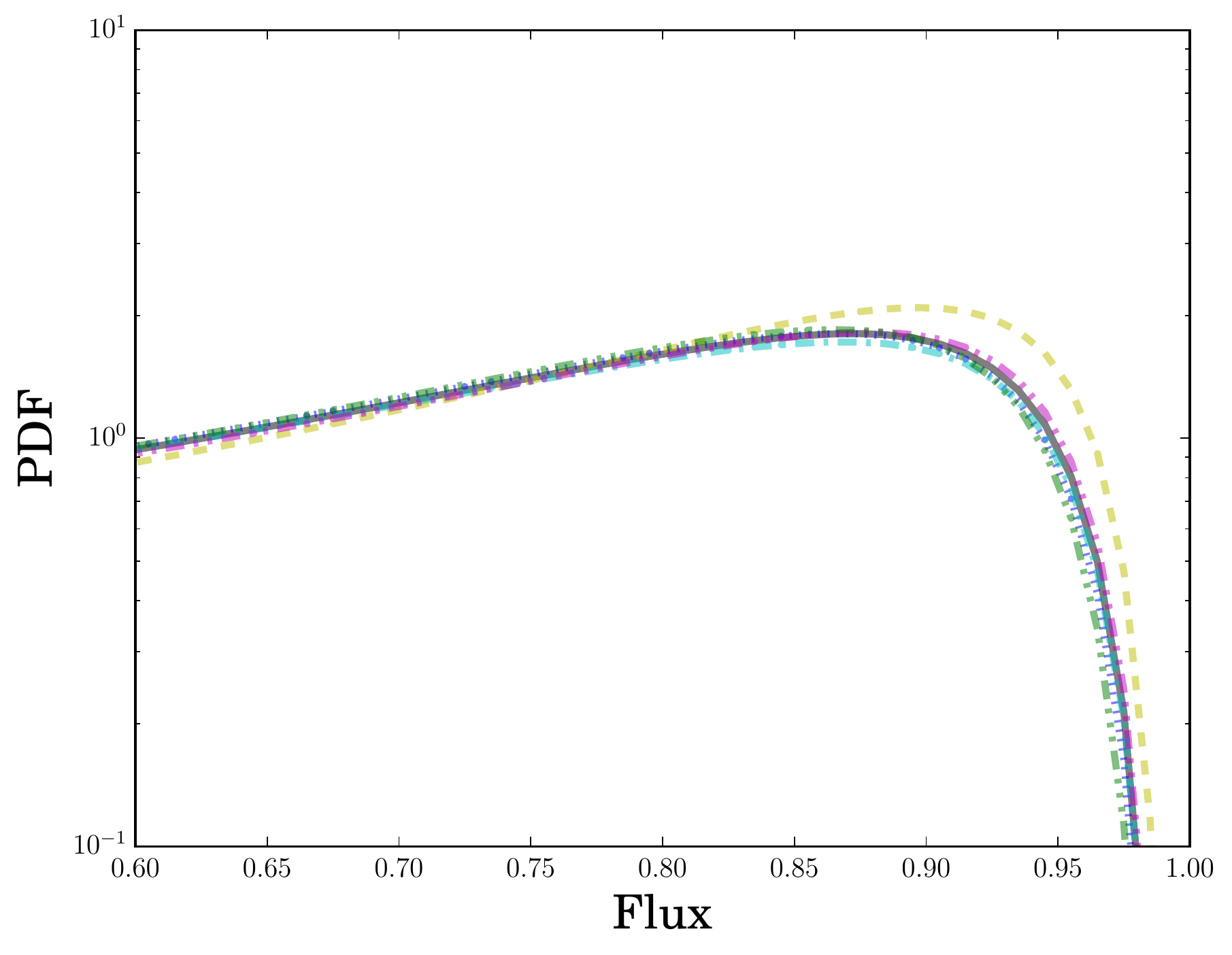}} \\
  \caption{Same plots as in Figure~\ref{fig:fluxpdf_HI}, but with binning
    performed at higher resolution. The redshifts chosen are $z \sim 2.5$ (top),
    $z \sim 3$ (bottom left), and $z \sim 3.5$ (bottom right), the same as in
    the previous figure. In addition, only flux values of $0.6 \leq F \leq 1.0$
    are plotted to emphasize the most different portions of the flux PDF. The
    flux PDF shows a much more gradual transition at $F \sim 1$ than the PDF
    conveyed in Figure~\ref{fig:fluxpdf_HI}, further emphasizing that the
    resolution and binning of the flux PDF are important for understanding
    it. The higher resolution also makes the differences between simulations
    more apparent, especially at $z \sim 3$.}
  \label{fig:hires_fluxpdf}
\end{figure*}

From a simulation point of view, the resolution of the gas grid (and to a lesser
extent, the radiation grid) affects the resolution of the Ly$\alpha$ forest. For
the default-resolution grid at $z \sim 3$, a single gas cell has an equivalent
velocity width of $\Delta v = 7.3$ km~s$^{-1}$. This resolution level is
significantly greater than that of BOSS ($\Delta v \sim 69$ km~s$^{-1}$,
\citealt{lee_etal2015}), though not as good as Keck-HIRES ($\Delta v \sim 6.6$
km~s$^{-1}$, \citealt{lu_etal1996}).

Figure~\ref{fig:fluxpdf_HI} shows the flux PDF of the \HI\ Ly$\alpha$ forest as
a function of redshift across the various simulations. The figure also includes
the measurements of \citet{calura_etal2012}. The spectra from
\citet{calura_etal2012} were taken at UVES, with a FWHM of 6.7 km~s$^{-1}$,
slightly better than the resolution of our simulations. As a result, the
different resolution may have a non-trivial impact on the shape of the resulting
flux PDF. The flux PDF in general has a similar shape for different simulations
at the same redshift although the simulations have different \HeIII\ ionization
fractions and thermal histories. This result implies that given the same
underlying gas structure, the flux PDF depends on having the same value of
$\tau_\mathrm{eff}$. Given the same large-scale structure and
$\tau_\mathrm{eff}$, our result shows that helium reionization is largely
undetectable in the hydrogen flux PDF.

Nevertheless, there are still several trends that are visible upon closer
inspection. After helium reionization is largely completed at $z \sim 2.5$, the
values of the flux PDF in the highest transmission bin of $F \sim 1$ are ordered
by the helium ionization fraction: Simulation \ampdown\ has the highest value in
this bin, and Simulation \uvb\ has the lowest.  Helium~\textsc{ii} reionization
is still ongoing for Simulation \ampdown, whereas for the other simulations,
reionization is largely completed (Figure~\ref{fig:fHe}).

We can understand this trend by employing the fluctuating Gunn-Peterson
approximation (FGPA, \citealt{croft_etal1998}). The FGPA assumes that the gas of
the IGM accurately follows a temperature-density relation of the form found in
Equation~(\ref{eqn:tdelta}) and is in photoionization equilibrium with a
uniform ionization background. Under these assumptions, the local optical depth
of the IGM $\tau_\mathrm{HI}$ can be expressed in terms of the gas density, mean
temperature of the IGM, and the \HI\ photoionization rate, along with other
cosmological parameters. In particular, it can be shown that the optical depth
is related to the temperature as $\tau_\mathrm{HI} \propto T^{-0.7}$. Thus, for
reionization histories with a higher average temperature, there is a decreased
local value of $\tau$, leading to an overall higher flux value everywhere, but
in low-density regions in particular. Therefore, the comparatively high value
for the flux PDF in the bin where $F \sim 1$ for Simulation \ampdown\ can be
interpreted as conveying information about the thermal state of the IGM. Indeed,
\citet{lee_etal2015} have proposed using the flux PDF to gain information about
the thermal state of the IGM at different redshifts.
 
One point to note is the visible difference between the observations of
\citet{calura_etal2012} at $z \sim 2.9$ and the results from the simulations at
$z \sim 3$, shown in the bottom left panel in the plot. The flux PDF at
intermediate flux values in the simulations is higher than that of the
observations, until the highest bin (where there is almost total flux
transmission). Part of the difference can be attributed to the fact that the
simulations and the observations are normalized to different values of
$\tau_\mathrm{eff}$: the simulations use the value from \citet{lee_etal2015},
whereas the observational results determine the parameters for
$\tau_\mathrm{eff}(z)$ based on their measurements. At $z \sim 3$, the results
for $\tau_\mathrm{eff}$ from \citet{lee_etal2015} are higher than those from
\citet{calura_etal2012} by about 30\%. This result accounts for some of the
difference in the flux PDF, but not all of it. (See Appendix
\ref{appendix:fluxpdf} for further discussion of the renormalization effect.)
Alternatively, as discussed in \citet{calura_etal2012}, the continuum-level
estimation of the observational Ly$\alpha$ forest can significantly affect the
shape of the flux PDF. As shown in Figure~8 of \citet{calura_etal2012},
increasing the continuum level by 5\% modifies the shape of the flux PDF to be
comparable to the levels seen in the simulations. Thus, a combination of
changing $\tau_\mathrm{eff}$ of the simulations and the continuum-level of the
observations can bring the simulations and observations into agreement.

Figure~\ref{fig:hires_fluxpdf} shows the flux PDF for the simulations, but
binned at higher resolution than in Figure~\ref{fig:fluxpdf_HI}. The
increased resolution in the binning shows a much more gradual transition at
$F \sim 1$ than the stark fall-off in Figure~\ref{fig:fluxpdf_HI}. This
higher-resolution binning also clarifies the differences between the
simulations. As mentioned above, these differences are likely due to the thermal
state of the IGM, since in the FGPA the absorption is proportional to the
temperature. High-resolution measurements of the flux PDF may therefore yield
information about the thermal state of the IGM, although as discussed in
Appendix~\ref{appendix:fluxpdf}, the determination of the continuum-level flux
for observations remains a significant systematic uncertainty. The continuum
level plays a significant role here as well, since it determines the
distribution of $F sim 1$ pixels when generating a PDF.

\subsection{One-dimensional flux power spectra}
\label{sec:1dps}

\begin{figure*}[t]
  \centering
  \resizebox{0.6\textwidth}{!}{\includegraphics{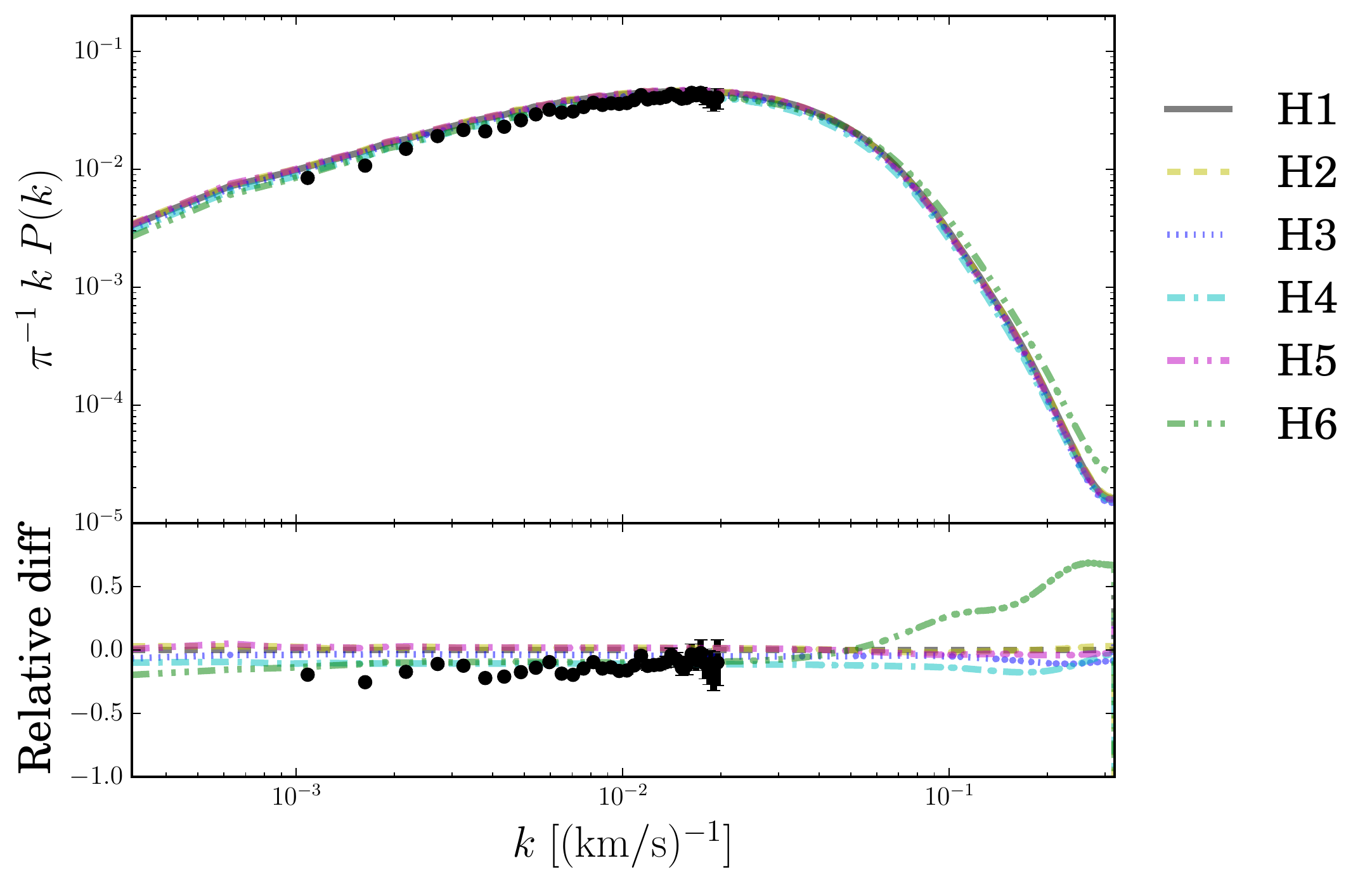}} \\
  \resizebox{0.48\textwidth}{!}{\includegraphics{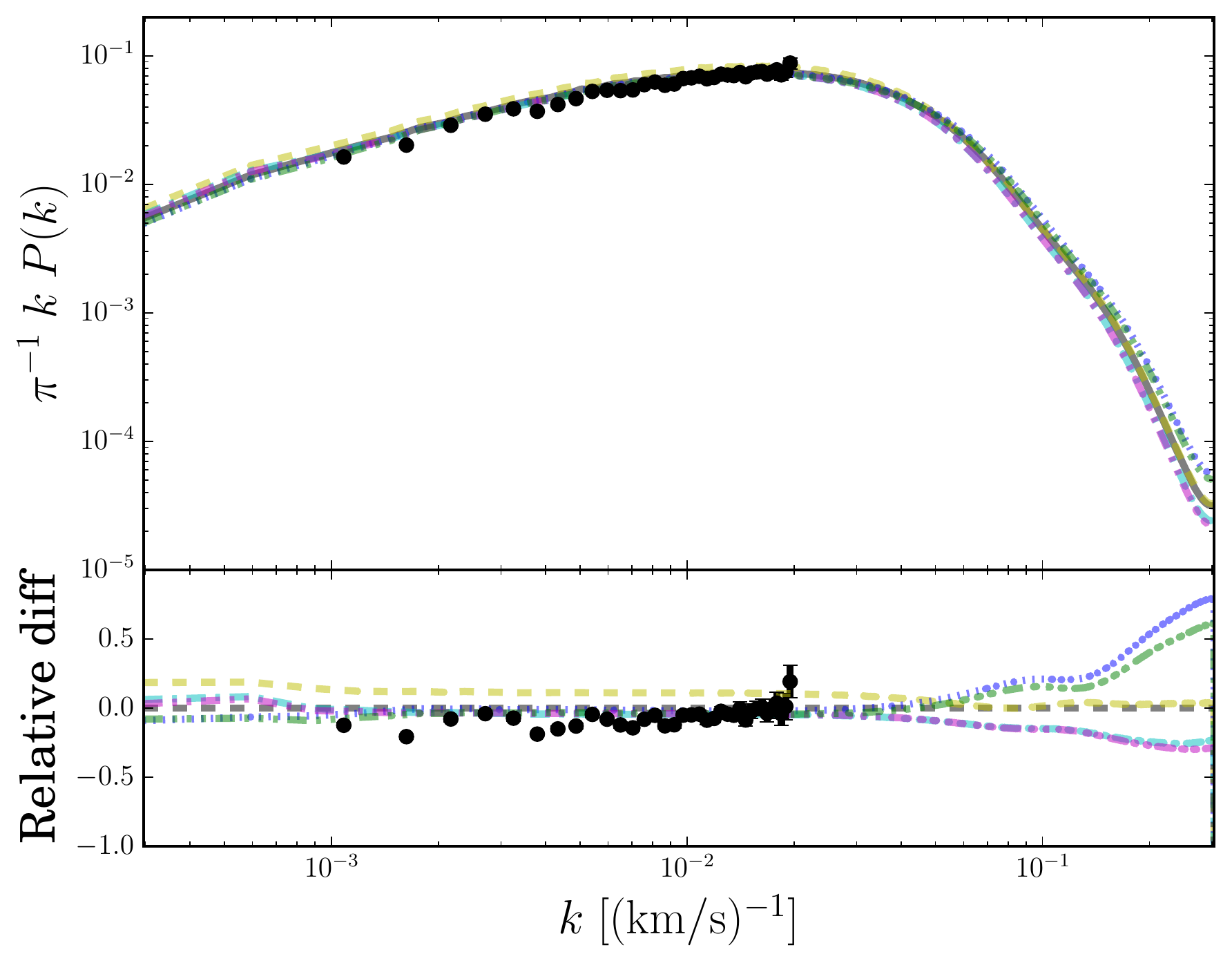}}%
  \resizebox{0.48\textwidth}{!}{\includegraphics{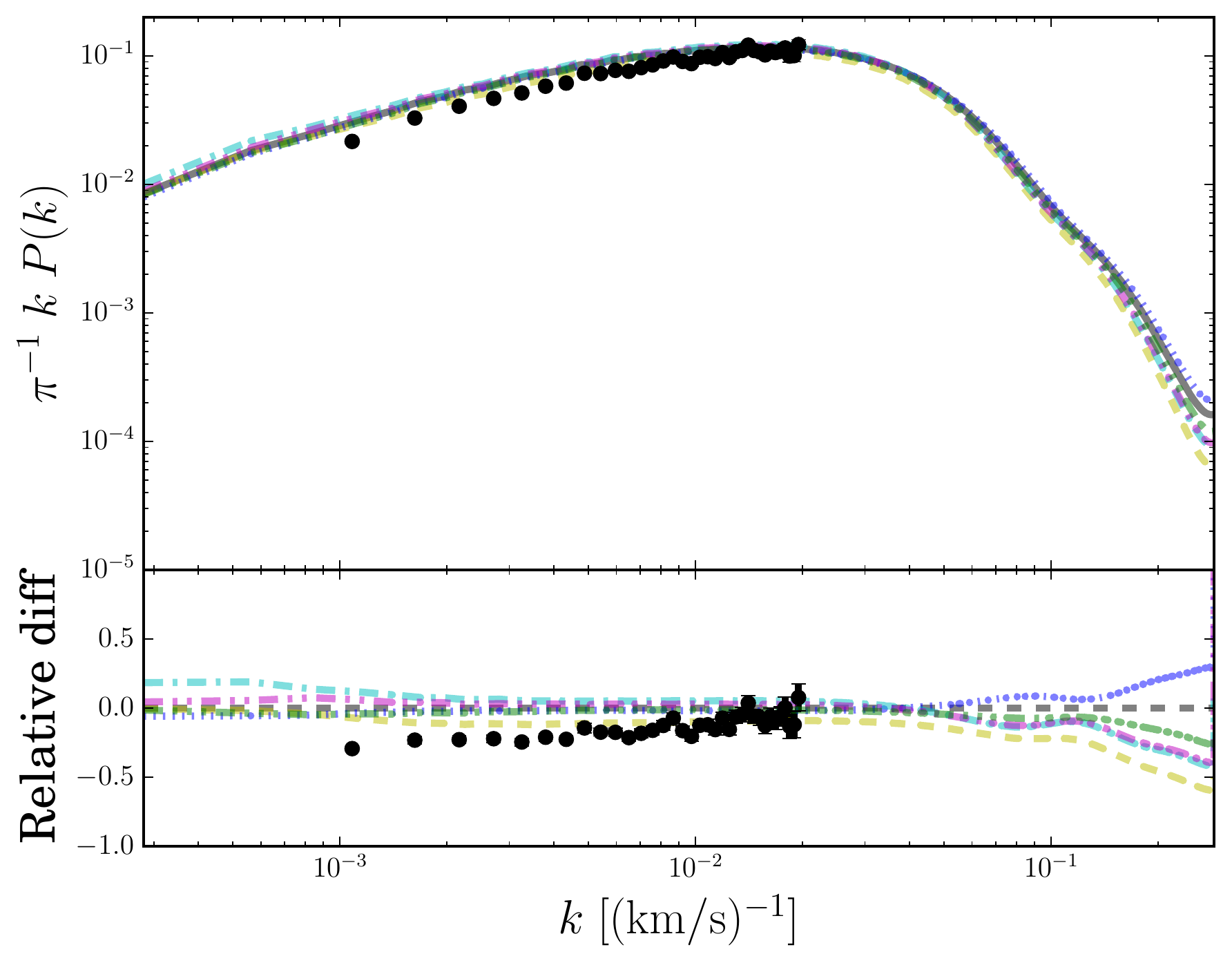}} \\
  \caption{One-dimensional flux power spectrum of the \HI\ Ly$\alpha$ forest at
    $z \sim 2.5$ (top), $z \sim 3$ (bottom left), and $z \sim 3.5$ (bottom
    right). Black dots with error bars are observational data measured from BOSS
    as reported by \citet{palanque-delabrouille_etal2013} as reported at
    $z \sim 2.4$, $z \sim 3$, and $z \sim 3.4$, respectively. In general, the
    data points agree with the simulations, although this agreement at these
    scales is likely due primarily to having comparable values of
    $\tau_\mathrm{eff}$. The effects of helium~\textsc{ii} reionization are
    visible in the small-scale power. Specifically, simulations in which the
    average temperature of the IGM is higher show lower power on small
    scales. This is due to the thermal motion of the gas, which washes out some
    of the small-scale structure. The overall amplitude of the power spectrum
    tends to decrease with redshift, since the total number density of hydrogen
    is decreasing. See the text for additional discussion.}
  \label{fig:oned_HI}
\end{figure*}

In addition to the statistics already discussed, the one-dimensional flux power
spectrum can provide valuable information about underlying dark matter density
distributions. To calculate the one-dimensional flux power spectrum, we first
define a ``flux overdensity'' $\delta_F$ for each pixel:
\begin{equation}
\delta_F \equiv \frac{F}{\ev{F}} - 1,
\end{equation}
where $\ev{F}$ is the average flux for all pixels in the volume (which is also
typically close to the average flux within a given sightline because of the
length of the sightlines). After defining this quantity, a Fourier transform is
applied to each sightline, so that we have $\delta_F(k)$. The one-dimensional
power spectrum $P_\mathrm{1D}(k)$ is the average power per $k$-mode:
$P_\mathrm{1D}(k) = \ev*{\abs{\delta_F(k)}^2}$. In the following analysis, we
primarily study the dimensionless power spectrum,
\begin{equation}
\Delta^2_\mathrm{1D}(k) = \frac{k}{\pi} P_\mathrm{1D}(k).
\end{equation}
Previous studies have shown that the one-dimensional power spectrum can be used
to measure the three-dimensional power spectrum
\citep{croft_etal1998,mcdonald_etal2005,mcdonald_eisenstein2007}, although here
we explore the one-dimensional power spectrum per se and treat the
three-dimensional power spectrum separately in Section~\ref{sec:3dps}. As with
the flux PDF, the amplitude of the one-dimensional power spectrum on large
scales is largely similar between the different reionization scenarios at the
same redshift. However, there are significant differences on small scales
($k \gtrsim 0.1\ (\mathrm{km/s})^{-1}$). This is likely due to the differences
in the thermal histories of the IGM. In particular at $z \sim 2.5$, Simulation
\uvb\ shows a greater amplitude than many of the other simulations and also has
a cooler temperature (see Figure~\ref{fig:temp}). The cooler temperature is
correlated with additional power at small scales, which is consistent with
additional structure as a result of cooler gas.

Figure~\ref{fig:oned_HI} shows the one-dimensional power spectrum of the
Ly$\alpha$ forest for redshifts $z \sim 2.5$, $z \sim 3$, and $z \sim
3.5$. As with the flux PDF in Figure~\ref{fig:fluxpdf_HI}, the simulations show
largely similar result. Nevertheless, key differences due to the effects of
helium~\textsc{ii} reionization are still visible. Most of the differences
between simulations are visible at small scales. In general, the simulations
that have a hotter average temperature of the IGM show less power at small
scales. This is due to the decrease in clumping that results from the increased
thermal motion of the gas. On large scales, the differences between the
simulations are typically smaller than 10\%.

The data points in Figure~\ref{fig:oned_HI} are the results from
\citet{palanque-delabrouille_etal2013} from BOSS. The data plotted show redshift
values of $z \sim 2.4$, $z \sim 3$, and $z \sim 3.4$, compared to the redshift
values of $z \sim 2.5$, $z \sim 3$, and $z \sim 3.5$ from the simulations. The
slight discrepancy in redshift for the plots may lead to some of the differences
seen, especially for the plot from the simulations at $z \sim
3.5$. Nevertheless, there is good agreement in general between the data and the
simulations. A primary driver of this agreement may be the similar values of
$\tau_\mathrm{eff}$ in the data and simulations. The value of
$\tau_\mathrm{eff}$ in the simulations matches that of \citet{lee_etal2015} is
based on SDSS DR7 data, whereas the results from
\citet{palanque-delabrouille_etal2013} include additional data from BOSS. On the
scales where the data are reported, the difference between the different
simulations is small. As such, the data are not able to break the degeneracy
between the simulations.

\begin{figure*}[t]
  \centering
  \resizebox{0.6\textwidth}{!}{\includegraphics{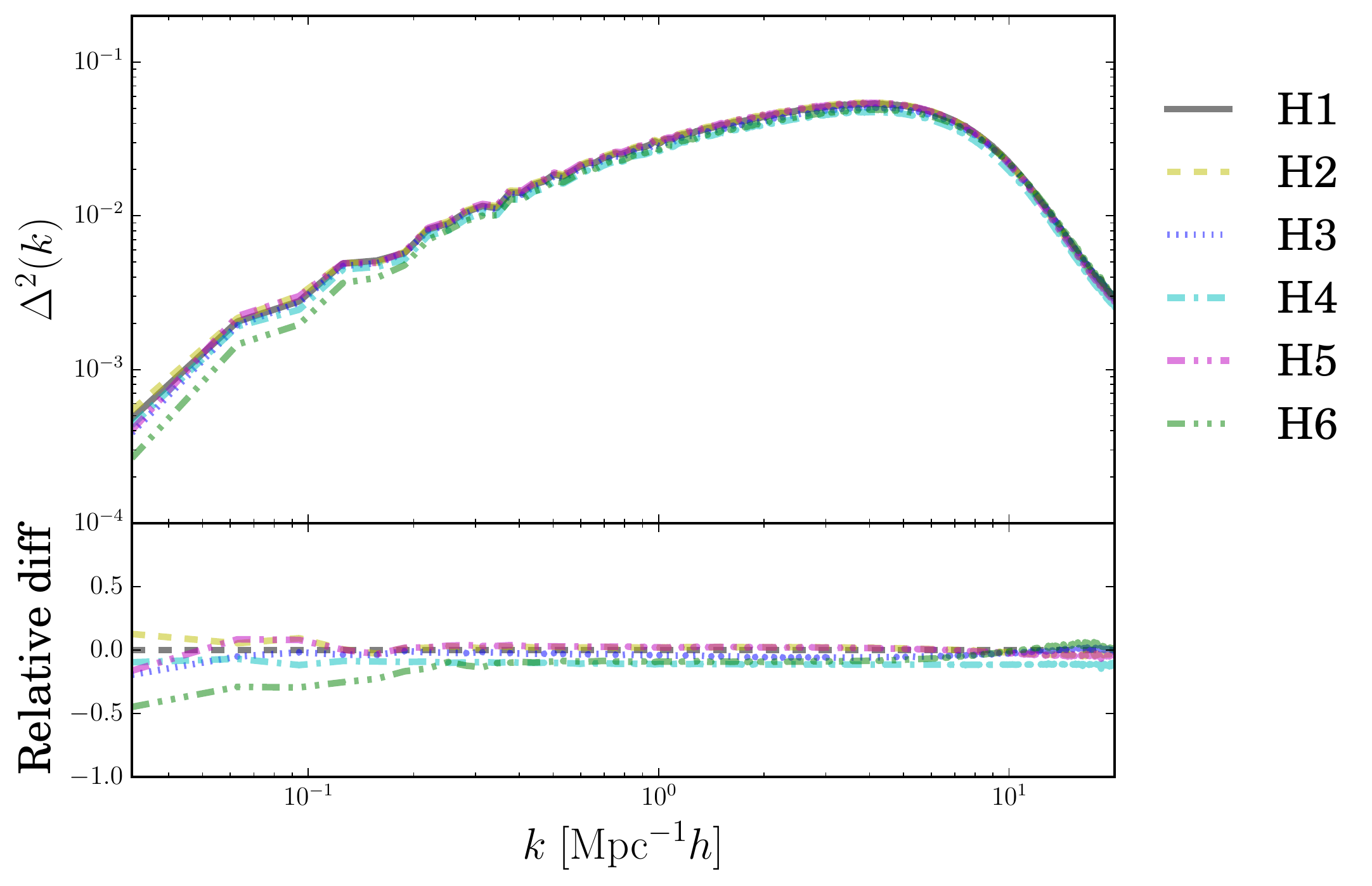}} \\
  \resizebox{0.48\textwidth}{!}{\includegraphics{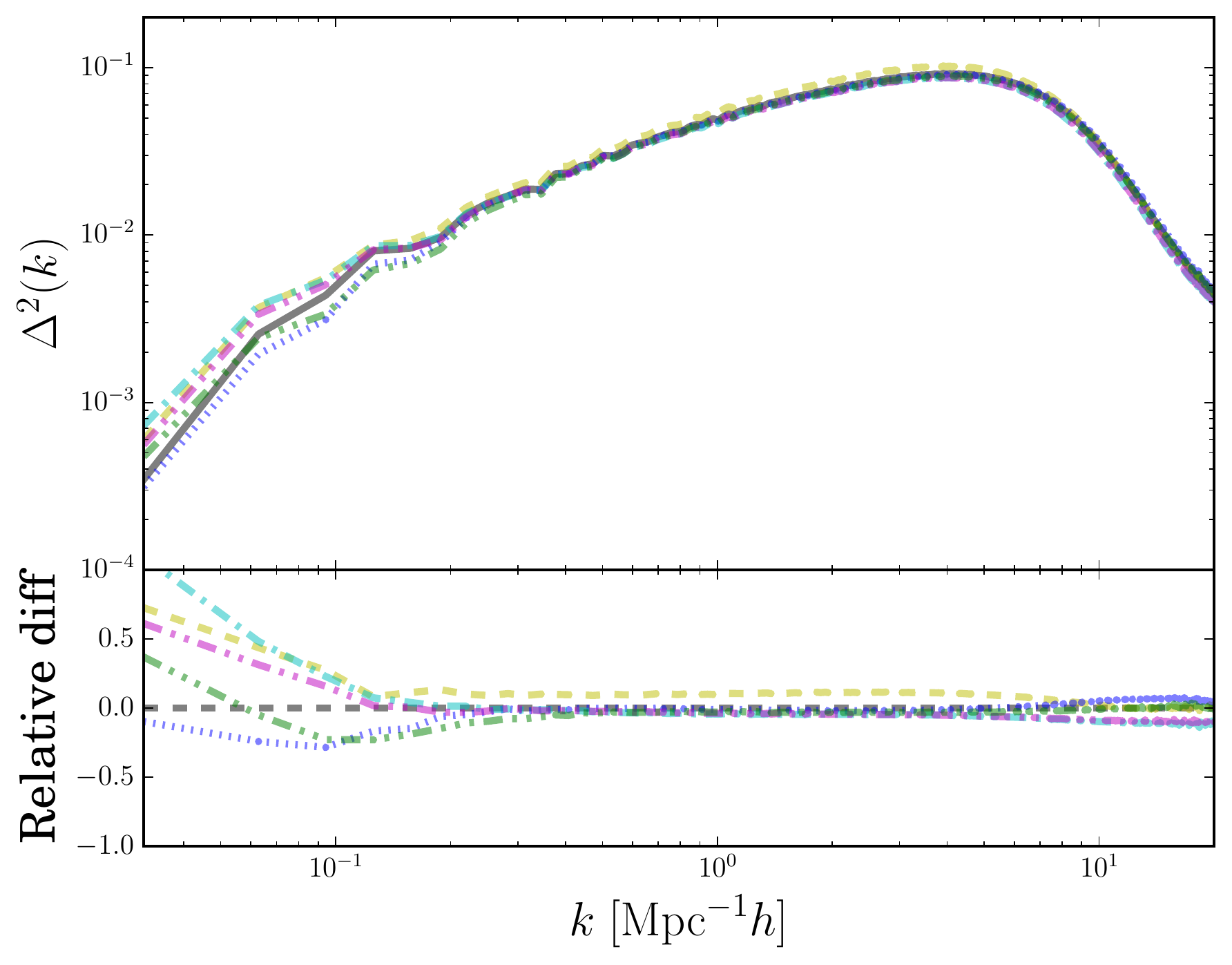}}%
  \resizebox{0.48\textwidth}{!}{\includegraphics{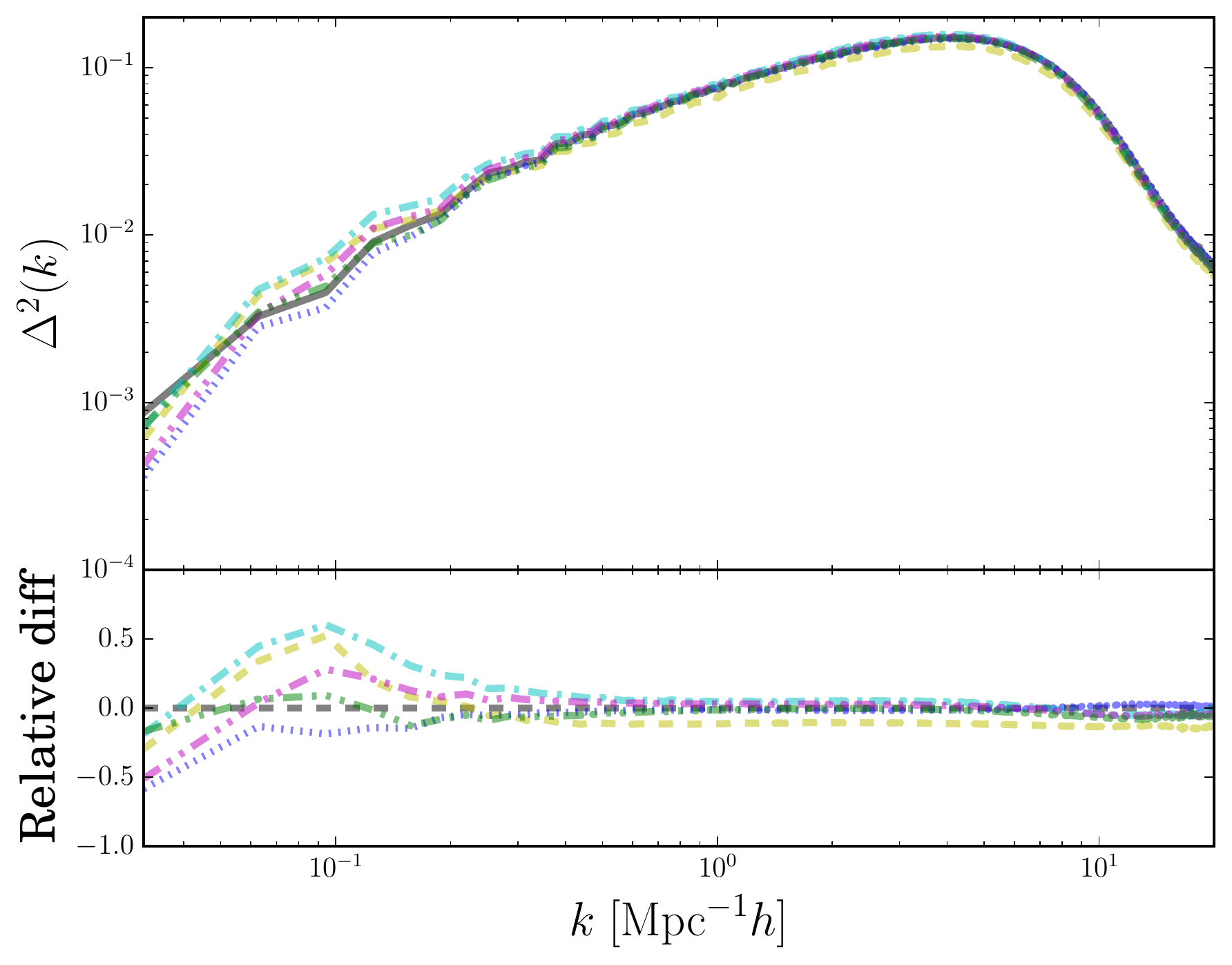}} \\
  \caption{Three-dimensional power spectrum of the \HI\ Ly$\alpha$ forest flux
    at $z \sim 2.5$ (top), $z \sim 3$ (bottom left), and $z \sim 3.5$ (bottom
    right). The most significant differences between the simulations emerge on
    large scales. In previous studies on the three-dimensional power spectrum
    \citep{mcdonald2003}, the deviations due to different thermal histories of
    the IGM led to differences of $\sim 10\%$, consistent with the results
    presented here. The dramatic differences on large scales may be due to
    correlations in the radiation field that are the result of quasar
    emission. See the discussion in Section~\ref{sec:3dps} for further details.}
  \label{fig:threed_HI}
\end{figure*}

One important point is that the small-scale structure of the one-dimensional
power spectrum is its dependence on the thermal history of the gas. The power
spectrum is sensitive not only to the current temperature of the IGM, but also
to its past temperature, a phenomenon first pointed out in
\citet{gnedin_hui1998}. The power on small scales is set by Jeans smoothing in
the gas, which is caused by the propagation of pressure waves in the gas and
hence depends on the sound speed in the gas. Because the sound speed depends on
the temperature of the gas (for an ideal gas, $c \propto T^{1/2}$), the thermal
history sets the maximum scale over which a pressure wave can travel in the
IGM. In the bottom left panel of Figure~\ref{fig:oned_HI} at $z \sim 3$, on
small scales the simulations with the most power are Simulation \uvb\ and
Simulation \ampdown. According to Figure~\ref{fig:temp}, the temperature of the
mean-density gas is similar between the two simulations. However, in the case of
Simulation \uvb, the temperature is decreasing after having reached an earlier
peak, whereas in Simulation \ampdown, the temperature is increasing from a
relatively cool phase after hydrogen reionization. Accordingly, there is
additional power in the smallest scales for Simulation \ampdown, which is
consistent with the findings of \citet{gnedin_hui1998}.

\subsection{Three-dimensional flux power spectra}
\label{sec:3dps}

We have also made predictions for the full three-dimensional flux power spectrum
of the \HI\ Ly$\alpha$ forest. To compute this quantity, we have generated the
full number of sightlines in the volume of $N_\mathrm{grid}^2$, which provides
the full three-dimensional information about the volume. Several previous
studies \citep{croft_etal1998} instead differentiated the one-dimensional power
spectrum to extract the three-dimensional information. Our approach of using the
full set of correlations present in the underlying density field, as well as
yielding the power spectrum at finer resolution in $k$-space. The information
contained in the three-dimensional flux power spectrum can contain information
about the state of the gas of the IGM
\citep{pichon_etal2001,mcdonald2003,caucci_etal2008,cisewski_etal2014,ozbek_etal2016},
which would provide an exciting window into the IGM at high
redshift. Additionally, several previous studies have started to measure the
full three-dimensional power spectrum using quasar sightlines from SDSS
\citep{slosar_etal2011,lee_etal2014}, which have provided important insight. In
principle, like the one-dimensional flux power spectrum, the three-dimensional
flux power spectrum can reveal important information about the thermal history
of the IGM \citep{gnedin_hui1998}, as well as the large-scale distribution of
matter.

Figure~\ref{fig:threed_HI} shows the three-dimensional power spectrum of the
\HI\ Ly$\alpha$ forest flux. The general shape of the power spectrum is similar
to that of the one-dimensional version seen in Figure~\ref{fig:oned_HI},
although the drop in power at high-$k$ is not as pronounced. More importantly,
there are observable differences on large scales between the different
reionization histories, which can differ by up to a factor of 2. Importantly,
the gas power spectrum of all of the simulations is essentially identical on
large scales, so all differences are due to the different ionization histories
of the IGM and not to the the underlying matter or gas distribution.

The differences in power at large scales are likely due to the correlations in
the radiation field in the IGM. As mentioned in \citet{mcdonald2003},
differences in the thermal state of the IGM (either the temperature $T_0$ or the
slope $\gamma$) only lead to differences at the $\sim 10\%$ level, which is
consistent with the results seen in Figure~\ref{fig:threed_HI}. The differences
on large scales are significantly larger than this and furthermore do not seem
to be correlated with particular values of $T_0$ and $\gamma$. Indeed, when we
compare this with the values in Figure~\ref{fig:t0gamma}, the power on large
scales does not seem to be correlated with either value, further demonstrating
that the thermal history alone is not responsible for the differences on large
scales.

Proper characterization of the full three-dimensional power spectrum is
important for measurements of the BAO from the Ly$\alpha$ forest
\citep{busca_etal2013,slosar_etal2013}. As can be seen in
Figure~\ref{fig:threed_HI}, there are differences on large scales, in some cases
as large as a factor of two between the different reionization scenarios. Thus,
properly understanding the impact that the reionization of helium has on the
three-dimensional power spectrum is important for systematic errors for the BAO
measurement.

\section{Conclusion}
\label{sec:conclusion}
In this paper, we have presented a new suite of simulations that couple $N$-body
methods, hydrodynamics, and radiative transfer simultaneously in order to study
helium~\textsc{ii} reionization. Some of the most important observational
implications that helium~\textsc{ii} reionization leaves on the low-density gas
of the IGM come from the dramatic increase in temperature from the photoheating
of the gas. Using the results of the simulations, we summarize here several
conclusions that we can make:
\begin{enumerate}
\item In addition to changing the ionization fraction of helium as a function of
  redshift $x_\mathrm{HeIII}(z)$, helium~\textsc{ii} reionization also leaves an
  important signature on the thermal history of the IGM. This finding is
  consistent with previous studies of helium~\textsc{ii} reionization, which
  suggest using the temperature of the IGM to learn about helium~\textsc{ii}
  reionization. We show that the peak in the temperature at mean density as a
  function of redshift $T(z)$ is a relatively robust signifier of
  helium~\textsc{ii} reionization, occurring when the volume is 90-95\% ionized
  by volume. The redshift interval over which the temperature of the IGM
  increases can be used to determine the duration of reionization, although this
  measurement is observationally less straightforward.
\item Observations of synthetic \HI\ Ly$\alpha$ sightlines show that many
  statistics concerning the forest are similar when we control the value of
  $\tau_\mathrm{eff}$, although important differences caused by different
  thermal histories may still be detectable. In particular, the one-dimensional
  power spectrum and the flux PDF show important differences that can be
  understood in terms of the thermal state of the IGM. These differences can be
  substantial, especially in the small-scale one-dimensional power spectrum.
\item The three-dimensional flux power spectrum shows significant differences
  between the simulations, with differences on large scales of up to a factor of
  2. Previous studies have attempted to measure this quantity
  \citep{slosar_etal2011,lee_etal2014}, although the error bars are still
  significant.
\end{enumerate}
In future studies, we plan to investigate the effect that anisotropic sources
have on helium~\textsc{ii} reionization. The effect was discussed briefly in
\citet{mcquinn_etal2009}, although we plan to explore this aspect more
thoroughly. In addition, we plan to detect observational signatures in the
Ly$\beta$ and Ly$\gamma$ forests. \citet{irsic_viel2014} showed that temperature
information of the IGM was better determined by studying the Ly$\beta$ forest
and the cross-correlation with the Ly$\gamma$ forest. Additionally, these
transitions saturate at much higher neutral hydrogen densities, and so they can
give additional information about the thermal state of the IGM at higher
densities. This type of comparison can provide an additional observational tool
for understanding helium~\textsc{ii} reionization, and provide another point of
comparison with observations.

\acknowledgements{We thank Francesco Haardt and Piero Madau for providing us
  with a version of their uniform UV background with contributions only from
  galaxies. We thank Matt Cary and all of the support staff at the NASA Ames
  Research Center for invaluable help with computing. This work was supported in
  part by NASA grants NNX14AB57G and NNX12AF91G, and NSF grants AST 1312724 and
  AST15-15389.}

\begin{appendix}

\section{Renormalizing $\tau_\mathrm{eff}$}
\label{appendix:hm}
In several previous observational studies of the \HI\ Ly$\alpha$ forest
\citep{theuns_etal2002,bernardi_etal2003,dallaglio_etal2008,faucher-giguere_etal2008a},
a dip in the effective optical depth $\tau_{\mathrm{eff}}$ at $z \sim 3.2$ was
reported. It was proposed that this dip might be related to helium
reionization. Several subsequent studies
\citep{bolton_etal2009a,bolton_etal2009b,mcquinn_etal2009,compostella_etal2013}
did not reproduce this feature. In particular, the functional form of
$\tau_{\mathrm{eff}} (z)$ from \citet{lee_etal2015} does not include this
feature. As explained in Section~\ref{sec:features}, the usual approach taken in
the simulations is to renormalize the photoionization rate of galaxies
$\Gamma_{\mathrm{gal}}$ in order to reproduce $\tau_{\mathrm{eff}}(z)$ by
construction. As a result, the potential dip at $z \sim 3.2$ would not
appear. To study whether this feature emerges from the simulations without
renormalization, we have run Simulation \hm, which uses the same simulation
parameters as \fid, but with $\Gamma_{\mathrm{gal}}$ provided by the model of
HM12. To isolate the contribution of the galaxies, the authors of HM12 have
furnished a series of photoionization rates and photoheating rates that only
include the contribution from galaxies, and do not include quasars (P. Madau
2017, private communication). Thus, we are able to determine if the dip in
$\tau_{\mathrm{eff}}$ can be reproduced in our simulations.

To explore the extent to which helium~\textsc{ii} reionization affects the
Ly$\alpha$ forest, we have not explicitly matched the value of
$\tau_\mathrm{eff}$ for this simulation. The value of $\tau_\mathrm{eff}$ for
Simulation \hm\ differs from that of the other simulations, although it does not
differ by more than a factor of 2. Initially, the value of $\tau_\mathrm{eff}$
is greater than the other simulations (showing lower overall flux), and then
crosses over to become lower than the other simulations around $z \sim 3$.
Furthermore, there is no significant dip in $\tau_\mathrm{eff}$ at $z \sim 3.2$,
or any other point in the evolution of the simulation. This lack of a feature in
$\tau_\mathrm{eff}$ is consistent with more recent findings
\citep{bolton_etal2009a,bolton_etal2009b,mcquinn_etal2009,compostella_etal2013}.

\section{Renormalizing the Ly$\alpha$ Flux PDF}
\label{appendix:fluxpdf}

\begin{figure}[t]
  \centering
  \includegraphics[width=0.45\textwidth]{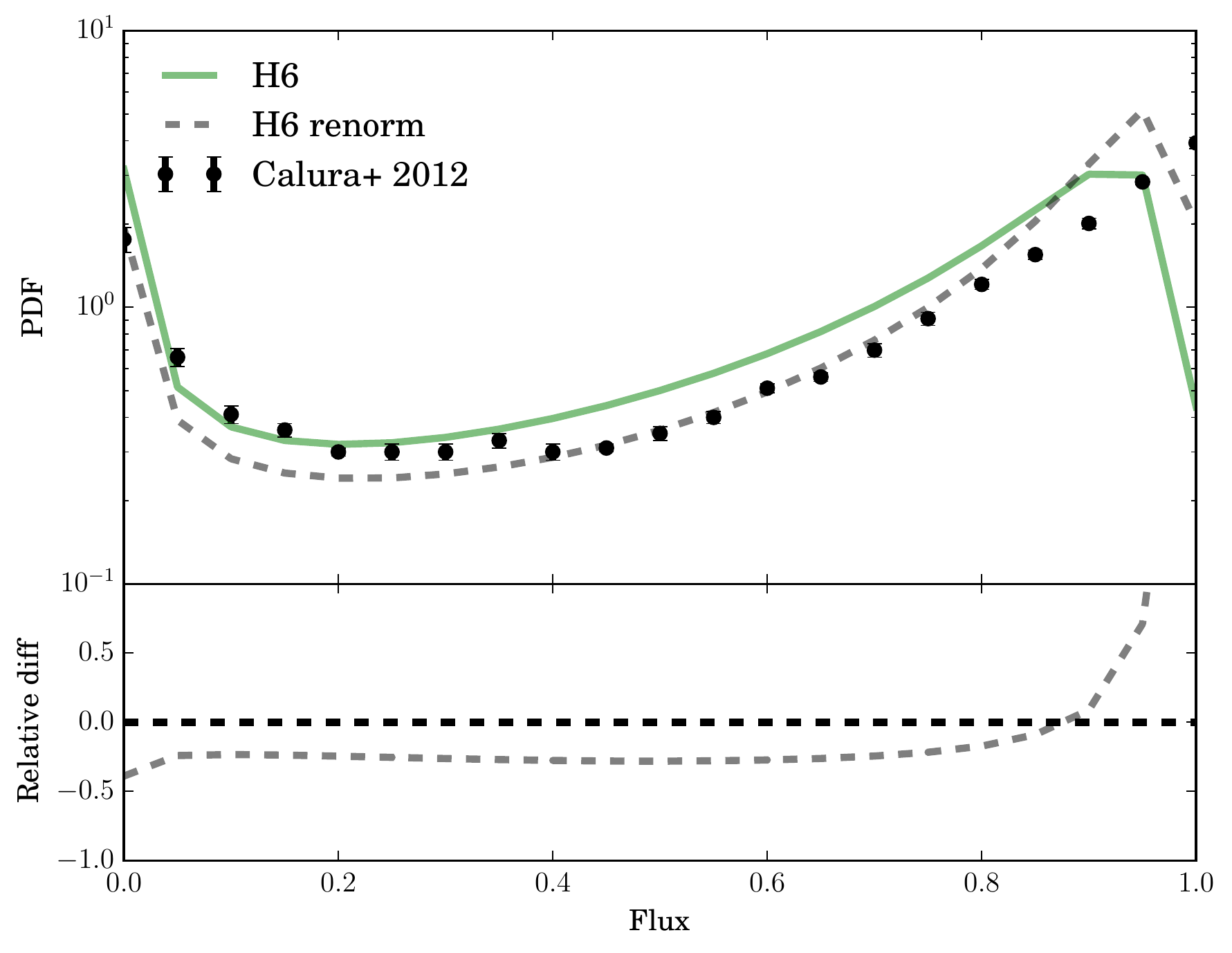}
  \caption{Comparison of the Ly$\alpha$ forest flux PDF of Simulation \uvb\
    using the default value of $\tau_\mathrm{eff}$ from \citet{lee_etal2015}
    (labeled ``fiducial'' in the Figure) and a renormalized value of
    $\tau_\mathrm{eff}$ from \citet{calura_etal2012}, as well as the data from
    \citet{calura_etal2012}. The discrepancy at high flux values is smaller in
    the case of the renormalized value of $\tau_\mathrm{eff}$, which is
    consistent with the fact that the value of $\tau_\mathrm{eff}$ is lower for
    \citet{calura_etal2012}. Nevertheless, the difference in the normalization
    cannot account for the whole discrepancy. As shown in Figure~8 of
    \citet{calura_etal2012}, the continuum uncertainty can have a significant
    effect on the shape of the flux PDF. Thus, a proper estimation of the
    continuum level for observations is essential for understanding the flux
    PDF.}
  \label{fig:tau_renorm}
\end{figure}

In Section~\ref{sec:fluxpdf} we discussed the results of measuring the
Ly$\alpha$ forest flux PDF for the different simulations. In order to compare
them against observation, Figure~\ref{fig:fluxpdf_HI} shows the measurement of
the flux PDF from \citet{calura_etal2012} at $z \sim 2.9$. There is a noticeable
difference in the shape between the observational and simulated results. As
explained in Section~\ref{sec:fluxpdf}, there is a difference in the measured
$\tau_\mathrm{eff}$ of the measurements compared to our simulations, which used
the more recent measurements of \citet{lee_etal2015} to define the value of
$\tau_\mathrm{eff}(z)$ that the simulations matched. In order to investigate
whether the difference in the flux PDF shape could be attributed entirely to the
different value of $\tau_\mathrm{eff}$, we adjusted the average flux absorption
$\ev{F}$ of the volume to match the lower value of $\tau_\mathrm{eff}$ from
\citet{calura_etal2012}.

Figure~\ref{fig:tau_renorm} shows the flux PDF of Simulation \uvb\ renormalized
to have the same value of $\tau_\mathrm{eff}$ as \citet{calura_etal2012}, and
the measurements. The figure also shows the original flux PDF for the default
normalization. The value of $\tau_\mathrm{eff}$ reported by \citet{lee_etal2015}
is greater than the value reported by \citet{calura_etal2012}. Accordingly, when
the spectra have been renormalized to have the same value of
$\tau_\mathrm{eff}$, some of the discrepancy between the simulations and the
measurements has been removed. Nevertheless, there is still some difference
between the measurements, especially for the bins of high flux ($F \sim
1$). Thus, this difference cannot be attributed entirely to the difference in
$\tau_\mathrm{eff}$. As discussed in \citet{calura_etal2012}, the placement of
the continuum-level can have a significant effect on the shape of the flux
PDF. Additionally, the difference in effective resolution between the
simulations and the observations may also play some role.

\section{The Quasar Luminosity Function}
\label{appendix:qlf}

\begin{figure}[t]
  \centering
  \includegraphics[width=0.45\textwidth]{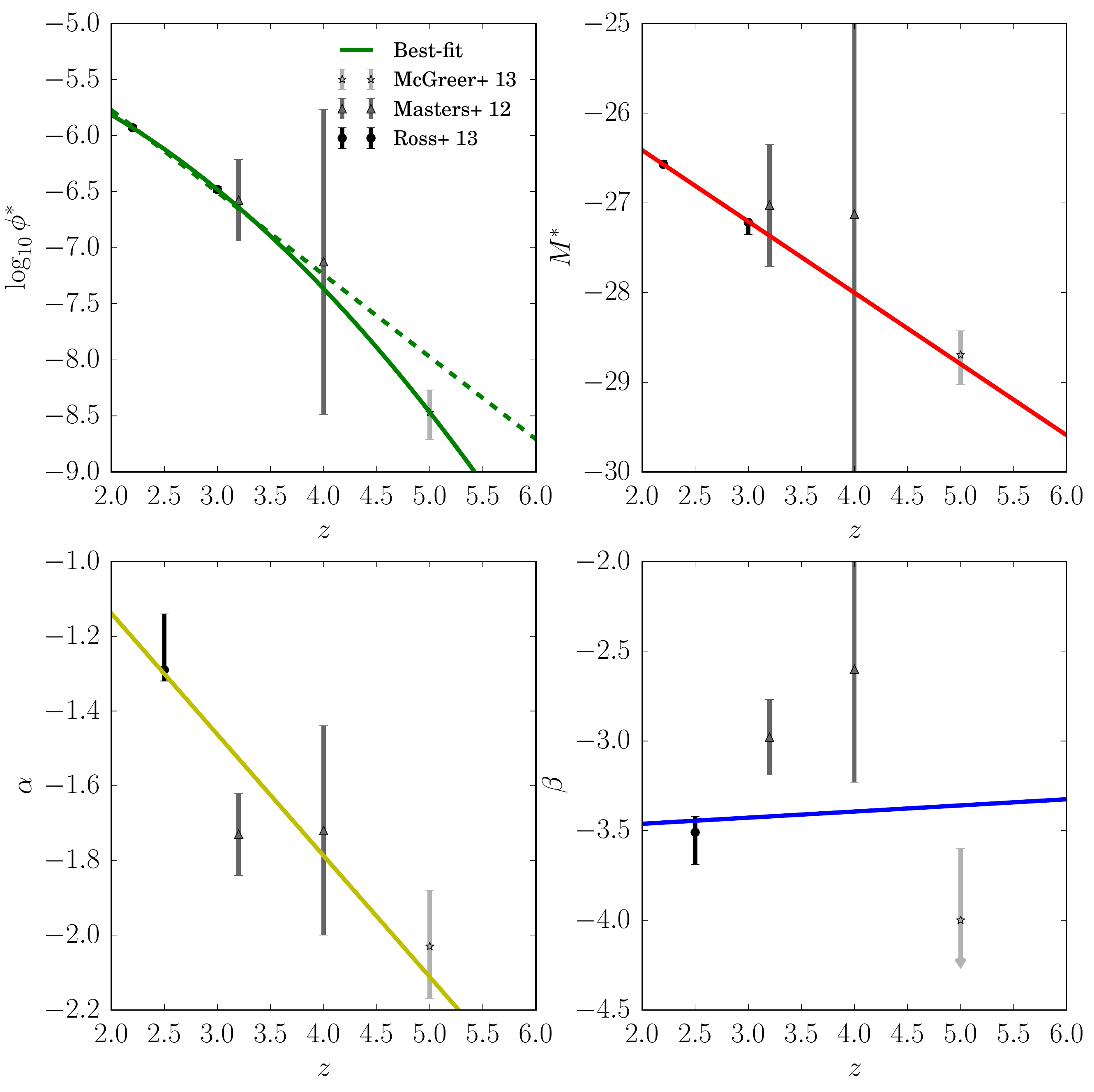}
  \caption{Evolution of the QLF parameters as a function of
    redshift: the base-ten logarithm of $\phi^*$ (top left), the break magnitude
    $M^*$ (top right), the faint-end slope $\alpha$ (bottom left), and the
    steep-end slope $\beta$ (bottom right). Best-fit values and associated
    1$\sigma$ errors from R13, M12, and M13 are represented as the black
    circles, dark gray triangles, and light gray stars, respectively. The solid
    lines show the parameterization of the parameters given by
    Equations~(\ref{eqn:logphiz}-\ref{eqn:betaz}) based on these data,
    reproduced in Table~\ref{table:parameters}. For the evolution of
    $\log_{10}\phi^*$, the dashed line shows the best-fit assuming only linear
    evolution in $z$ instead of quadratic, motivating an empirical need for
    quadratic evolution. See the text in Appendix~\ref{appendix:qlf} for further
    details.}
  \label{fig:parameters}
\end{figure}

In Paper~I, we provide a method for parameterizing the QLF as a function of
redshift that combines measurements from \citet{ross_etal2013},
\citet{masters_etal2012}, and \citet{mcgreer_etal2013} (hereafter referred to as
R13, M12, and M13). These observations provide fits for the QLF at redshifts
$2.2 \lesssim z \lesssim 3.5$, $z \sim 3.2$ and $z \sim 4$, and $z \sim 5$,
respectively. All three works parameterize the QLF as a double-power law,
defined by the same four parameters: $\phi^*$, the overall amplitude of the QLF
with units of Mpc$^{-1}$ mag$^{-1}$; $\alpha$, the slope of the faint end of the
QLF; $\beta$, the slope of the bright end; and $M^*$, the so-called break
magnitude where the QLF transitions between the slopes $\alpha$ and
$\beta$. Mathematically, the QLF can be written as
\begin{equation}
\phi(M) = \frac{\phi^*}{10^{0.4(\alpha+1)(M-M^*)} + 10^{0.4(\beta+1)(M-M^*)}}.
\label{eqn:qlf}
\end{equation}
To combine the R13, M12, and M13 data sets into a single set of quantities, we
cast the four parameters of the QLF ($\phi^*$, $M^*$, $\alpha$, and $\beta$) as
quantities that have evolution in redshift. We define these parameters as
\begin{subequations}
  \begin{align}
    \log_{10} \phi^*(z) &= \log_{10}\phi^*_0 + c_1 (z-3) + c_2 (z-3)^2, \label{eqn:logphiz} \\
    M^*(z) &= M^*_0 + c_3(z-3), \label{eqn:miz} \\
    \alpha(z) &= \alpha_0 + c_4(z-3), \label{eqn:alphaz}\\
    \beta(z) &= \beta_0 + c_5(z-3). \label{eqn:betaz}
  \end{align}
\end{subequations}
For the case of the overall normalization $\log_{10} \phi^*$, we include
quadratic evolution with redshift. The comoving number density of quasars is not
monotonic and peaks at $z \sim 2$ (\textit{e.g.}, Figure~20 of
\citealt{richards_etal2006}). Accordingly, there is a significant decrease in
the overall amplitude in the QLF at high redshifts, and the redshift evolution
is not well fit by a single linear term. Thus, observations suggest that the
redshift evolution of this parameter is not purely linear over such a large span
in redshift. (See Figure~\ref{fig:parameters} for a comparison between a linear
and quadratic fit.) The other parameters have redshift evolutions that are fit
adequately with simple linear evolution in redshift, and so we only include
linear terms to avoid overfitting.

\begin{deluxetable}{cl}
  \tablecaption{Best-fit Parameters in
    Equations~(\ref{eqn:logphiz}-\ref{eqn:betaz}) Given the Data Listed in R13,
    M12, and M13. \label{table:parameters}} \tablewidth{0.35\textwidth}
  \tablehead{\colhead{Parameter} & \colhead{Best-fit Value}} \startdata
  $\log_{10}\phi^*_0$ & $-6.48$ \\
  $c_1$               & $-0.776$ \\
  $c_2$               & $-0.109$ \\
  $M^*_0$             & $-27.2$ \\
  $c_3$               & $-0.795$ \\
  $\alpha_0$           & $-1.46$ \\
  $c_4$               & $-0.324$ \\
  $\beta_0$           & $-3.43$ \\
  $c_5$ & $\phantom{-}0.0342$
  \enddata
  \tablecomments{These parameters provide a fit to the luminosity function
    through redshift and ensure that the abundance of quasars matches
    observations as best possible. For additional details on the parameters and
    the fitting procedure, see the text in Appendix~\ref{appendix:qlf}.}
\end{deluxetable}

We now briefly summarize the relevant findings of R13, M12, and M13. In all
three results, the QLF is parameterized as a double-power law, according to
Equation~(\ref{eqn:qlf}). R13 uses quasars identified from SDSS-III Data Release
9 (DR9), and provides a luminosity-evolution density-evolution (LEDE) model in
which the base-10 logarithm of the QLF normalization, $\log_{10}\phi^*$, and the
break magnitude $M^*$, evolve linearly with redshift. The parameters $\alpha$
and $\beta$ are fixed as a function of redshift. Nominally, the LEDE fit is
valid over the redshift range $2.2 \leq z \leq 3.5$. M12 uses data from the
COSMOS survey, and measures the four QLF parameters at $z \sim 3.2$ and
$z \sim 4$.  M13 uses quasars identified in SDSS data in Stripe 82 (S82), and
reports the four QLF parameters at $z \sim 5$. For all three results, the
parameters themselves and their associated 1$\sigma$ uncertainties are
reported. The one exception to this is the value of $\beta$ from the M13
measurements, which was fixed to a value of $\beta=-4$. The authors report that
the value was fixed during the fits since allowing the bright-end slope to take
on any value would result in arbitrarily steep value of $\beta$. The authors of
M13 state that this is due to the low number count of objects at very bright
magnitudes. In order to prevent against the value from being fixed in our
composite QLF, we parameterize $\beta$ as being an upper limit, with 1$\sigma$
scatter above the value of $\beta=-4$ of $\sigma = 0.4$. This value is inferred
from Figure~18 of M13, which shows the joint likelihood of $\beta$ and $M^*$,
the break magnitude. At 68\% confidence, the authors report $\beta < -3.6$.

It should also be pointed out that M12 and M13 use different magnitude
conventions from the data in R13. Rather than reporting $M_i (z=2)$, the
absolute $i$-band magnitude at $z=2$, M12 and M13 report magnitudes as
$M_{1450}$, the absolute magnitude at 1450 \AA. In order to convert between
these two systems, we follow the convention of R13 and use
$M_i (z=2) = M_{1450} - 1.486$ \citep[Appendix B]{ross_etal2013}. We should
mention, however, that this conversion assumes a power-law slope of
$\alpha = 0.5$ ($f_\nu \propto \nu^{-\alpha}$) and changes slightly for
different spectral indices. Ultimately, the conversion between different
magnitude systems is not important for our overall conclusions because for most
of our simulation models, the observables we are most interested in (especially
the peak in the IGM temperature, Figure~\ref{fig:temp}) are dominated by the QLF
at redshifts $z \leq 3.5$. At these redshifts, the QLF is determined with very
small statistical uncertainty by the measurements of R13, and thus no conversion
between magnitude systems is necessary.

\subsection{Model Q1}
To combine the data from the different data sets, we fit for the four QLF
parameters independently as a function of redshift. The parameters are assumed
to vary linearly in redshift, except for the base-10 logarithm of the
normalization, which includes quadratic evolution. As explained above, we would
expect that a purely linear fit of this quantity is not adequate over such a
large range in redshift, since the total quasar number density peaks around
$z \sim 2$ and turns over. The equations for the parameters are given in
Equations~(\ref{eqn:logphiz}-\ref{eqn:betaz}), and the resulting best-fit values
for the parameters and uncertainties given in
Table~\ref{table:parameters}. Instead of fitting for the evolution of the four
parameters independently, it would be better to find a simultaneous fit to all
of the data spanning the entire redshift range. However, many degeneracies exist
between these parameters, and finding a simultaneous fit to adequately describe
all of the data over a very large redshift range is difficult to achieve.

Figure~\ref{fig:parameters} shows the measured parameters as a function of
redshift, as well as the best-fit line for each parameter. As explained above,
for the QLF normalization $\log_{10}\phi^*$ and break magnitude $M^*$, we
include the parameters from R13 at $z=2.2$, where the parameters are determined
best, and at $z=3$, in order to provide good constraints on the overall
normalization at a slightly higher redshift. Combined with the two points from
M12 ($z = 3.2$ and $z=4$) and the single point from M13 ($z=5$), there are five
total data points that are fit. We include $\alpha$ and $\beta$ from R13 at
$z=2.5$, since there is no explicit redshift dependence included in the R13
fits. Nevertheless, the reported parameters from higher redshift data (and even
when we compare the binned data from $z \sim 3$ in the R13 data) apparently show
no redshift evolution particularly for $\alpha$. Therefore, our model includes
redshift evolution in these parameters. At $z \sim 2.5$, there is very good
agreement between the binned QLF and the fit model of R13. However, the fit
values are ultimately not very sensitive to the choice of redshift. Combined
with the results from M12 and M13, this creates four data points to fit. The fit
for all of the parameters is reasonably good, with the notable exception of the
steep-end slope $\beta$. As mentioned earlier, constraining $\beta$ is
observationally difficult because of the low number count of objects. It is also
worth noting that the fits of M12 do not directly constrain $\beta$ with their
data. Their measurements from the COSMOS field are primarily for faint objects
and are fainter than the break magnitude $M^*$. In order to determine $\beta$ in
their fits, M12 use measurements from \citet{richards_etal2006} to provide
observations of bright objects. The overall result is little evolution in
$\beta$ over the redshift interval $2.5 \lesssim z \lesssim 5$, with perhaps a
slight steepening at lower redshifts. This trend is opposite to the trend of
$\alpha$, which shows a very clear trend of becoming shallower at lower
redshifts. Nevertheless, owing to the low overall amplitude of the luminosity
function at high magnitudes, the precise value of $\beta$ does not significantly
affect the predictions for reionization.

\subsection{Model Q2}
An alternative to finding the best-fit parameterizations is to simply
interpolate between the values reported in R13, M12, and M13. To this end, we
assume that the values for the parameters $\phi^*$, $M^*$, $\alpha$, and $\beta$
reported by the different studies are accurate for their respective redshift
ranges. Specifically, we use the values reported by R13 for redshifts
$z \leq 3.5$, the values of M12 at $z \sim 4$, and the values of M13 for
redshifts of $z \geq 5$. In order to determine values of the parameters at
intermediate redshifts, we linearly interpolate in redshift. This method
produces a QLF that is consistent with the different measurements by
construction, but can introduce some features into the QLF evolution through the
na\"{i}ve linear interpolation method. We therefore regard Model Q1 as our
fiducial model and present this alternative merely as a point of comparison.

\section{Galaxy Photoionization Rates}
\label{appendix:gamma_gal}
When presenting the methods of the simulations in Section~\ref{sec:features}, we
note that this work introduces the concept of renormalizing the photoionization
rate of galaxies $\Gamma_\mathrm{gal}$ on-the-fly by computing the
Ly$\alpha$ forest effective optical depth of the volume to match
observational constraints. In this appendix, we directly compare directly the
photoionization rates of the different simulations.

As explained in Section~\ref{sec:features}, the value of $\Gamma_\mathrm{gal}$
is increased or decreased to agree by construction with the value of
$\tau_\mathrm{HI}$ as determined by \citet{lee_etal2015}. However, if the value
of $\Gamma_\mathrm{gal}$ were to decrease below a value of
$\Gamma_\mathrm{gal} = 10^{-13}$~s$^{-1}$, then the photon output of quasars is
decreased instead. This decrease prevents $\Gamma_\mathrm{gal}$ from dropping
below levels that are far below the levels determined theoretically
\citep{haardt_madau2012} or observationally \citep{faucher-giguere_etal2008b}.

\begin{figure}[t]
  \centering
  \includegraphics[width=0.45\textwidth]{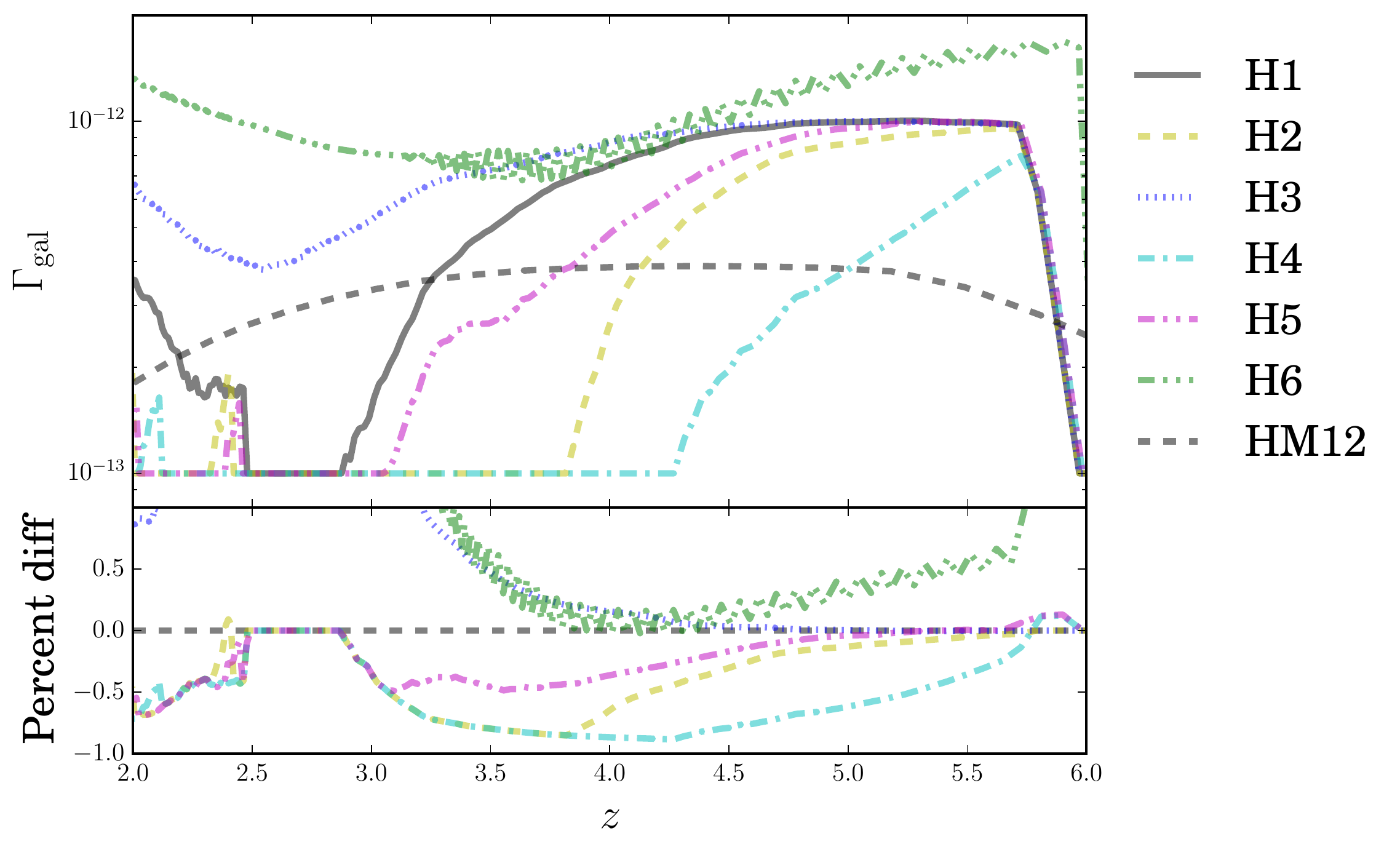}\\%
  \includegraphics[width=0.45\textwidth]{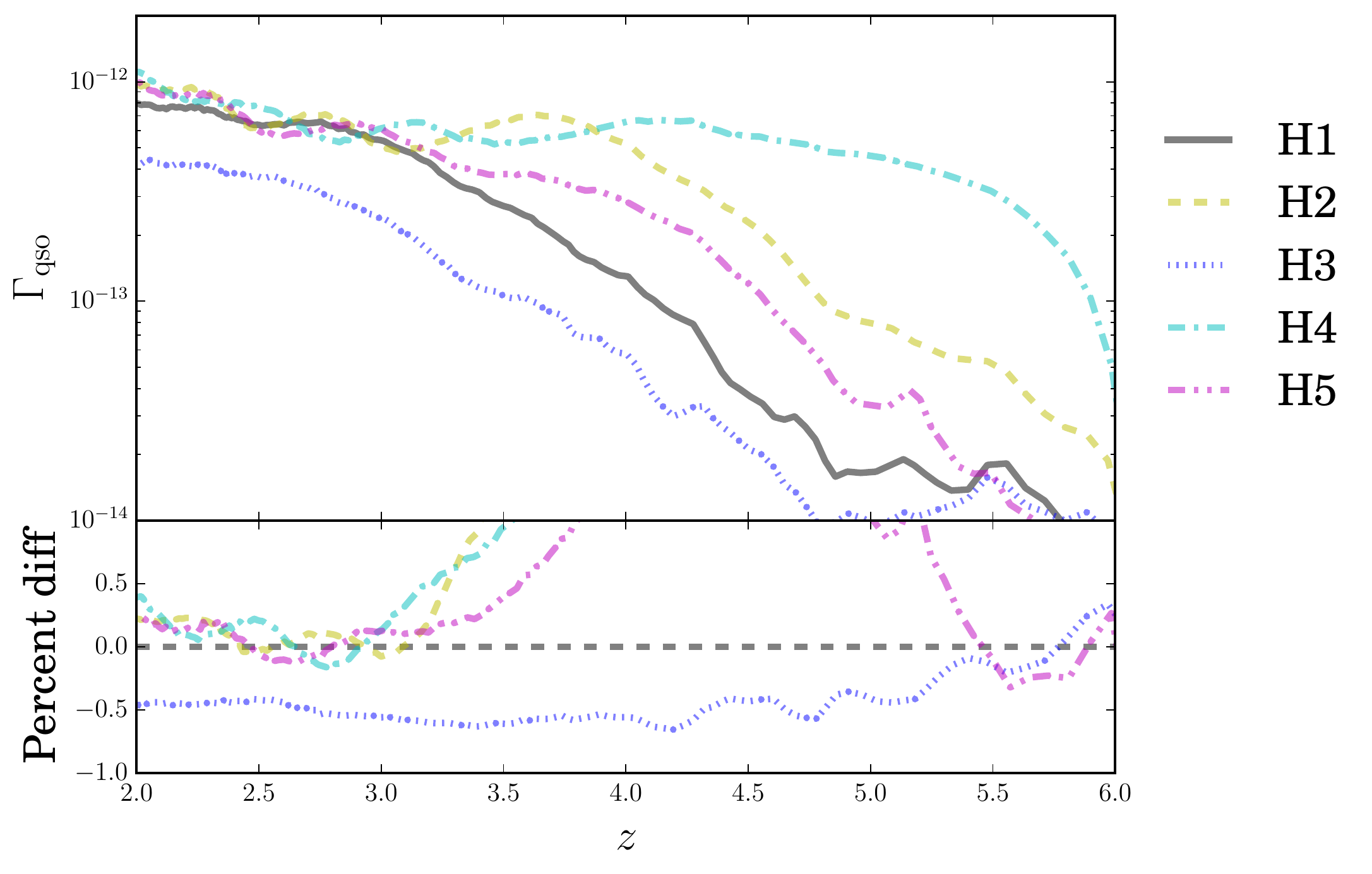}\\%
  \includegraphics[width=0.45\textwidth]{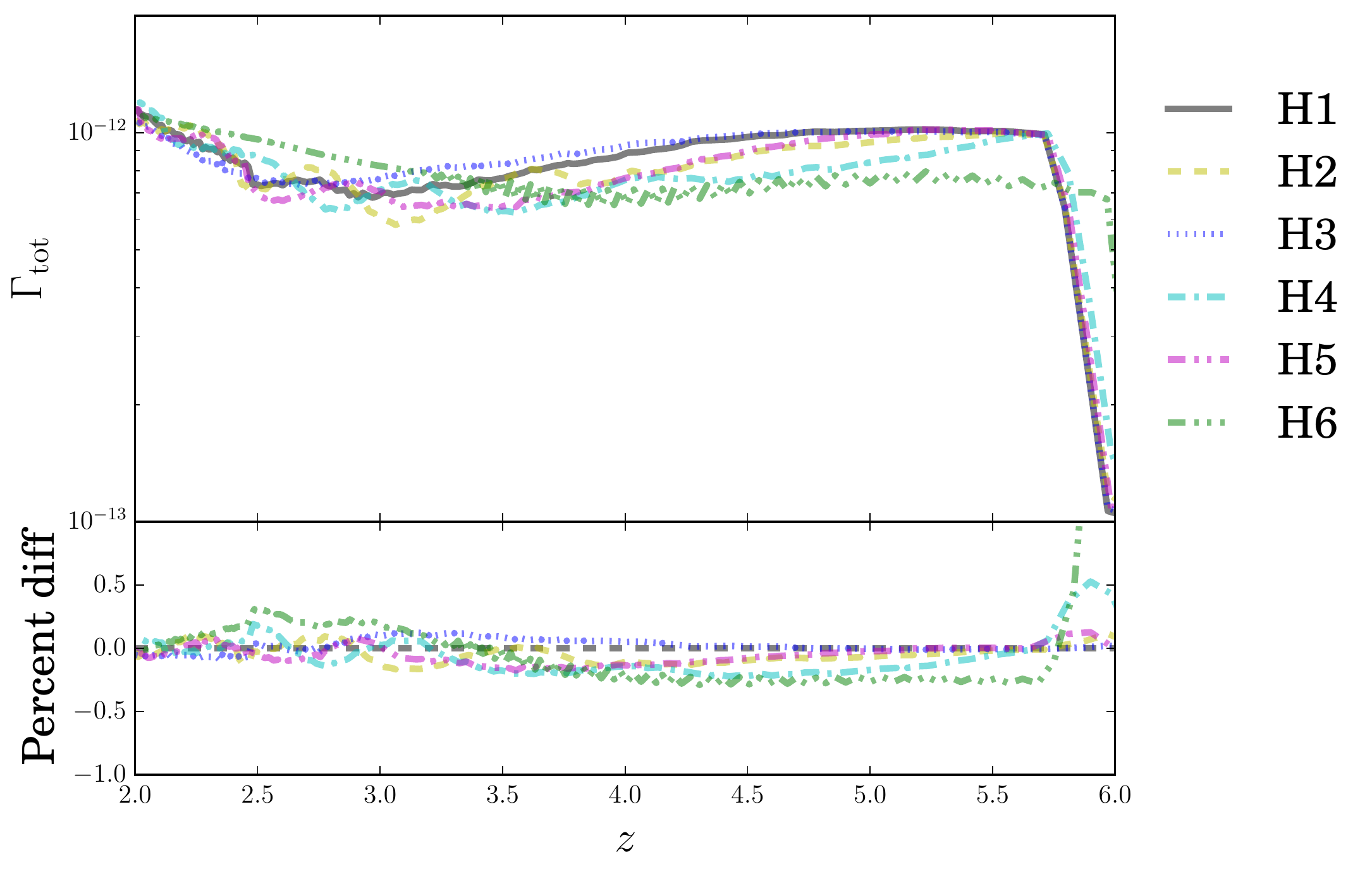}\\%
  \includegraphics[width=0.45\textwidth]{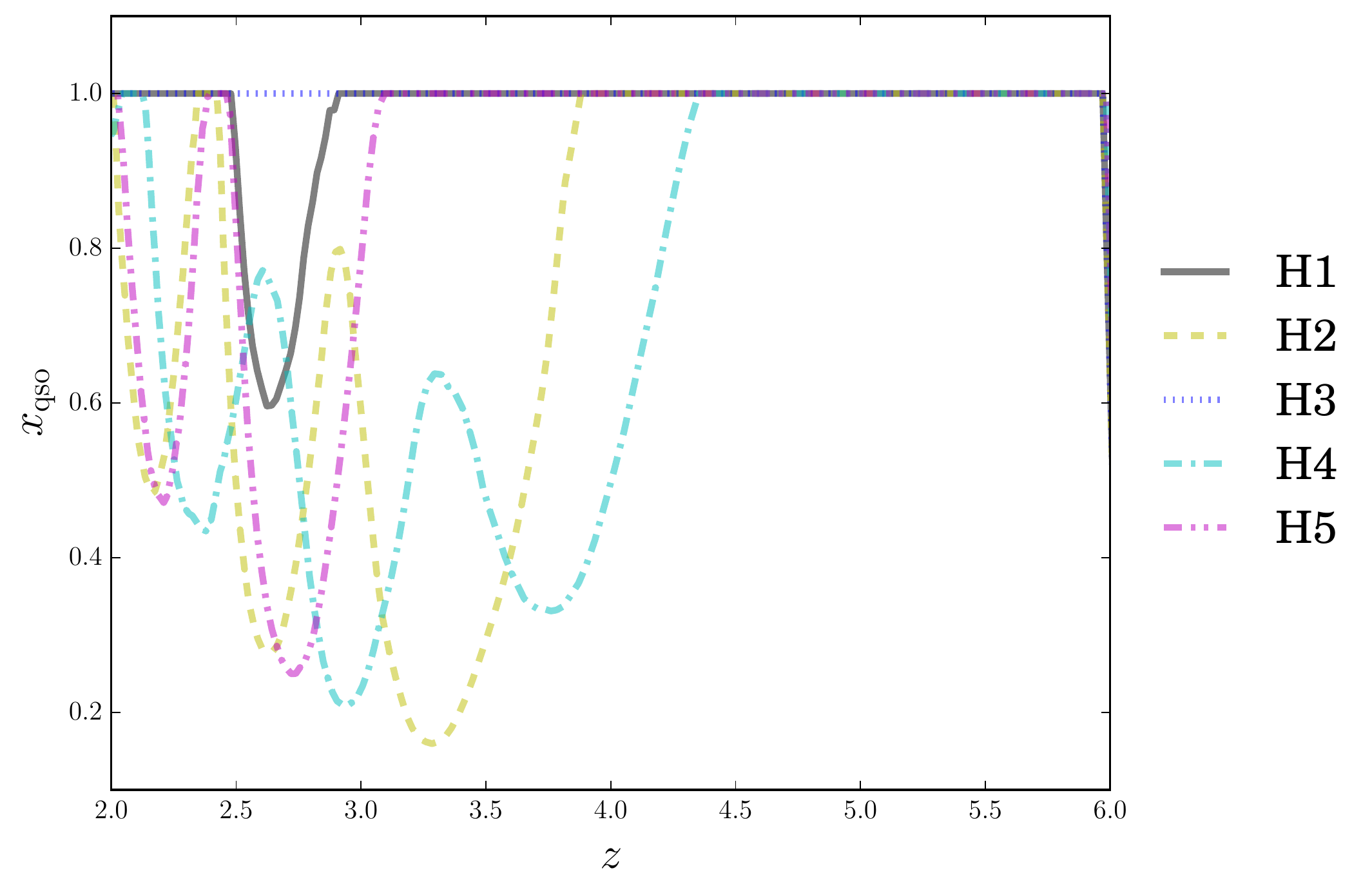}
  \caption{From top to bottom: the combined galaxy photoionization rate
    $\Gamma_\mathrm{gal}$, quasar photoionization rate $\Gamma_\mathrm{qso}$,
    total photoionization rate $\Gamma_\mathrm{tot}$, and quasar photon
    production rate $x_\mathrm{qso}$ as a function of redshift for each of the
    simulations presented in this work. In the top plot we also plot the value
    of $\Gamma_\mathrm{gal}$ from \citet{haardt_madau2012}. Simulation \uvb\
    does not have explicit quasar sources and so does not have values for
    $\Gamma_\mathrm{qso}$ or $x_\mathrm{qso}$. }
  \label{fig:gamma_gal}
\end{figure}

Figure~\ref{fig:gamma_gal} shows four different quantities as a function of
redshift: the photoionization rate from galaxies $\Gamma_\mathrm{gal}$, the
photoionization rate from quasars $\Gamma_\mathrm{qso}$, the total
photoionization rate
$\Gamma_\mathrm{tot} = \Gamma_\mathrm{gal} + \Gamma_\mathrm{qso}$, and the
multiplicative factor for quasar photon production $x_\mathrm{qso}$. This last
factor is applied to each quasar source at each time step in the simulation as
\begin{equation}
N_{\gamma,\mathrm{actual}} = x_\mathrm{qso} N_{\gamma,\mathrm{catalog}}.
\end{equation}
A value of $x_\mathrm{qso} = 1$ represents quasars producing as many photons as
dictated by the na\"{i}ve calculation of the catalog, with the value
$x_\mathrm{qso} < 1$ to ensure that
$\Gamma_\mathrm{gal} \geq 10^{-13}$~s$^{-1}$, while still being able to match
the overall Ly$\alpha$ forest optical depth (see Figure~\ref{fig:teffHI}).

The top panel of Figure~\ref{fig:gamma_gal} shows the value of
$\Gamma_\mathrm{gal}$ for all of the simulations, as well as for the
semi-analytic calculation of \citet{haardt_madau2012}. A minimum value of
$\Gamma_\mathrm{gal} = 10^{-13}$~s$^{-1}$ is imposed, which corresponds to
values of $x_\mathrm{qso} < 1$ in the bottom panel. Moreover, most of the
simulations have comparable values of $\Gamma_\mathrm{tot}$, to within about
10\%. This makes sense, since the value of $\Gamma_\mathrm{tot}$ is closely
related to $\tau_\mathrm{eff,HI}$, which is matched between the simulations by
construction.

\section{Number of Frequency Bins}
\label{sec:nfreq}

\begin{deluxetable}{cccc}
  \tablecaption{Frequency Bins Used in the Radiative Transfer
    Calculations. \label{table:freq}} \tablewidth{0pt}
  \tablehead{\colhead{Frequency Bin} & \colhead{Left Edge\tablenotemark{a}} &
    \colhead{Right Edge} & \colhead{Central Value}} \startdata
  \enddata
  1 & 13.6 & 24.6 & 17.6 \\
  2 & 24.6 & 54.4 & 34.2 \\
  3 & 54.4 & 65 & 59.3 \\
  4 & 65 & 75 & 69.7 \\
  5 & 75 & 125 & 94.2 \\
  6 & 125 & 250 & 168 \\
  7 & 250 & 1000 & 410 \\
  \tablenotetext{a}{All values are in eV.}  \tablecomments{This choice for the
    distribution of frequency bins shows agreement with $N=50$ frequency bins to
    about 2\% in observed temperature in a test simulation. This translates into
    a precision of about 400 K.}
\end{deluxetable}

An important parameter for radiative transfer calculations is the number of
frequency bins used in the calculation. This creates a trade-off in accuracy of
certain quantities in the simulation, most notably the temperature as a result
of photoheating, and computational resources required for
computation. Increasing the number of frequency bins requires additional memory
and computation time, which tend to scale linearly with the number of frequency
bins. Thus, finding the fewest number of bins for a given accuracy is of great
importance.

\begin{figure}[t]
  \centering
  \includegraphics[width=0.45\textwidth]{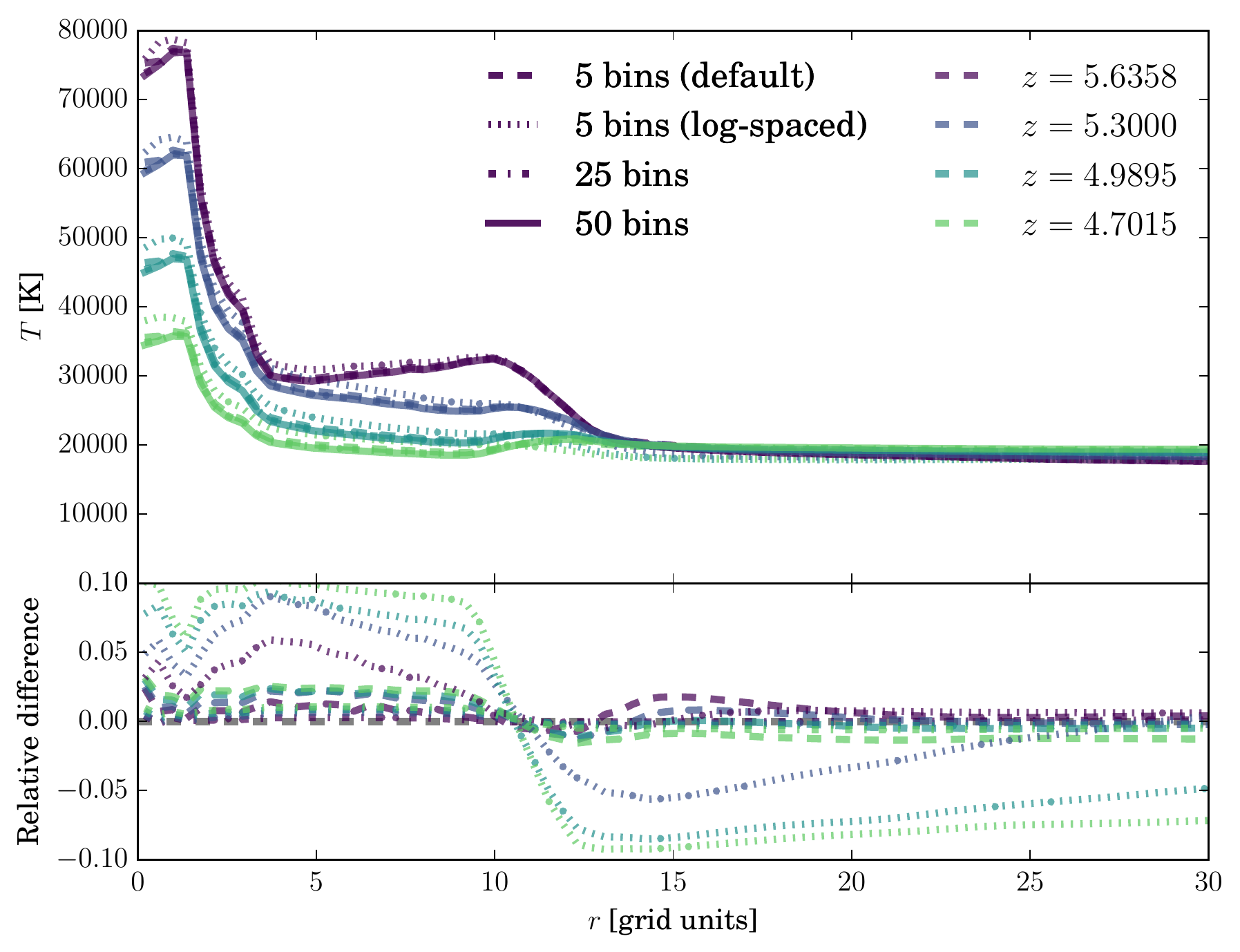} \\
  \includegraphics[width=0.45\textwidth]{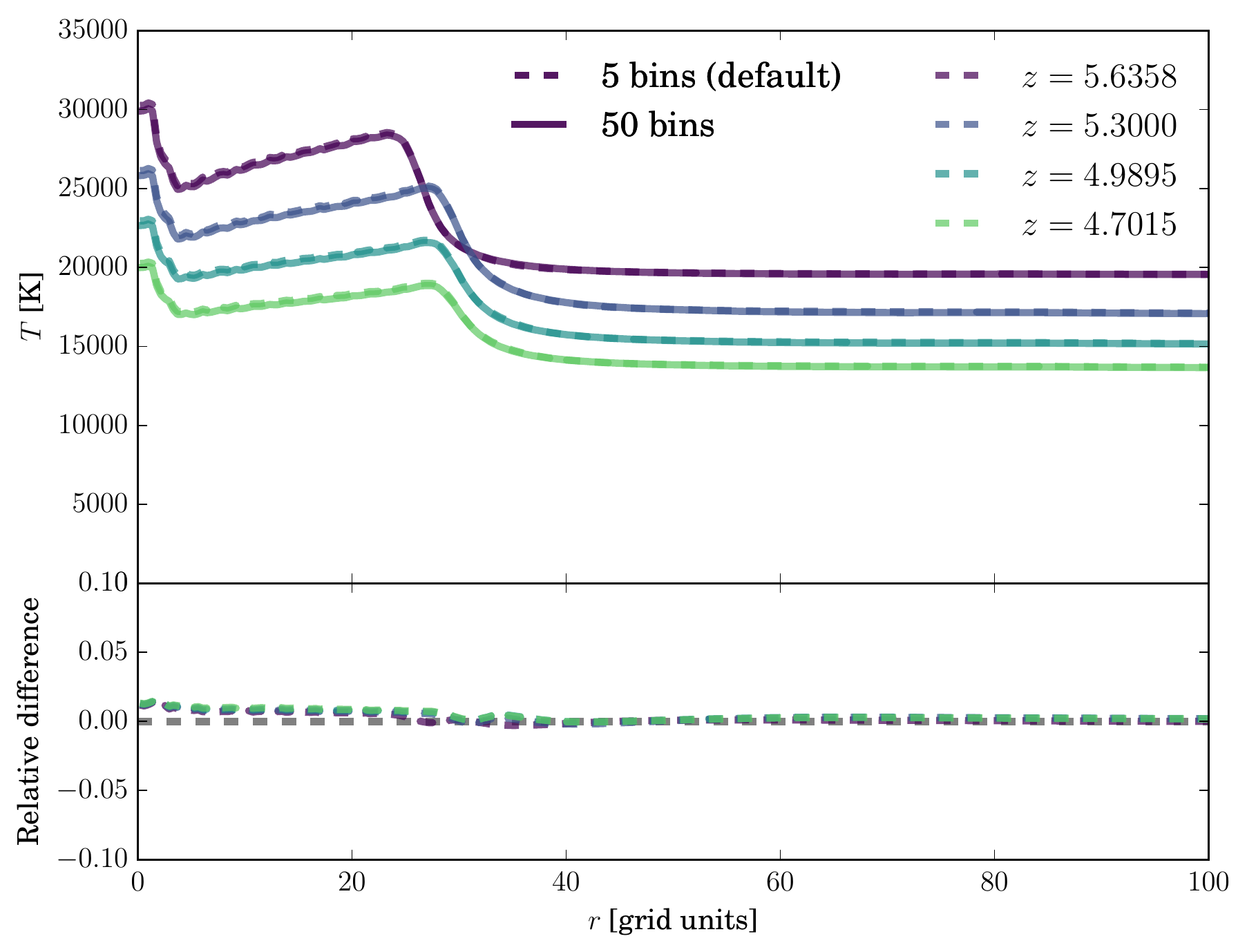}
  \caption{The temperature in spherically averaged shells at a distance of $r$
    (in grid units) from a single quasar source versus the computed temperature
    (in K) for different choices of the number of frequency bins. Top: a uniform
    medium of density $\Delta = 10$; bottom: density $\Delta = 1$. The different
    colors show different times of the simulation, and different line styles
    show the number of frequency bins used to compute helium reionization
    ($54.4~\mathrm{eV} \leq h\nu \leq 1~\mathrm{keV}$). The relative difference
    in the lower panel shows that the scheme chosen in the main body of the
    text, listed explicitly in Table~\ref{table:freq}, exhibits significantly
    better convergence than selecting bins which are more regularly spaced. The
    differences shown here correspond to absolute temperature differences of
    about 400 K for $\Delta \sim 10$, and less than 100 K for $\Delta \sim 1$.}
  \label{fig:freq_temp}
\end{figure}

In the simulations presented here, we have used a total of seven frequency bins
to accurately capture the thermal state of the IGM: one bin between the 13.6 and
24.6 eV (photons only energetic enough to ionize H~\textsc{i}), one bin between
24.6 and 54.6 eV (which ionize both H~\textsc{i} and He~\textsc{i}), and five
frequency bins between 54.4 eV and 1 keV (which are capable of ionizing
H~\textsc{i}, He~\textsc{i}, and He~\textsc{ii}). The distribution of the bins
is not regularly spaced but instead more finely samples the frequencies closer
to the He~\textsc{ii} edge (as a result of the steeply dropping cross-section of
He~\textsc{ii} with frequency).

A further consideration is which choice to make for the ``center'' of the
frequency bin, since this affects the actual amount of heat deposited in the gas
when we compute photoheating rates. For the bin centers, we choose the
energy-weighted photon number produced by quasars, which depends only on the
spectral index $\alpha$ and not the overall normalization. If the luminosity of
a quasar as a function of frequency is written as $L(\nu) = L_0 \nu^{-\alpha}$,
then the number of photons produced is
$N_\gamma = \int L_0 \nu^{-\alpha}/(h\nu) \dd{\nu}$. The energy-weighted number
of photons from quasars is therefore
\begin{equation}
\ev{N_\gamma}_{E} = \frac{\int h\nu \frac{L_0 \nu^{-\alpha}}{h\nu} \dd{\nu}}{\int \frac{L_0 \nu^{-\alpha}}{h\nu} \dd{\nu}}.
\end{equation}
By choosing this convention, the bin centers are optimal for calculating the
photoheating of gas. Table~\ref{table:freq} summarizes the properties of the
photon bins used in the body of the paper.

In order to test for the numerical convergence of temperature given the number
of bins we used, we ran a series of test simulations while increasing the number
of frequency bins we used. These simulations featured a single source placed at
the center of a uniform cubic volume. We compare the five bin edges used in this
simulation with five bins whose edges are distributed on logarithmically even
intervals. We also use a simulation with 25 logarithmically spaced intervals and
50 logarithmically spaced ones. These simulations with 50 frequency bins are
treated as the ``ground truth'' simulations for the purposes of comparison. This
approach is similar to the approaches discussed in the appendices of
\citet{mcquinn_etal2009} and \citet{compostella_etal2013}.

Figure~\ref{fig:freq_temp} shows the spherically averaged temperature as a
function of radius for the different number of frequency bins we used. The
different line colors show the simulations at different times (given as
redshifts), and different line styles represent different numbers of frequency
bins. The top panel shows the difference in temperature in a uniform medium of
constant density $\Delta = 10$ and the lower panel shows the same for
$\Delta = 1$. The lower axis in each panel shows the relative difference
compared to the $N = 50$ simulation. In general, the number and distribution of
the frequency bins used in the simulation show fairly good agreement with the
$N = 50$ simulations. In particular, a more na\"{i}ve choice of $N=5$
logarithmically spaced bins shows poor numerical convergence, differing from the
$N=50$ case by as much as 10\% at some places. The choice of bins used in the
body of the work deviates by at most about 2\%, a much more modest amount. In
terms of absolute temperature, this corresponds to a difference of about 400
K. The relative difference is much smaller in low-density regions, and so we
expect the temperature calculation to remain accurate for the majority of
regions probed by the Ly$\alpha$ statistics discussed in the body of the
paper. In summary, by choosing bins that oversample the frequency bins near the
He~\textsc{ii} edge, much better temperature convergence can be achieved than by
selecting a distribution that is more equitably sampled. At the same time, the
shape of the SED of quasars also has a significant effect on the inferred
temperature induced by photoheating and represents a much larger systematic
error than the number of frequency bins used in the radiative transfer scheme.

\end{appendix}

\bibliography{mybib}
\bibliographystyle{apj}

\end{document}